\documentclass[12pt]{article}
\usepackage{amssymb}

\usepackage{amsmath}
\usepackage{pst-all}
\usepackage{pstcol}
\usepackage[dvips]{graphicx}

\numberwithin{equation}{section}
\renewcommand{\baselinestretch}{1.4}
\begin{document}
\begin{titlepage}
\renewcommand{\baselinestretch}{1.1}
\small\normalsize
\begin{flushright}
%hep-th/\\
MZ-TH/10-32
\end{flushright}

\vspace{0.1cm}

\begin{center}

{\LARGE \textsc{Quantum gravity Effects\\[2mm]in the Kerr Spacetime}} % \\[6mm]Conformally Reduced Gravity}}

\vspace{1.4cm}
{\large M.~Reuter$^1$ and E.~Tuiran$^2$}\\

\vspace{0.7cm}
\noindent
\textit{$^1$Institute of Physics, University of Mainz\\
Staudingerweg 7, D--55099 Mainz, Germany, E-mail: reuter@thep.physik.uni-mainz.de}\\
\textit{$^2$Departamento de F\'{\i}sica, Universidad del Norte\\
Km 5 v\'{\i}a a Puerto Colombia, AA--1569 Barranquilla, Colombia, \\ E-mail: etuiran@uninorte.edu.co}\\
\end{center}

\vspace*{0.6cm}
\begin{abstract}
We analyze the impact of the leading quantum gravity effects on the
properties of black holes with nonzero angular momentum by performing a
suitable renormalization group improvement of the classical Kerr metric
within Quantum Einstein Gravity (QEG). In particular we explore the structure of
the horizons, the ergosphere, and the static limit surfaces as well as the
phase space avilable for the Penrose process. The positivity properties of
the effective vacuum energy momentum tensor are also discussed and the
``dressing'' of the black hole\'{}s mass and angular momentum are investigated by computing the corresponding
Komar integrals. The pertinent Smarr formula turns out to retain its
classical form. As for their thermodynamical properties, a modified
first law of black hole thermodynamics is found to be satisfied by the
improved black holes (to second order in the angular momentum); the
corresponding Bekenstein-Hawking temperature is not proportional to the
surface gravity.
\bigskip
\end{abstract}
\end{titlepage}

\section{Introduction}

During the past decade the gravitational average action \cite{mr} has been
used both as a framework within which the asymptotic safety scenario for a
consistent microscopic quantum theory of gravity can be tested \cite{mr}-\cite{livrev} and
as a convenient tool for finding the leading quantum gravity corrections to
various classical spacetimes. The latter investigations exploited the
effective field theory properties of the average action $\Gamma _{k}$ in an
essential way. It can be regarded as a one parameter family of effective
field theories, one for each value of the built-in infrared cutoff $k$ \cite
{avact}-\cite{ymrev}. In single scale problems involving a typical covariant
momentum scale $k$ a tree-level evaluation of $\Gamma _{k}$ encapsulates the
leading quantum effects at this scale. Thanks to this property the running
couplings contained in $\Gamma _{k}$ can be used in order to
``renormalization group improve'' the classical field equations or solutions
thereof \cite{bh1}-\cite{mof}.
The possibility of interpreting $\Gamma _{k}$ as a ``running effective field
theory'' distinguishes the effective average action \cite{avact} from
alternative functionals satisfying exact renormalization group (RG)
equations. The functional evolved by Polchinski\'{}s equation, for instance,
has the interpretation of a bare action. Therefore
it cannot be used for ``improvement'' purposes in the same way \cite{avactrev}.

\bigskip

Knowing the gravitational average action with some accuracy (i.e. , in some
truncation) means that we know the scale dependence of a set of generalized
gravitational couplings; typically it includes Newton\'{}s constant, for instance.
These running couplings can be used in order to ``RG improve'' classical spacetimes.
The basic idea is as follows. One starts by picking a solution of the classical field equation.
This solution will in general depend on the classical gravitational couplings. Then one
replaces the classical ones by their $k$-dependent counterparts and tries to
express the value of $k$ by means of a ``cutoff identification'' in terms of
the relevant geometrical or dynamical scale.

\bigskip

In refs. \cite{bh2,evap} this approach has been applied to stationary and
spherically symmetric, uncharged black holes. The classical starting point
was the Schwarzschild metric which involves the classical Newton\'{}s constant $%
G_{0}$ in the familiar way. The improvement consisted in replacing the
classical $G_{0}$ by the running Newton\'{}s constant $G\left( k\right) $
obtained from the functional \ RG equation for the effective average action.
A subtle point is finding a suitable cutoff identification. It should be
chosen in such a way that higher values of $k$ correspond to a ``zooming''
into the details of the black hole. One can try to find a meaningful
identification in the form $k=k\left( \mathcal{P}\right) $ which associates scales to
spacetime points $\mathcal{P}$. It is plausible that this map should be such that $k$
is smaller (larger) at larger (smaller) distances from the center of the
black hole. In the analogous situation in flat space one would set $%
k \propto 1/r$ where $r$ is the radial distance; with this identification
one can obtain the quantum corrected Coulomb potential from the $k$%
-dependence of the fine structure constant, for instance. In gravity the
assignment of scales to points should be diffeomorphism invariant, i.e. upon
introducing coordinates $x^{\mu }$ the relationship $k=k\left( \mathcal{P}\right) $
should be represented by a \textit{scalar} function $x^{\mu }\mapsto k\left(
x^{\mu }\right) $. In \cite{bh2,evap} the following class of cutoff
identification was considered:
\begin{equation}
k\left( \mathcal{P}\right) =\xi /d\left( \mathcal{P}\right)  \label{1.1}
\end{equation}
\begin{equation}
d\left( \mathcal{P}\right) =\int_{\mathcal{C}}\sqrt{\left| ds^{2}\right| }  \label{1.1B}
\end{equation}
Here $\xi $ is a constant of order unity and $d\left( \mathcal{P}\right) $ is a
distance scale typical of the point $\mathcal{P}$. According to (\ref{1.1B}) it is
given by the length of a certain curve $\mathcal{C}$. This curve is supposed to end at
$\mathcal{P}$, and to start at some reference point $\mathcal{P}_{0}$. The line element $ds^{2}$
refers to the classical metric. While diffeomorphism invariant by
construction, the above ansatz is still very general and different choices
are possible for $\mathcal{C}$. They correspond to different ways of ``re-focusing''
the ``microscope'' with which spacetime is observed when one goes from one
point to another. In refs. \cite{bh2, evap} a straight radial line from the
center to the point $\mathcal{P}$ has been employed, and this choice has been
motivated in detail. The only running parameter considered in this analysis
was Newton\'{}s constant. Its $k$-dependence had been assumed to be given by the formula
\begin{equation}
G\left( k\right) =\frac{G_{0}}{1+wG_{0}k^{2}}  \label{1.2}
\end{equation}
Here $G_{0}$ is the classical (macroscopic) Newton\'{}s constant, and $w$ is a
positive constant. This equation is a rather precise approximation to $%
G\left( k\right) $ as obtained from the Einstein-Hilbert truncation \cite{mr}
for all RG trajectories with a negligible cosmological constant in the
classical regime. According to (\ref{1.2}), the running Newton\'{}s constant
interpolates between $G_{0}$ for $k\rightarrow 0$ and the non-Gaussian fixed
point behavior $G\left( k\right) \propto 1/k^{2}\rightarrow 0$ for $%
k\rightarrow \infty $. With (1.1) inserted into (\ref{1.2}) we obtain the
position dependent Newton\'{}s constant
\begin{equation}
G\left( \mathcal{P}\right) =\frac{G_{0}d^{2}\left( \mathcal{P}\right) }{d^{2}\left( \mathcal{P}\right) +%
\bar{w}G_{0}}  \label{1.3}
\end{equation}
with $\bar{w}=w\xi ^{2}$.

\bigskip

The RG improved Schwarzschild metric was obtained by replacing $%
G_{0}\rightarrow G\left( \mathcal{P}\right) $ in the classical metric. It has been
analysed in great detail in \cite{bh2}. In particular, its horizon structure
was investigated. One finds that besides the usual Schwarzschild horizon
there exists a new inner horizon which merges with the (standard) outer one
at a critical value of the mass. An ``extremal'' black hole of this kind has
vanishing Hawking temperature. In fact the improvement suggests a very
attractive scenario for the final state of black hole evaporation: In the
early stages the temperature increases with decreasing mass, as predicted by
the conventional semiclassical analysis. However, once the mass approaches
the Planck mass, the quantum gravity effects reduce the temperature, and
ultimately ``switch off'' the Hawking radiation. For further details on the
RG improved Schwarzschild black hole we refer to \cite{bh2} and to \cite{evap}
where a dynamical picture of the evaporation process by means of a quantum
corrected Vaidya metric has been developed. The generalization to higher dimensions was considered in \cite{Falls}.

\bigskip

The purpose of the present paper is to perform a similar analysis for
rotating black holes. We shall construct and analyse an RG-improved version
of the Kerr metric. In Boyer-Lindquist coordinates the classical Kerr metric
reads \cite{BoyerL}
\begin{equation}
ds_{\text{class}}^{2}=-\left( 1-\frac{2MG_{0}r}{\rho }\right) dt^{2}+\frac{%
\rho ^{2}}{\Delta }dr^{2}+\rho ^{2}d\theta ^{2}+\frac{\Sigma \sin ^{2}\theta
}{\rho ^{2}}d\varphi ^{2}-\frac{4MG_{0}ra\sin ^{2}\theta }{\rho ^{2}}%
dtd\varphi  \label{1.4}
\end{equation}
Here we used the traditional abbreviations
\begin{equation}
\rho ^{2}\equiv r^{2}+a^{2}\cos ^{2}\theta  \label{1.5}
\end{equation}
\begin{equation}
\Delta \equiv r^{2}+a^{2}-2MG_{0}r  \label{1.6}
\end{equation}
\begin{equation}
\Sigma \equiv \left( r^{2}+a^{2}\right) ^{2}-a^{2}\Delta \sin ^{2}\theta
\label{1.7}
\end{equation}
Kerr black holes are characterized by two parameters, their mass $M$ and
angular momentum $J=aM$ \cite{Kerr63,Cohen,Carter}.

\bigskip

Applying the method outlined above we shall ``improve'' $ds_{\text{class}%
}^{2}\equiv g_{\mu \nu }^{\text{class}}dx^{\mu }dx^{\nu }$ by replacing $%
G_{0}\rightarrow G\left( k\right) $ and using a cutoff identification of the
type (\ref{1.1}). To start with, we are going to analyse various plausible curves $\mathcal{C}$, including
a straight radial line again, and discuss their physical properties.

\bigskip

The classical Kerr spacetime has two spherical horizons $H_{\pm }$ at the
radii \cite{Adler,Taylor}
\begin{equation}
r_{\pm }=m\pm \sqrt{m^{2}-a^{2}}  \label{1.8}
\end{equation}
and two static limit surfaces $S_{\pm }$ at
\begin{equation}
r_{S_{\pm }}\left( \theta \right) =m\pm \sqrt{m^{2}-a^{2}\cos ^{2}\theta }
\label{1.9}
\end{equation}
(Here
\begin{equation}
m\equiv MG_{0}  \label{1.10}
\end{equation}
denotes the ``geometric mass'' which actually has the dimension of a
length.) We shall discuss in detail the analogous critical surfaces
(horizons and static limit surfaces) of the improved metric. In particular
we demonstrate that, contrary to the Schwarzschild case, the improvement
does not lead to the formation of additional horizons.

\bigskip

As compared to the Schwazschild metric, the Kerr spacetime displays several
new features which are interesting from a conceptual point of view. One of
them is the existence of an ergosphere and the possibility of extracting
energy from the black hole via the Penrose process \cite{Taylor,MTW,Christ1}. We shall analyse in
detail how the quantum gravity effects influence the structure of the
ergosphere and the ``phase space'' available for the Penrose process.

Another new feature of the Kerr spacetime becomes apparent when one asks
whether the improved black holes still satisfy a set of (quantum corrected) laws of
black hole thermodynamics. In full generality this is an extremely difficult
question. Here we can only analyze whether there exists an entropy-like
state function satisfying a modified version of the first law. In the case
of Kerr black holes the space of states, labeled by $M$ and $J$, is
2-dimensional. As a result, it turns out that the mere \textit{existence} of an
entropy is a non-trivial issue. (For the Schwarzschild metric the space is
1-dimensional and so the existence of an entropy for the improved black hole
is guaranteed.) We shall see that, within the present approach, a state
function with the interpretation of an entropy can exist only if the
corresponding Hawking temperature is no longer proportional to the surface
gravity, as it is semiclassically. At least in the limit of small angular
momentum we shall find unambiguously defined relations $T=T\left( J,M\right)
$ and $S=S\left( J,M\right) $ for the temperature and entropy of the
improved rotating black holes.

\bigskip

The remaining sections of this paper are organized as follows. In section 2
we discuss the cutoff identification we are going to employ, and in section
3 we introduce the RG improved Kerr metric and analyse some of its general
properties; in particular we derive formulas for the modified static
limit and horizon surfaces, we reexpress the metric in a set of
appropiately generalized Eddington-Finkelstein coordinates, and compute the
surface gravity of the rotating quantum black holes. Then, in section 4 and
5 we analyse the detailed structure of the critical surfaces and the phase
space of the Penrose mechanism (negative energy states), respectively, usign
both analytical and numerical methods. In section 6 we reinterprete the
improved vacuum black hole as a classical one in presence of a certain kind
of fictitious matter which mimicks the quantum effects, and we investigate
the positivity properties of this matter system. In section 7 we show how
the ``bare'' mass and angular momentum of these black holes get ``dressed''
by the quantum effects in according with the antiscreening character of
Quantum Einstein Gravity. Finally, in section \ref{Seccion 8} we take a first step towards
an RG improved black hole thermodynamics; in particular we derive a modified
first law satisfied by the improved Kerr black holes. Section 9 contains a
summary of the results.

\bigskip

\section{The cutoff identification}

After replacing $G_{0}\rightarrow G\left( k\right) $ we would like to
express the scale $k$ as a scalar function on spacetime so that Newton\'{}s constant becomes position dependent:
\begin{equation}
G\left( r,\theta \right) \equiv G\Bigl( k=k\left( r,\theta \right) \Bigr)
\label{2.1}
\end{equation}
Here we have indicated that for symmetry reasons $k$ and $G$ can depend on
the Boyer-Lindquist coordinates $r$ and $\theta $ only. The classical
spacetime is stationary and invariant under rotations about the $z$-axis ;
we require that the corresponding Killing vectors \cite{Poisson,MTW}
\begin{equation}
\mathbf{t}\equiv t^{\mu }\partial _{\mu }=\frac{\partial }{\partial t},\;%
\boldsymbol{\varphi}\equiv \varphi ^{\mu }\partial _{\mu }=\frac{\partial }{%
\partial \varphi }  \label{2.2}
\end{equation}
are Killing vectors of the improved metric, too. If $G\left( x^{\mu }\right) $
is annihilated by $\mathbf{t}$ and $\boldsymbol{\varphi}$ this is indeed the case. In the
Boyer-Lindquist (BL) system this means that $G=G\left( r,\theta \right) $.

\bigskip

When an explicit form of the ``RG trajectory'' $G=G\left( k\right) $ is
needed we shall use the relationship (\ref{1.2}). However, for our mostly
qualitative discussion the precise details of this function are not
important. What matters is only that it smoothly interpolates between $G= const$ in the infrared $\left( k\rightarrow 0\right) $ and $G\left( k\right)
\propto 1/k^{2}$ in the ultraviolet $\left( k\rightarrow \infty \right) $.
Furthermore, we assume, as in the previous analyses \cite{bh2, evap} that $%
k\left( \mathcal{P}\right) =\xi /d\left( \mathcal{P}\right) $, which is given by the integral (\ref{1.1B}). In the case at hand it reads
\begin{equation}
d\left( r,\theta \right) =\int_{\mathcal{C}\left( r,\theta \right) }\sqrt{\left|
ds^{2}\right| }  \label{2.3}
\end{equation}
where $\mathcal{C}\left( r,\theta \right) $ is a path associated to the point $\mathcal{P}$ with
BL coordinates $\left( t,r,\theta ,\varphi \right) $. By stationarity and
axial symmetry, $\mathcal{C}$ and $d$ must not depend on $t$ and $\varphi$. The line
element $ds^{2}$ in (\ref{2.3}) is the one of the classical Kerr metric.

\bigskip

The choice for $\mathcal{C}$ which appears most natural is a radial path from the
origin to $\mathcal{P}$. Along this path, $dt=d\theta =d\varphi =0$ and, by (\ref{1.4}), 
$ds^{2}=\left( \rho ^{2}/\Delta \right) \;dr^{2}$. Hence we have in this
case
\begin{equation}
d\left( r,\theta \right) =\int_{0}^{r}d\bar{r}\sqrt{\left| \frac{\bar{r}%
^{2}+a^{2}\cos ^{2}\theta }{\bar{r}^{2}+a^{2}-2m\bar{r}}\right| }
\label{2.4}
\end{equation}
This integral is easy to perform only in the equatorial plane, i.e. for $%
\theta =\pi /2$. One obtains
\begin{equation}
d\left( r\right) \equiv d\left( r,\pi /2\right) =
\begin{cases}
d_{1}\left( r\right)  & \text{ if }r<r_{-} \\
d_{2}\left( r\right)  & \text{ if\ }r_{-}<r<r_{+} \\
d_{3}\left( r\right)  & \text{ if\ }r_{+}<r
\end{cases}
\label{2.5}
\end{equation}
where $r_{\pm }$ are the radii of the classical horizons given in (\ref{1.8}), and \cite{Tesis}
\begin{eqnarray}
d_{1}\left( r\right)  &=&\sqrt{r^{2}+a^{2}-2mr}+m\ln \left( \frac{-r+m-\sqrt{%
r^{2}+a^{2}-2mr}}{\left| a-m\right| }\right) -a  \label{2.6} \\
d_{2}\left( r\right)  &=&\frac{m}{2}\ln \left| \frac{m+a}{m-a}\right| -a-%
\sqrt{2mr-r^{2}-a^{2}} \\
&&+m\arctan \left( \frac{r-m}{\sqrt{2mr-r^{2}-a^{2}}}\right) +\frac{m\pi }{2} \nonumber
\\
d_{3}\left( r\right)  &=&\sqrt{r^{2}+a^{2}-2mr}+m\ln \left( r-m+\sqrt{%
r^{2}+a^{2}-2mr}\right) + \\
&&\pi m-a-m\ln \left| m-a\right| \nonumber
\end{eqnarray}
\centerline{}
\begin{center}
\begin{pspicture}(-1.2,2)(4.2,7.5)
\includegraphics{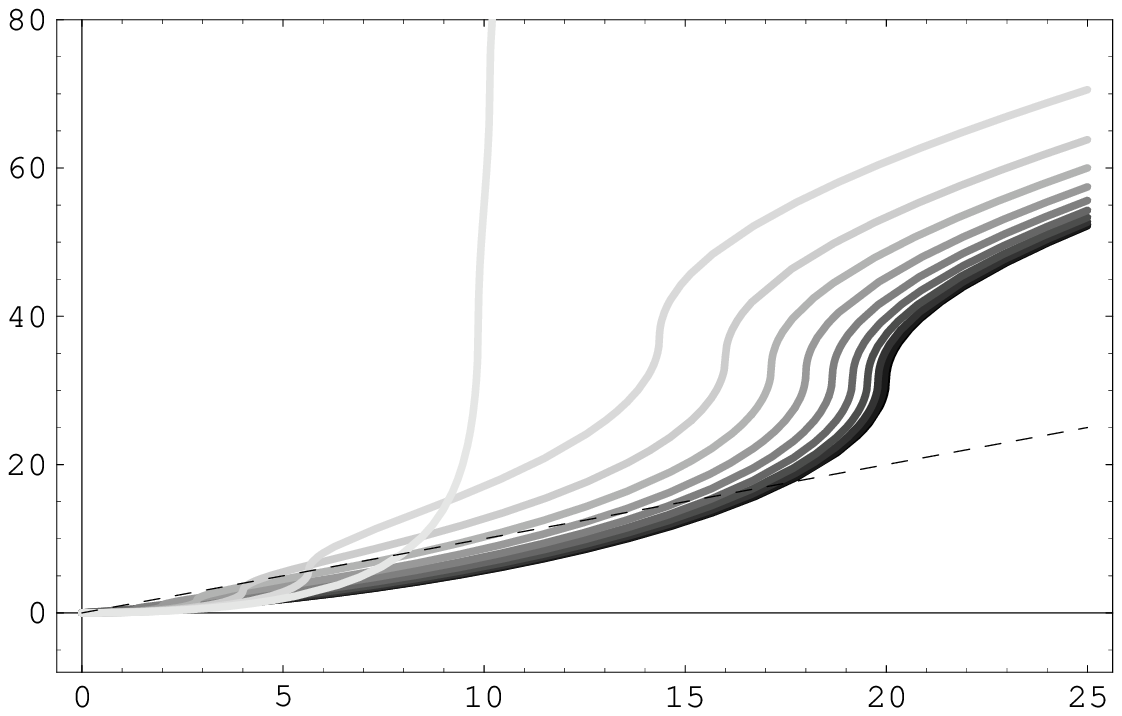}
\rput[l]{90}(-11.6,5.6){$d\left(r\right)$}
\rput[l](-10.25,8.6){$\theta=90^{\circ}$}
\rput[l](-8.25,8.6){$m=10$}
\rput[l](-6.85,7.6){$a \approx 10$}
\rput[l](-7.2,7.2){\tiny{(Extremal Case)}}
\rput[l](-6.95,6.9){\tiny{($r_-=r_+$)}}
\psline[linestyle=dashed,linewidth=.5pt](-8.18,3.1)(-8.18,3.5)
\psline[linestyle=dashed,linewidth=.5pt](-4.6,3.1)(-4.6,6)
\rput[l](-8.18,2.8){$r_-$ \tiny{(for $a=9$)}}
\rput[l](-4.6,2.8){$r_+$ \tiny{(for $a=9$)}}
\rput[l](-4.75,6.6){$a= 9$}
\rput[l](-4.45,5.6){$a=8$}
\rput[l](-2,5.6){$a=0$}
\rput[l](-3.3,4.1){$d\left(r\right)=r$}
\rput[l](-5.5,1.9){$r$}
%\rput[l](-6.3,1.3){Fig. 1.}
\rput[l](-12.5,0.8){Fig. 1:}
\rput[l](-10.8,0.8){The radial distance $d\left(r\right)$ in the equatorial plane for $m=10$}
\rput[l](-10.8,0.3){and various values of $a$. All quantities are expressed in Planck}
\rput[l](-10.8,-.2){units. The grayscale runs from black to gray for increasing $a$.}
%\rput[l](-10.8,-.7){for increasing $a$.}
\end{pspicture}
\end{center}
\centerline{}
\centerline{}
\centerline{}
\centerline{}
The function $d\left( r\right) $ for the equatorial plane is displayed in
Fig. 1 for a black hole with a mass of $10m_{\rm Pl}$ and for various values of
the angular momentum parameter $a$. (In this and the following figures all
dimensionful quantities are expressed in units of the Planckian quantities
formed with the infrared value of Newton\'{}s constant, $\ell_{\rm Pl}=m_{\rm Pl}^{-1}=\sqrt{G_{0}}$. Since $a,m,r$ and $d\left(
r\right) $ have the dimension of a length they are measured in units of $%
\ell_{\rm Pl}$. As $m\equiv G_{0}M$ by definition, the geometric mass $m$ equals
the actual mass $M$ when Planck units are used.)

The main features of the $d\left( r\right) $ curves are as follows. For $a<m$
not too close to the extreme case $a=m$, the curves run essentially parallel
to the dashed line in Fig. 1, representing the function $d\left( r\right) =r$%
. At their respective values of $r_{-}$ and $r_{+}$, all curves have a
vertical tangent. Near the classical horizon radii $r_{\pm }$ the functions $%
d\left( r\right) $ shift away from the $d\left( r\right) =r$ - line by a
kind of smoothed-out step function. At a sufficient distance from $r_{\pm }$
they run parallel to $d\left( r\right) =r$. In particular for $r>>r_{+}$ the
exact $d\left( r\right) $ is approximately of the form $d\left( r\right)
\approx r+\Delta d$ where $\Delta d$ is a constant independent of $r$.
Obviously, for $r$ large enough so that $\Delta d/r<<1$, we can approximate
the $d\left( r\right) $ curves simply by $d\left( r\right) =r$. For smaller $%
r$ there is the step-like behavior near $r_{-}$ and $r_{+}$, but in most of
our qualitative investigations it will not play a role. The deviations from $%
d\left( r\right) =r$ become significant when $a$ approaches $m$ which
corresponds to the situation of an extremal classical black hole.

For $\theta \neq \pi /2$ it is easy to evaluate the integral (\ref{2.4})
numerically. It turns out that $d\left( r,\theta \right) $ has a similar $r$%
-dependence for all values of $\theta $. For $\theta <\pi /2$ the shift $%
\Delta d$ is somewhat larger than at the equator, but nevertheless all
curves are essentially parallel to $d\left( r\right) =r$ again.

A more precise asymptotic analysis of the integral (\ref{2.4}) reveals that $%
d\left( r,\theta \right) $ has the following structure for $r\rightarrow
\infty $:
\begin{equation}
d\left( r,\theta \right) =r+m\ln \left( r\right) +F\left( \theta \right)
+O\left( \frac{1}{r}\right)  \label{2.7}
\end{equation}
There are three types of terms which do not vanish for $r\rightarrow \infty $%
: a linearly increasing one, a logarithmically increasing one, and one which
is $r$-independent. Among the three, only the $r$-independent one depends on
the angle $\theta $. Since $F\left( \theta \right) $ is subdominant we see
that, to logarithmic accuracy, $d\left( r,\theta \right) $ is actually
independent of $\theta $ at large $r$.

An alternative definition of the distance scale $d\left( r,\theta \right) $
could be as follows \cite{Taylor}. Let $\mathcal{C}\left( r,\theta \right) $ be a circular path of
coordinate radius $r$, contained in the $\theta =$const plane and centered
about the origin. In this case we define $d\left( r,\theta \right) $ to be
the \textit{reduced circumference} of this path, i.e.
its proper length divided by $2\pi $. In flat space the reduced
circumference would equal $r$; in the Kerr background there are
corrections. A detailed numerical analysis \cite{Tesis} shows that, for $r$ not too
small, and $a$ not too close to $m$, the resulting distance functions $%
d\left( r,\theta \right) $ have similar qualitative properties as those fom
the radial path.

For concreteness we shall use the distance function obtained from the radial
path whenever a concrete expression is needed. Since our analysis is mostly
at a qualitative or ``semi-quantitative'' level we shall be concerned with
leading order effects only. For this reason we shall neglect the subdominant
$\theta $-dependence of $d\left( r,\theta \right) $ and assume that $d\equiv
d\left( r\right) $ and, as a result $G,$ depends on $r$ only:
\begin{equation}
G\left( r\right) \equiv G\Bigl( k=\xi /d\left( r \right) \Bigr)  \label{2.8}
\end{equation}
The implications of the $\theta $-dependence are presumably too weak to be
accesible by our present method.

\bigskip

\section{General properties of the improved Kerr metric}

From now on we assume that we are given a $r$-dependent Newton\'{}s constant, $%
G=G\left( r\right) $. It may arise by inserting the cutoff identification $%
k\propto 1/d\left( r\right) $ into a solution of the RG equation such as (%
\ref{1.2}), but for most parts of our discussion the actual
origin of the $r$-dependence is irrelevant.

\subsection{The quantum corrected metric}

Substituting $G_{0}\rightarrow G\left( r\right) $ in (\ref{1.4}) we arrive
at the improved Kerr metric in BL coordinates
\begin{equation}
ds_{I}^{2}=g_{tt}dt^{2}+2g_{t\varphi }dtd\varphi +g_{rr}dr^{2}+g_{\theta
\theta }d\theta ^{2}+g_{\varphi \varphi }d\varphi ^{2}  \label{3.1.a}
\end{equation}
with the components
\begin{eqnarray}
g_{tt} &=&-\left( 1-\frac{2MG\left( r\right) r}{\rho ^{2}}\right)
\;,\;g_{rr}=\frac{\rho ^{2}}{\Delta _{I}\left( r\right) }\;,\;g_{\varphi
\varphi }=\frac{\Sigma _{I}\left( r,\theta \right) \sin ^{2}\theta }{\rho
^{2}}  \label{3.1.b} \\
g_{\theta \theta } &=&\rho ^{2}\;,\;g_{t\varphi }=-\frac{2MG\left( r\right)
ra\sin ^{2}\theta }{\rho ^{2}}
\end{eqnarray}
Here $\rho ^{2}\equiv r^{2}+a^{2}\cos ^{2}\theta $ is unchanged, but $\Delta
$ and $\Sigma $ contain $G\left( r\right) $ now:
\begin{equation}
\Delta _{I}\left( r\right) \equiv r^{2}+a^{2}-2MG\left( r\right) r
\label{3.2}
\end{equation}
\begin{equation}
\Sigma _{I}\left( r,\theta \right) \equiv \left( r^{2}+a^{2}\right)
^{2}-a^{2}\Delta _{I}\left( r\right) \sin ^{2}\theta   \label{3.3}
\end{equation}
For later use we also note the components of the inverse metric tensor:
\begin{eqnarray}
g^{tt} &=&-\frac{\Sigma _{I}}{\rho ^{2}\Delta _{I}}\;,\;g^{rr}=\frac{\Delta
_{I}}{\rho ^{2}}\;,\;g^{\varphi \varphi }=\frac{\Delta _{I}-a^{2}\sin
^{2}\theta }{\rho ^{2}\Delta _{I}\sin ^{2}\theta }  \label{3.4} \\
g^{\theta \theta } &=&\frac{1}{\rho ^{2}}\;,\;g^{t\varphi }=-\frac{2MG\left(
r\right) ra}{\rho ^{2}\Delta _{I}}
\end{eqnarray}
In the rest of this section we shall describe various general properties of
the metric (\ref{3.1.a}), (\ref{3.1.b}). The discussion parallels the
classical case to some extent \cite{Poisson}, but the results collected here
will be needed for an analysis of the quantum effects.

\subsection{Killing vectors and conserved quantitites}

We mentioned already that the improved metric has the Killing vector $\boldsymbol{t}$ and $\boldsymbol{\varphi}$ of eq. (\ref{2.2}). If we employ BL coordinates its components
are obviously
\begin{equation}
t^{\mu }=\delta _{t}^{\mu }\;,\;\varphi ^{\mu }=\delta _{\varphi }^{\mu
}\;\;\;\;\;\;\;\text{(BL)}  \label{3.5}
\end{equation}

Considering a point particle of mass $m$ which moves along the trajectory $%
x^{\mu }\left( \tau \right) $ with four-velocity $u^{\mu }\equiv dx^{\mu
}/d\tau $ and momentum $p^{\mu }=mu^{\mu }$ these Killing vectors imply a
conserved energy and angular momentum about the symmetry axis \cite{MTW}:
\begin{eqnarray}
E &=&-t_{\mu }p^{\mu }\equiv -mt_{\mu }u^{\mu }  \label{3.6} \\
L &=&-\varphi _{\mu }p^{\mu }\equiv -m\varphi _{\mu }u^{\mu }  \notag
\end{eqnarray}

\subsection{Zero angular momentum, static, and \\ stationary observers}

We consider three classes of special ``observers'' (actually point
particles) following a world line $x^{\mu }\left( \tau \right) $,
parametrized by the proper time $\tau $, with the velocity $u^{\mu }=dx^{\mu
}\left( \tau \right) /d\tau \equiv \dot{x}^{\mu }$, $u^{\mu }u_{\mu }=-1$.
(A dot will always denote the derivative with respect to $\tau $.)

\subsubsection{Zero angular momentum observers}

By definition zero angular momentum observers (or ``ZAMOs'') are particles
with vanishing $L$: $0=L=mg_{\mu \nu }\dot{x}^{\mu }\varphi ^{\nu }$. When
evaluated in BL coordinates, this condition reads $g_{t\varphi }\dot{t}+g_{\varphi \varphi }\dot{\varphi}=0$. Parametrizing the ZAMO\'{}s world
line by the coordinate time $t$ rather than the proper time $\tau $,
the condition assumes the form $g_{t\varphi }+g_{\varphi \varphi }\left(
d\varphi /dt\right) =0$. Therefore, introducing the angular velocity with
respect to the coordinate time,
\begin{equation}
\Omega \equiv \frac{d\varphi }{dt},  \label{3.7}
\end{equation}
as well as the convenient abbreviation
\begin{equation}
\omega \equiv \omega \left( r,\theta \right) \equiv -\frac{g_{t\varphi }}{%
g_{\varphi \varphi }}=-\frac{2G\left( r\right) Mar}{\Sigma _{I}}  \label{3.8}
\end{equation}
we conclude that even though they have no angular momentum, the ZAMOs rotate
around the $z$-axis with the angular velocity
\begin{equation}
\Omega ^{\text{ZAMO}}=\omega  \label{3.9}
\end{equation}
The quantity $\omega \geq 0$ is the coordinate angular velocity with which
inertial frames are dragged along \cite{Taylor,Lense,Adler,MTW}. It is affected by the $r$-dependence of $%
G $ on which it depends both explicitly and via $\Sigma _{I}$.

\subsubsection{Static observers}

By definition, the four-velocity of static observers is proportional to the
Killing vector $\boldsymbol{t}$, i.e. $u^{\mu }=\gamma t^{\mu }$ where $\gamma $ is chosen as $\gamma =\left[ -g_{\mu \nu }t^{\mu
}t^{\nu }\right] ^{-\frac{1}{2}}$ in order to achieve $u_{\mu }u^{\mu }=-1$.
The motion of static observers is not geodesic. To follow their world line
they will need a rocket engine say. Static observers exist only in those
portions of the improved Kerr spacetime in which $\boldsymbol{t}$ is timelike. The \textit{``static limit''}
is reached when $\boldsymbol{t}$ becomes null, i.e. when $\gamma ^{-2}=-g_{\mu \nu }t^{\mu }t^{\nu }=0$. In BL coordinates this is the
case where $g_{tt}=0$, or explicitly,
\begin{equation}
r^{2}-2G\left( r\right) Mr+a^{2}\cos ^{2}\theta =0  \label{3.10_}
\end{equation}
In the classical case the solution to this condition are two static limit
surfaces $S_{\pm }$ which can be parametrized as $r=r_{S_{\pm }}\left(
\theta \right) $ with $r_{S_{\pm }}\left( \theta \right) $ given in (\ref
{1.9}). For the improved metric the situation will be more complicated;
depending on the values of $M$ and $a$ there can be two or one, or no static
limit surface $S$ at all. Also in the improved case, since $g_{tt}=0$ on $S$%
, static limit surfaces are surfaces of infinite redshift.

\subsubsection{Stationary observers} \label{3.3.3}

A way of defining event horizons, different from their characterization as
one-way surfaces, is related to stationary observers. By definition a
stationary observer moves with a constant angular velocity $\Omega =d\varphi
/dt$ in the $\varphi $-direction. Its four-velocity is proportional to the
Killing vector $\boldsymbol{\xi}=\boldsymbol{t}+\Omega \boldsymbol{\varphi}$, i.e. $u^{\mu }=\gamma
\left( t^{\mu }+\Omega \varphi ^{\mu }\right) =\gamma \xi ^{\mu }$. This
class of observers is stationary in the sense that they perceive no time
variation of the gravitational field. They exist only if $\Omega $ and the
\textit{constant} parameters of their orbit, $r$ and $%
\varphi $, are such that $\gamma ^{-2}=-g_{\mu \nu }\xi ^{\mu }\xi ^{\nu }>0$%
. In BL coordinates this condition boils down to
\begin{equation}
q\left( \Omega \right) \equiv \Omega ^{2}-2\omega \Omega +g_{tt}/g_{\varphi
\varphi }<0  \label{3.11}
\end{equation}
If
\begin{equation}
\Omega _{\pm }=\omega \pm \sqrt{\omega ^{2}-g_{tt}/g_{\varphi \varphi }}
\label{3.12}
\end{equation}
is real, the function $q$ has two zeros on the real axis, and (\ref{3.11})
is satisfied if $\Omega _{-}<\Omega <\Omega _{+}$. Depending on whether $%
g_{tt}$, evaluated at the $\left( r,\theta \right) $-values of the orbit, is
negative, zero, or positive qualitatively different situations can
occur. The corresponding graph of $q\left( \Omega \right) $ is sketched in
Fig. 2. Let us discuss the 4 cases depicted there in turn.

\newpage

\begin{minipage}[t]{.55\linewidth}
\begin{center}
\begin{pspicture}(2.3,0.5)(5.3,6)
%\begin{pspicture}(2,0.5)(5,6)
\includegraphics[width=\linewidth]{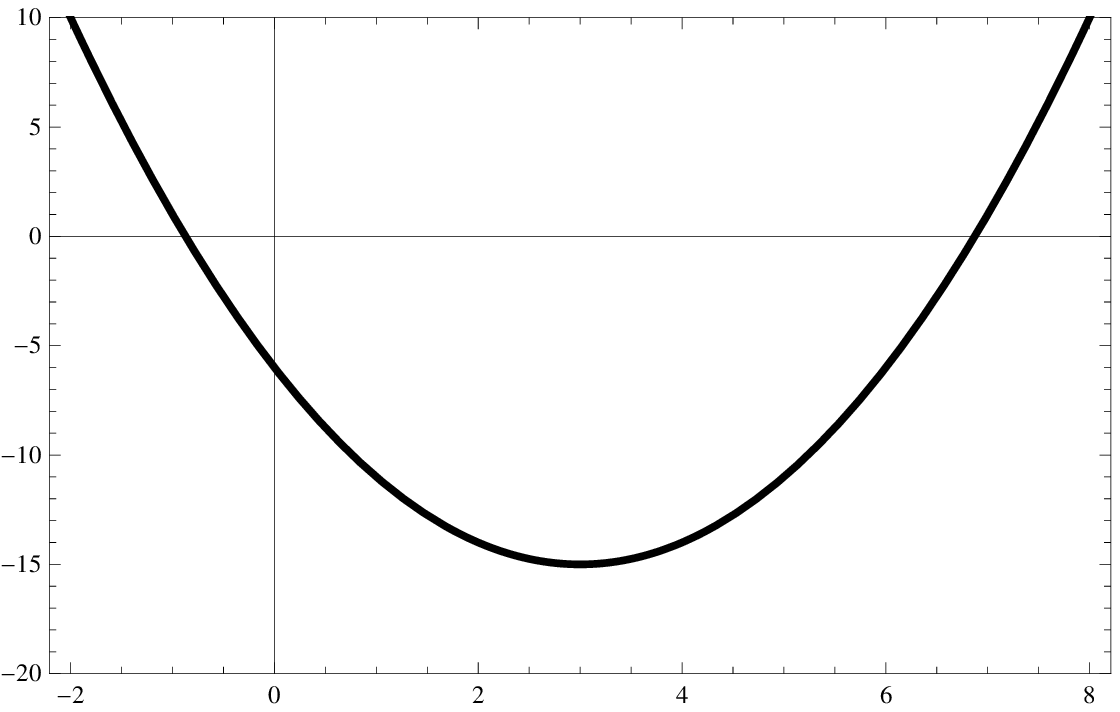}
%\rput[l]{90}(-8.3,2.3){$\Omega^2-2\omega\Omega+g_{tt}/g_{\phi\phi}$}
%\rput[l]{90}(-9,2.8){$\Omega^2-2\omega\Omega+g_{tt}/g_{\phi\phi}$}
\rput[l]{90}(-9,3.8){$q\left(\Omega\right)$}
\rput[l](-0.5,1.4){$\Omega$}
\rput[l](-8,5.55){$\Omega_{-}$}
\rput[l](-4.5,1.4){a)}
\end{pspicture}
\end{center}
\end{minipage}\hfill
\begin{minipage}[t]{.55\linewidth}
\begin{center}
\begin{pspicture}(5.3,0.5)(1.3,6)
%\begin{pspicture}(5,0.5)(1,6)
\includegraphics[width=\linewidth]{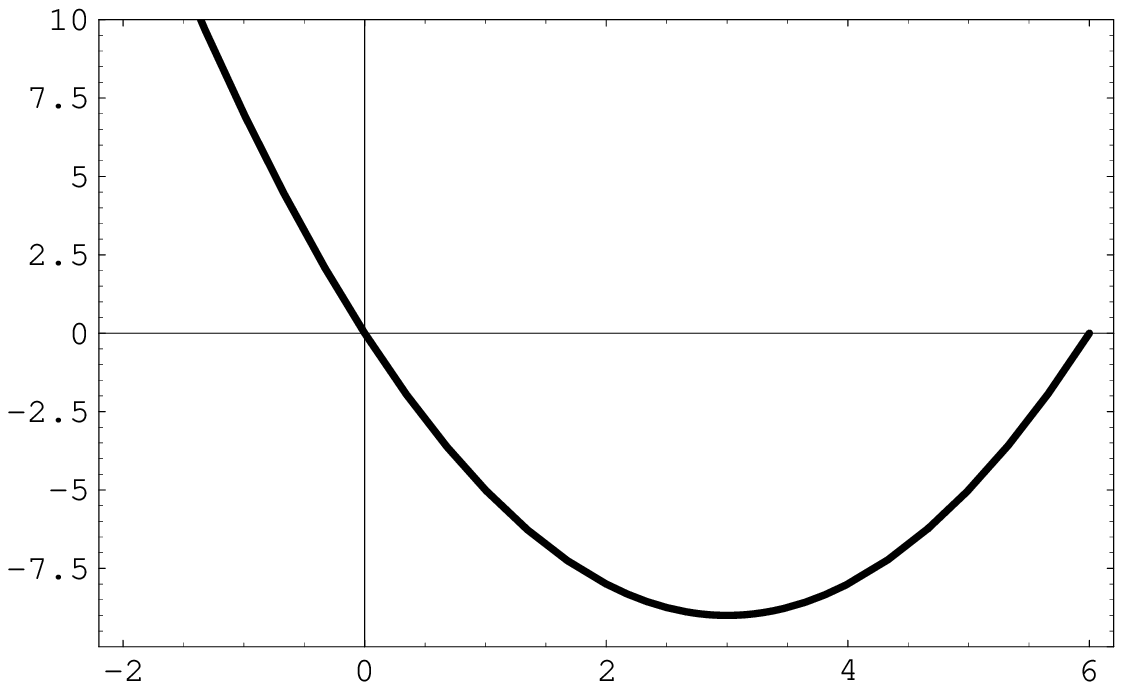}
%\rput[l]{90}(-8.3,2.3){$\Omega^2-2\omega\Omega+g_{tt}/g_{\phi\phi}$}
%\rput[l]{90}(-9,2.8){$\Omega^2-2\omega\Omega+g_{tt}/g_{\phi\phi}$}
\rput[l]{90}(-9,3.8){$q\left(\Omega\right)$}
\rput[l](-0.5,1.4){$\Omega$}
\rput[l](-6.7,4.75){$\Omega_{-}$}
\rput[l](-0.8,4.75){$\Omega_{+}$}
\rput[l](-4.5,1.5){b)}
\end{pspicture}
\end{center}
\end{minipage}\hfill
\begin{minipage}[t]{.55\linewidth}
\begin{center}
%\begin{pspicture}(1.2,0.5)(4.4,6)
\begin{pspicture}(1.2,1.5)(3.4,7)
\includegraphics[width=\linewidth]{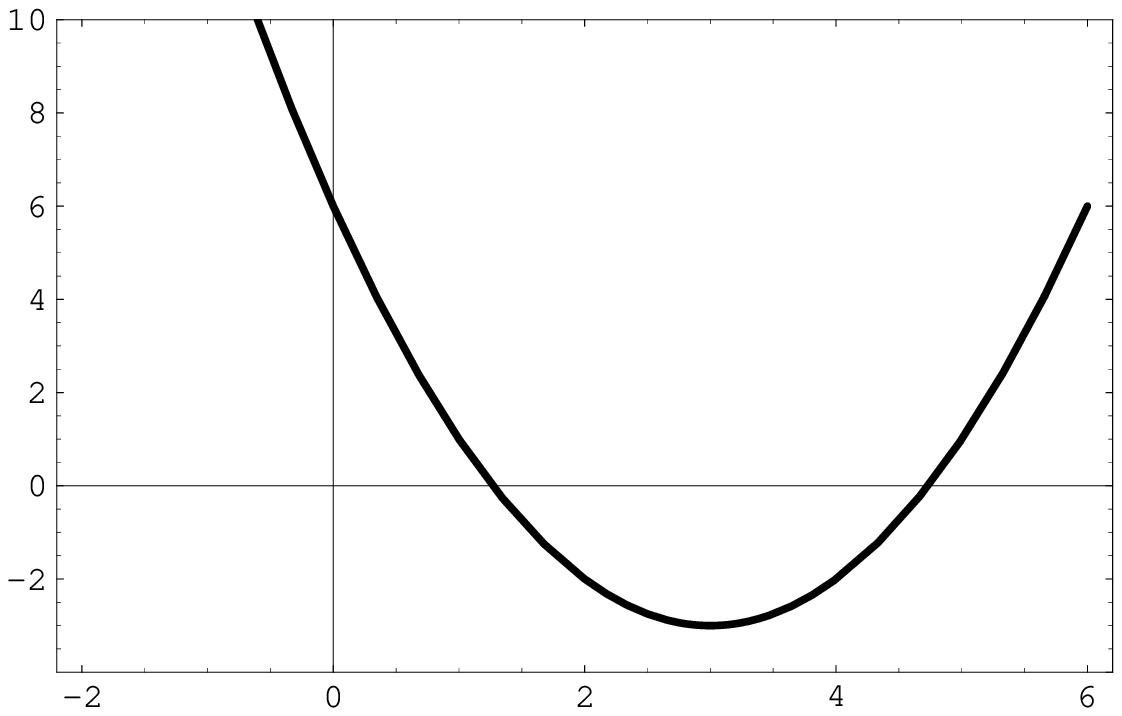}
%\rput[l]{90}(-8.3,2.3){$\Omega^2-2\omega\Omega+g_{tt}/g_{\phi\phi}$}
%\rput[l]{90}(-9,2.8){$\Omega^2-2\omega\Omega+g_{tt}/g_{\phi\phi}$}
\rput[l]{90}(-9,3.8){$q\left(\Omega\right)$}
\rput[l](-0.5,1.2){$\Omega$}
\rput[l](-5.9,3.62){$\Omega_{-}$}
\rput[l](-2,3.62){$\Omega_{+}$}
\rput[l](-4.5,1.2){c)}
\end{pspicture}
\end{center}
\end{minipage}\hfill
\begin{minipage}[t]{.55\linewidth}
\begin{center}
%\begin{pspicture}(.3,0.5)(3.3,6)
\begin{pspicture}(.3,1.5)(3.3,7)
\includegraphics[width=\linewidth]{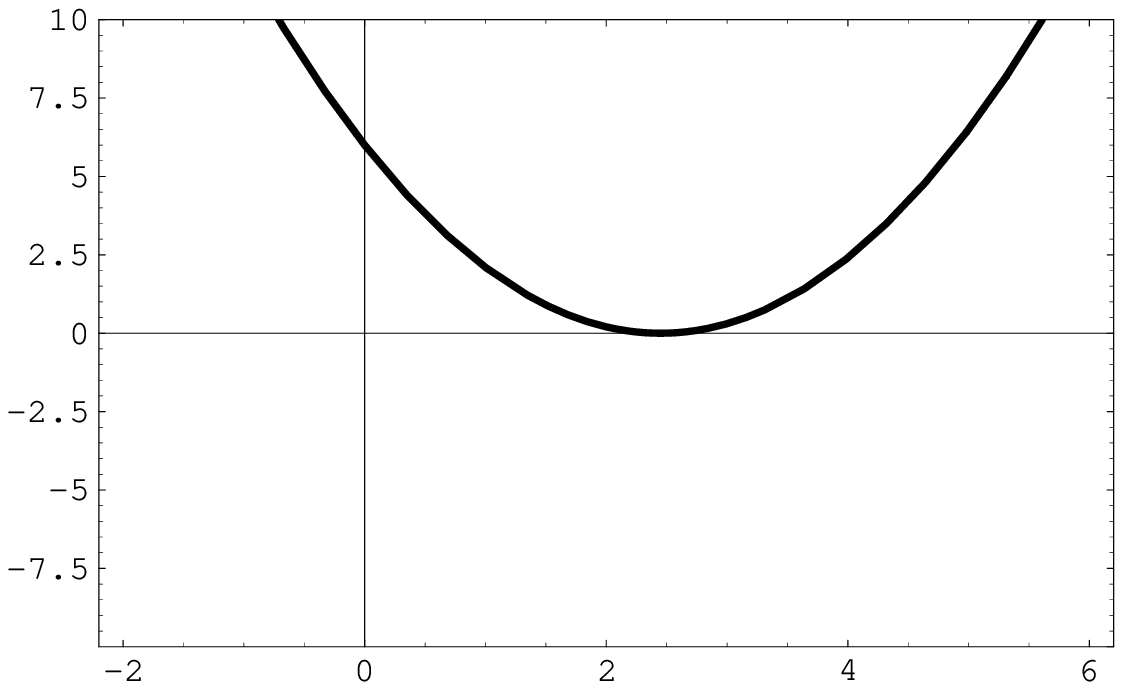}
%\rput[l]{90}(-8.3,2.3){$\Omega^2-2\omega\Omega+g_{tt}/g_{\phi\phi}$}
%\rput[l]{90}(-9,2.8){$\Omega^2-2\omega\Omega+g_{tt}/g_{\phi\phi}$}
\rput[l]{90}(-9,3.8){$q\left(\Omega\right)$}
\rput[l](-0.5,1.2){$\Omega$}
\rput[l](-4.85,4.25){$\Omega_{-}=\Omega_{+}=\Omega_{\text{H}}$}
\rput[l](-4.5,1.2){d)}
\end{pspicture}
\end{center}
\end{minipage}
%\centerline{}
%\centerline{}
%\centerline{}
%\centerline{}
\begin{center}
%\rput[l](0,-0.5){Fig. 2.}
\rput[l](-5.4,-1){Fig. 2:}
\rput[l](-3.7,-1){The function $q\left(\Omega\right)$ in the 4 cases discussed in the text.}
\end{center}
\centerline{}
\begin{enumerate}
\item The case $g_{tt}<0$: In this case $\sqrt{\omega
^{2}-g_{tt}/g_{\varphi \varphi }}=\sqrt{\omega ^{2}+\left| g_{tt}/g_{\varphi
\varphi }\right| }>\omega $ since $g_{\varphi \varphi }>0$ for all $r>0$ and
$\theta \neq 0,\pi $. Therefore since $\omega \geq 0$, it follows that $%
\Omega _{-}<0$ and $\Omega _{+}<0$. Stationary observers exist for $\Omega
\in \left( \Omega _{-},\Omega _{+}\right) $. Those with $\Omega \in \left(
\Omega _{-},0\right) $ are rotating in the opposite direction as the black
hole, those with $\Omega \in \left( 0,\Omega _{+}\right) $ rotate in the
same direction. Static observers correspond to the special case $\Omega =0$.
The case $g_{tt}<0$ is depicted in Fig. 2a.

\item  The case $g_{tt}=0$: In this case $\Omega _{-}=0$ and $\Omega
_{+}=2\omega >0$. There are no counter-rotating $\left( \Omega <0\right) $
observers any more; stationary observers are necesarily co-rotating with the
black hole. Counter-rotating light rays are bound to stay static with $\Omega \equiv
\Omega _{-}=0$. (see Fig.2b).

\item  The case $g_{tt}>0$: Here $\sqrt{\omega ^{2}-g_{tt}/g_{\varphi
\varphi }}<\omega $ and therefore $\Omega _{-}>0$ and $\Omega _{+}>0$. All
stationary observers are co-rotating with strictly positive angular velocity
$\Omega \in \left( \Omega _{-},\Omega _{+}\right) $. There are no static
observers.

\item  The case $\Delta _{I}=0$: Using the explicit form of the metric
components, the frequencies $\Omega _{\pm }$ can always be written as \cite{Poisson,Tesis}
\begin{equation}
\Omega _{\pm }=\omega \pm \frac{\Delta _{I}^{1/2}\rho ^{2}}{\Sigma _{I}\sin
\theta }  \label{3.13}
\end{equation}
This implies that when $\Delta _{I}=0$ the two frequencies become equal: $%
\left. \Omega _{+}\right| _{\Delta _{I}=0}=\left. \Omega _{-}\right|
_{\Delta _{I}=0}=\left. \omega \right| _{\Delta _{I}=0}=0$. At a radius $r$
such that $\Delta _{I}\left( r\right) =0$ stationary observers are forced to
rotate precisely with the angular velocity $\omega $ about the black hole.
The condition $\Delta _{I}=0$ is equivalent to $g^{rr}=0$. Therefore using
the same argument as classically \cite{Poisson}, one sees that it defines an
event horizon of the improved Kerr spacetime.
\end{enumerate}

We shall find that under the condition $M\gg m_{\rm Pl}$ the improved spacetime
has two spherical horizons $H_{\pm }$ and two limit surfaces $S_{\pm }$ exactly like the classical one. The radii of the static limit
surfaces, $r_{S_{\pm }}^{I}\left( \theta \right) \equiv r_{S}^{I}\left(
\theta \right) $, satisfy
\begin{equation}
g^{rr}=0\Leftrightarrow \left( r_{S}^{I}\right) ^{2}-2G\left(
r_{S}^{I}\right) Mr_{S}^{I}+a^{2}\cos ^{2}\theta =0  \label{3.14}
\end{equation}
while the radii of the horizons, $r_{_{\pm }}^{I}\equiv r_{H}^{I}$, are such
that
\begin{equation}
\Delta _{I}\left( r_{H}^{I}\right) =0\Leftrightarrow \left( r_{H}^{I}\right)
^{2}-2G\left( r_{H}^{I}\right) Mr_{H}^{I}+a^{2}=0  \label{3.15}
\end{equation}
The 4 surfaces can be ordered by increasing radius:
\begin{equation}
r_{S_{-}}^{I}\left( \theta \right) \leq r_{_{-}}^{I}\leq r_{_{+}}^{I}\leq
r_{S_{+}}^{I}\left( \theta \right)  \label{3.16}
\end{equation}
Here as always, the label ``$I$'' stands for ``improved''.

\bigskip

If one decreases $r$ at fixed $\theta $ the 4 cases occur in the above
order: For $r>r_{S_{+}}^{I}\left( \theta \right) $, outside the static
limit, case (1) is realized. At $r=r_{S_{+}}^{I}\left( \theta \right) $ we
have $g_{tt}=0$ and case (2) applies. Between $S_{+}$ and $H_{+}$, in the
ergosphere, we have (3), and all stationary observers with $r\in \left(
r_{_{+}}^{I},r_{S_{+}}^{I}\left( \theta \right) \right) $ necessarily rotate
in the direction of the black hole. When we approach $r=r_{_{+}}^{I}$ from
above the only allowed angular velocity is
\begin{equation}
\Omega _{+}=\Omega _{-}=\omega \left( r_{_{+}}^{I},\theta \right) \equiv
\Omega _{H}  \label{3.17}
\end{equation}
For $r<r_{_{+}}^{I}$ , there exist no stationary observers any longer: once
it has crossed the horizon $H_{+}$, a particle necessarily falls into the
black hole. We shall refer to $\Omega _{H}$ as ``the angular velocity of the
black hole''. Noting that $\Omega _{H}=2G\left( r_{+}^{I}\right)
Mar_{+}^{I}/\Sigma _{I}\left( r_{+}^{I},\theta \right) $ with $\Sigma
_{I}\left( r_{+}^{I},\theta \right) =\left[ \left( r_{+}^{I}\right)
^{2}+a^{2}\right] ^{2}-a^{2}\Delta _{I}\left( r_{+}^{I}\right) \sin
^{2}\theta =\left[ \left( r_{+}^{I}\right) ^{2}+a^{2}\right] ^{2}$ we
observe that $\Omega _{H}$ is actually independent of the angle $\theta $
and depends only on the parameters $M$ and $a$:
\begin{equation}
\Omega _{H}\left( M,a\right) =\frac{a}{r_{+}^{I}\left(M,a\right) ^{2}+a^{2}}
\label{3.18}
\end{equation}
This formula looks like its classical counterpart \cite{Poisson}; however, the improvement
changes the $M$ and $a$ dependence of $r_{+}^{I}$.

\subsection{Generalized Eddington-Finkelstein coordinates}

The systems of Boyer-Lindquist coordinates $\left( t,r,\theta ,\varphi
\right) $ breaks down when $\Delta _{I}=0$ i.e. on a possible horizon. In
order to reexpress the improved Kerr metric in a system of coordinates which
remains regular there, we define a generalization of the familiar advanced
time (or ingoing) Eddington-Finkelstein (EF) coordinates \cite{Poisson,MTW}:
\begin{eqnarray}
v =t+r^{\ast }\left( r\right)\;,\; r =r\;,\;\theta =\theta\;,\;
\psi  =\varphi +r^{\#}\left( r\right) \label{3.19}
\end{eqnarray}
Here the functions $r^{\ast }$ and $r^{\#}$ are given by
\begin{eqnarray}
r^{\ast }\left( r\right)  &\equiv &\int^{r}dr^{\prime }\frac{r^{\prime
2}+a^{2}}{\Delta \left( r^{\prime }\right) }=\int^{r}dr^{\prime }\frac{%
r^{\prime 2}+a^{2}}{r^{\prime 2}-2Mr^{\prime }G\left( r^{\prime }\right)
+a^{2}}  \label{3.20} \\
r^{\#}\left( r\right)  &\equiv &\int^{r}dr^{\prime }\frac{a}{\Delta \left(
r^{\prime }\right) }=\int^{r}dr^{\prime }\frac{a}{r^{\prime 2}-2Mr^{\prime
}G\left( r^{\prime }\right) +a^{2}}  \notag
\end{eqnarray}
For a constant $G\left( r\right) $ these integrals can be performed in
closed form. For the improved metric this is not possible in general.
Luckily the explicit forms of $r^{\ast }$ and $r^{\#}$ are not needed in
order to express the metric in terms of the new coordinates $x^{\mu }=\left(
v,r,\theta ,\psi \right) $. It is enough to use that by (\ref{3.20}) $%
dt=dv-\left( r^{\prime 2}+a^{2}\right) \Delta _{I}^{-1}dr$ and $d\varphi
=d\psi -a\Delta _{I}^{-1}dr$. Inserting these differentials into (\ref{3.1.a}) we obtain the following line element for the improved Kerr metric in
ingoing EF coordinates:
\begin{eqnarray}
ds_{I}^{2} &=&-\left( 1-\frac{2G\left( r\right) Mr}{\rho ^{2}}\right)
dv^{2}+2drdv-2a\sin ^{2}\theta d\psi dr+  \label{3.21} \\
&&-\frac{4G\left( r\right) Mar\sin ^{2}\theta }{\rho ^{2}}d\psi dv+
\frac{\Sigma _{I}\sin ^{2}\theta }{\rho ^{2}}d\psi ^{2}+\rho ^{2}d\theta ^{2} \notag
\end{eqnarray}

We shall also need the Killing vector $\boldsymbol{\xi} =\boldsymbol{t}+\Omega _{H}\boldsymbol{\varphi}$ in EF
coordinates. It is trivial to see that \ $\boldsymbol{\xi} =\frac{\partial }{\partial v}%
+\Omega _{H}\frac{\partial }{\partial \varphi }$ , i.e.
\begin{equation}
\xi ^{v}=1\;,\;\xi ^{r}=0\;,\;\xi ^{\theta }=0\;,\;\xi ^{\psi }=\Omega _{H}
\label{3.22}
\end{equation}
Using the metric (\ref{3.21}) one obtains the following expression for the
square $\boldsymbol{\xi} ^{2}=g_{\mu \nu }\xi ^{\mu }\xi ^{\nu }$:
\begin{equation}
\boldsymbol{\xi} ^{2}=\frac{\Sigma _{I}\sin ^{2}\theta }{\rho ^{2}}\left( \omega -\Omega
_{H}\right) ^{2}-\frac{\rho ^{2}\Delta _{I}}{\Sigma _{I}}  \label{3.23}
\end{equation}
This scalar function is well defined both away from and directly on $H_{+}$.
In fact, it vanishes on the horizon, $\left. \boldsymbol{\xi} ^{2}\right| _{H_{+}}=0$,
since $\Delta _{I}=0$ and $\omega =\Omega _{H}$ there. This is exactly as it
should be: In subsection \ref{3.3.3} we saw that $\gamma ^{-2}=-$ $\boldsymbol{\xi} ^{2}\propto q\left( \Omega _{H}\right) $ , and since $q\left( \Omega _{H}\right) =0$,
the Killing vector becomes null on the horizon.

\subsection{Quantum corrections to the surface gravity}

As the improved Kerr metric admits a Killing vector which is null at the
event horizon and tangent to the horizon\'{}s null generators we may define the surface gravity $\kappa $ in the usual
way \cite{Poisson}:
\begin{equation}
-D_{\mu }\boldsymbol{\xi} ^{2}\left( r_{+}^{I}\right) =2\kappa \xi _{\mu }\left(
r_{+}^{I}\right)  \label{3.24_}
\end{equation}
To determine $\kappa $ we shall evaluate (\ref{3.24_}) in the generalized EF
coordinates introduced in the previous subsection. On the RHS of (\ref{3.24_}%
) we insert $\xi _{\mu }=g_{\mu v}+\Omega _{H}g_{\mu \psi }$ which, in EF
coordinates, evaluates to
\begin{eqnarray}
\xi _{\mu }\left( r_{+}^{I}\right) &=&\left[ 1-a\Omega _{H}\sin ^{2}\theta %
\right] \partial _{\mu }r  \label{3.25} \\
&=&\frac{\left( r_{+}^{I}\right) ^{2}+a^{2}\cos ^{2}\theta }{\left(
r_{+}^{I}\right) ^{2}+a^{2}}\partial _{\mu }r  \notag
\end{eqnarray}
In deriving (\ref{3.25}) we made repeated use of the horizon condition (\ref
{3.15}). On the LHS of (\ref{3.24_}) we need the derivative $D_{\mu }\boldsymbol{\xi}^{2}\equiv \partial _{\mu }\boldsymbol{\xi} ^{2}$ of the function $\boldsymbol{\xi} ^{2}$ given in eq.
(\ref{3.23}), evaluated at $r=r_{+}^{I}$. Since $\Delta _{I}=0$ and $\left(
\omega -\Omega _{H}\right) =0$ there, one easily finds
\begin{equation}
-D_{\mu }\boldsymbol{\xi} ^{2}\left( r_{+}^{I}\right) =\frac{\left( r_{+}^{I}\right)
^{2}+a^{2}\cos ^{2}\theta }{\left[ \left( r_{+}^{I}\right) ^{2}+a^{2}\right]
^{2}}\Delta _{I}^{\prime }\left( r_{+}^{I}\right) \partial _{\mu }r
\label{3.26}
\end{equation}
As a result, the surface gravity is given by
\begin{equation}
\kappa =\frac{1}{2}\frac{\Delta _{I}^{\prime }\left( r_{+}^{I}\right) }{%
\left( r_{+}^{I}\right) ^{2}+a^{2}}  \label{3.27}
\end{equation}
where the prime, as always, denotes a derivative with respect to the
argument. More explicitly,
\begin{equation}
\kappa =\frac{r_{+}^{I}-G\left( r_{+}^{I}\right) M-r_{+}^{I}G^{\prime
}\left( r_{+}^{I}\right) M}{\left( r_{+}^{I}\right) ^{2}+a^{2}}  \label{3.28}
\end{equation}

Several comments are in order here.\\
\textbf{(a)} For $G\left( r\right) =$const, eq. (\ref{3.28}) coincides with
the classical result. The quantum corrections modify $\kappa $ both
explicitly, by the $G^{\prime }\left( r_{+}^{I}\right) $-term, and
implicitly, via the shift in the radius $r_{+}^{I}$.\\
\textbf{(b) }The surface gravity of the improved metric has turned out
independent of $\theta $. It is constant on $H_{+}$ therefore. This is
nontrivial since the symmetry assumptions imply only $\varphi $-, but no $%
\theta $-independence. Classically, $\kappa =$const constitutes the zeroth
law of black hole thermodynamics where $\kappa $ is related to the
Bekenstein-Hawking temperature via $T=\kappa /2\pi $ \cite{Hawking-Bardeen-C,Wald-Racz}. In section 8 we shall
address the question whether a similar interpretation can hold in the
improved case.\\
\textbf{(c)} As in the classical case, $\kappa $ vanishes for extremal
black holes. Their $\Delta \left( r\right) $ has a double zero at the
horizon, implying $\Delta =\Delta ^{\prime }=0$ there.\\
\textbf{(d)} Sometimes it is convenient to rewrite $\kappa $ in a way which
removes any explicit $a$-dependence. Again exploiting $\Delta _{I}\left(
r_{+}^{I}\right) =0$ yields
\begin{equation}
\kappa =\frac{1}{2G\left( r_{+}^{I}\right) M}-\frac{1}{2r_{+}^{I}}-\frac{%
G^{\prime }\left( r_{+}^{I}\right) }{2G\left( r_{+}^{I}\right) }
\label{3.29}
\end{equation}
Of course $\kappa $ continues to be implicitly $a$-dependent via $r_{+}^{I}$.\\
\textbf{(e) }For $a=0$ the horizon condition is $r_{+}^{I}=2G\left(
r_{+}^{I}\right) M$. Using this relation in (\ref{3.29}) we obtain the
surface gravity for the improved Schwarzschild metric:
\begin{equation}
\kappa =\frac{1}{4G\left( r_{+}^{I}\right) M}-\frac{G^{\prime }\left(
r_{+}^{I}\right) }{2G\left( r_{+}^{I}\right) }  \label{3.30_}
\end{equation}
Assuming the validity of $T=\kappa /2\pi $ for the Schwarzschild black hole,
eq. (\ref{3.30_}) implies exactly the Hawking temperature which had been found in
ref. \cite{bh2} using a rather different argument.

\section{Horizons and static limit surfaces}

\subsection{Critical surfaces}

In this section we determine the horizons and the static limit surfaces of
the improved Kerr metric. We shall collectively refer to them as ``critical surfaces''. In Section 3 we saw that the radii $r_{S}^{I}\left( \theta \right)
$ and $r_{H}^{I}$ of a static limit surface $S$ and a horizon $H$ are given
by eqs. (\ref{3.14}) and (\ref{3.15}), respectively. By defining
\begin{equation}
b\equiv\left\{
\begin{tabular}{ll}
$a\cos \theta $ & for $S$ \\
$a$ & for $H$%
\end{tabular}
\right.  \label{4.1}
\end{equation}
those two equations can be combined into one, namely
\begin{equation}
r^{2}-2G\left( r\right) Mr+b^{2}=0  \label{4.2}
\end{equation}
With a $G\left( r\right) $ of the form (\ref{1.3}), i.e.
\begin{equation}
G\left( r\right) =\frac{G_{0}d^{2}\left( r\right) }{d^{2}\left( r\right) +%
\bar{w}G_{0}}  \label{4.3}
\end{equation}
this condition becomes
\begin{equation}
\tilde{d}^{2}\left( \tilde{r}\right) \left( \tilde{r}^{2}+\tilde{b}^{2}-2%
\tilde{m}\tilde{r}\right) +\bar{w}\left( \tilde{r}^{2}+\tilde{b}^{2}\right)
=0  \label{4.4}
\end{equation}
Here and in the following the tilde means that the corresponding quantity is
expressed in terms of the Planck units related to $G_{0}$. In particular, $%
\tilde{r}=r/\ell_{\rm Pl}$, $\tilde{m}=m/\ell_{\rm Pl}$, $\tilde{M}=M/m_{\rm Pl}$, $\tilde{a}%
=a/\ell_{\rm Pl}$, $\tilde{b}=b/\ell_{\rm Pl}$ and $\tilde{d}=d/\ell_{\rm Pl}$ where $G_{0}$ $%
\equiv m_{\rm Pl}^{-2}$ $\equiv \ell_{\rm Pl}^{2}$. Thus we are led to investigate
possible zeros of the family of functions
\begin{equation}
Q_{\tilde{b}}^{\bar{w}}\left( \tilde{r}\right) \equiv \tilde{d}^{2}\left(
\tilde{r}\right) \left( \tilde{r}^{2}+\tilde{b}^{2}-2\tilde{m}\tilde{r}%
\right) +\bar{w}\left( \tilde{r}^{2}+\tilde{b}^{2}\right)  \label{4.5}
\end{equation}
Depending on our choice for the parameters $\tilde{b}$ and $\bar{w}$ the
equation (\ref{4.5}) describes the critical surfaces of the following
metrics:

\begin{enumerate}
\item  Classical Schwarzschild metric: $\bar{w}=0$, $\tilde{b}=0$

\item  Classical Kerr metric: $\bar{w}=0$, $\tilde{b}\neq 0$

\item  Improved Schwarzschild metric: $\bar{w}\neq 0$, $\tilde{b}=0$

\item  Improved Kerr metric: $\bar{w}\neq 0$, $\tilde{b}\neq 0$
\end{enumerate}

\bigskip

We shall analyse (\ref{4.5}) for the distance function $d\left( r\right) $
obtained from the straight radial path $\mathcal{C}$ discussed in section 2. We
proceed in two steps: We first employ the simple approximation $d\left(
r\right) =r$ for an analytic discussion of the problem and then in a second
step, we use numerical methods to show that, qualitatively, the results
obtained analytically are indeed representative and provide us with a
correct picture of the new features which are due to the nonzero angular
momentum of the black hole.

\subsection{The approximation $d\left( r\right) =r$}

For $d\left( r\right) =r$ the function $Q_{\tilde{b}}^{\bar{w}}$ becomes a
quartic polynomial:
\begin{equation}
Q_{\tilde{b}}^{\bar{w}}\left( \tilde{r}\right) \equiv \tilde{r}^{4}-2\tilde{m%
}\tilde{r}^{3}+\left( \tilde{b}^{2}+\bar{w}\right) \tilde{r}^{2}+\bar{w}%
\tilde{b}^{2}  \label{4.6}
\end{equation}
Before turning to the general case of the improved Kerr metric it is
instructive to see how the critical surfaces arise in the special cases (1),
(2), and (3):

\begin{enumerate}
\item  \textit{The classical Schwarzschild metric}: In this
case the polynomial simplifies to
\begin{equation}
Q_{0}^{0}\left( \tilde{r}\right) \equiv \tilde{r}^{3}\left( \tilde{r}-2%
\tilde{m}\right)   \label{4.7}
\end{equation}
It has a triple zero at $\tilde{r}=0$ and a simple zero at $\tilde{r}=2%
\tilde{m}$, or $r=2G_{0}M$.

\item  \textit{The classical Kerr metric}: Here the
function (\ref{4.6}) becomes
\begin{equation}
Q_{\tilde{b}}^{0}\left( \tilde{r}\right) \equiv \tilde{r}^{2}\left( \tilde{r}%
^{2}-2\tilde{m}\tilde{r}+\tilde{b}^{2}\right)   \label{4.8}
\end{equation}
It has a double zero at $\tilde{r}=0$ and two simple zeros at
\begin{equation}
\tilde{r}_{\pm }=\tilde{m}\pm \sqrt{\tilde{m}^{2}-\tilde{b}^{2}}  \label{4.9}
\end{equation}
if $\tilde{m}\neq \tilde{b}$, or one double zero at $\tilde{r}=\tilde{m}$ if $%
\tilde{m}=\tilde{b}$. These zeros give rise to the familiar static limit
surfaces $S_{\pm }$ and horizons $H_{\pm }$ at
\begin{equation}
r_{S_{\pm }}\left( \theta \right) =G_{0}M\pm \sqrt{\left( G_{0}M\right)
^{2}-a^{2}\cos ^{2}\theta }  \label{4.10}
\end{equation}
\begin{equation}
r_{\pm }\equiv r_{H_{\pm }}=G_{0}M\pm \sqrt{\left( G_{0}M\right) ^{2}-a^{2}}
\label{4.11}
\end{equation}
In the case $\tilde{m}=\tilde{a}$ the two horizons $H_{+}$ and $H_{-}$ merge
to a simple one with the ``critical'' radius $\tilde{r}=\tilde{m}$. We then
have an extremal black hole with $a=G_{0}M$, or $J\equiv aM=G_{0}M^{2}$, and
$r_{\text{crit}}=G_{0}M_{\text{crit}}=a$ \cite{Bardeen}.

\item  \textit{The improved Schwarzschild metric}: In this
case (\ref{4.6}) reads
\begin{equation}
Q_{0}^{\bar{w}}\left( \tilde{r}\right) =\tilde{r}^{2}\left( \tilde{r}^{2}-2%
\tilde{m}\tilde{r}+\bar{w}\right)   \label{4.12}
\end{equation}
This function has a double zero at $\tilde{r}=0$ and two simple zeros at
\begin{equation}
\tilde{r}_{\pm }^{\text{I}}=\tilde{m}\pm \sqrt{\tilde{m}^{2}-\bar{w}}
\label{4.13}
\end{equation}
if $\tilde{m}^{2}\neq \bar{w}$, or one double zero at $\tilde{r}^{\text{I}}=%
\tilde{m}$ if $\tilde{m}^{2}=\bar{w}$. As a result, the quantum-corrected
Schwarzschild spacetime hast two spherical horizons $H_{\pm }$ at
\begin{equation}
r_{\pm }^{\text{I}}=G_{0}M\pm \sqrt{\left( G_{0}M\right) ^{2}-\bar{w}G_{0}}
\label{4.14}
\end{equation}
If $\tilde{m}^{2}=\bar{w}$ the two horizons coalesce to a single one at the
critical radius $r_{\text{cr}}=\sqrt{\bar{w}}\ell{\rm Pl}=G_{0}M_{\text{cr}}$.
This new type of an extremal black hole is realized when the mass equals the
critical mass $M_{\text{cr}}=\sqrt{\bar{w}}\;m_{\rm Pl}$. Since $\bar{w}=O\left(
1\right) $ extremal black holes have a mass of the order of $m_{\rm Pl}$. For $%
M<M_{\text{cr}}$ the improved Schwarzschild metric has no horizon at all.
%\end{description}

The improved Schwarzschild metric has been discussed in detail in ref. \cite
{bh2} to which the reader is refered for further details.

As for the existence of horizons it is also interesting to note that there
is a close analogy between the \textit{classical} Kerr
metric and the \textit{improved} Schwarzschild metric.
The above formulae are identical if one identifies $\tilde{a}^{2}$ with $%
\bar{w}$ or, for the dimensionful quantities $a^{2}$ with $\bar{w}G_{0}$.

\bigskip

Note that in going from case (1) to either case (2) or case (3) the
triple zero at $\tilde{r}=0$ turns into a double zero at $\tilde{r}=0$, plus
a simple zero at $\tilde{r}>0$.

%\begin{description}
\item  \textit{The improved Kerr metric}: Finally we
discuss the zeros of $Q_{\tilde{b}}^{\bar{w}}$ with both $\bar{w}$ and $%
\tilde{b}$ nonzero. In principle their dependence on $\tilde{m}$, $\tilde{b}$
and $\bar{w}$ could be written down in closed form but the formulas are not
very instructive. The following indirect reasoning shows the essential
points more clearly.

The first and second derivatives of $Q_{\tilde{b}}^{\bar{w}}$ are
\begin{equation}
\frac{d}{d\tilde{r}}Q_{\tilde{b}}^{\bar{w}}\left( \tilde{r}\right) =2\tilde{r%
}\left[ 2\tilde{r}^{2}-3\tilde{m}\tilde{r}+\left( \tilde{b}^{2}+\bar{w}%
\right) \right]  \label{4.15}
\end{equation}
\begin{equation}
\frac{d^{2}}{d\tilde{r}^{2}}Q_{\tilde{b}}^{\bar{w}}\left( \tilde{r}\right)
=12\tilde{r}^{2}-12\tilde{m}\tilde{r}+2\left( \tilde{b}^{2}+\bar{w}\right)
\label{4.16}
\end{equation}
The derivative (\ref{4.15}) vanishes at the $\tilde{r}$-values $\tilde{r}%
_{0} $, $\tilde{r}_{1}$, and $\tilde{r}_{2}$ given by
\begin{eqnarray}
\tilde{r}_{0} &=&0  \label{4.17} \\
\tilde{r}_{1} &=&\frac{3}{4}\tilde{m}\left[ 1-\sqrt{1-\frac{8}{9}\frac{%
\tilde{b}^{2}+\bar{w}}{\tilde{m}^{2}}}\right]  \notag \\
\tilde{r}_{2} &=&\frac{3}{4}\tilde{m}\left[ 1+\sqrt{1-\frac{8}{9}\frac{%
\tilde{b}^{2}+\bar{w}}{\tilde{m}^{2}}}\right]  \notag
\end{eqnarray}
Provided
\begin{equation}
\frac{8}{9}\left( \tilde{b}^{2}+\bar{w}\right) \leq \tilde{m}^{2}
\label{4.18}
\end{equation}
the square roots in (\ref{4.17}) are real so that $\tilde{r}_{1}$, and $%
\tilde{r}_{2}$ are real and positive. As a result, $Q_{\tilde{b}}^{\bar{w}}$
has 3 different extrema for $\tilde{r}\geq 0$, except when the equality sign
holds in (\ref{4.18}). Then two extrema merge to an inflection point.
Inserting (\ref{4.17}) into (\ref{4.16}) one finds that the second
derivative is negative at $\tilde{r}_{1}$ and positive at $\tilde{r}_{0}$,
and $\tilde{r}_{2}$. Therefore in the nondegenerate case, $\tilde{r}_{0}$,
and $\tilde{r}_{2}$ are minima, and $\tilde{r}_{1}$ is a maximum of $Q_{%
\tilde{b}}^{\bar{w}}$. If $\frac{8}{9}\left( \tilde{b}^{2}+\bar{w}\right) =%
\tilde{m}^{2}$ there is a minimum at $\tilde{r}_{0}=0$ and an inflection
point at $\tilde{r}_{1}=\tilde{r}_{2}=3\tilde{m}/4$, and if $\frac{8}{9}%
\left( \tilde{b}^{2}+\bar{w}\right) >\tilde{m}^{2}$ the only critical point
is the minimum at $\tilde{r}_{0}=0$.
\end{enumerate}

Let us come back to the zeros of $Q_{\tilde{b}}^{\bar{w}}$. Regarded a
function of the complex variable $\tilde{r}\in\mathbb{C}$, it has 4 zeros on the complex plane; only those on the positive
real axis are physically relevant though. Furthermore, regarded a function
on the full real line, $Q_{\tilde{b}}^{\bar{w}}\left( \tilde{r}\right) $ is
the sum of 4 terms all of which are positive if $\tilde{r}<0$. As a
consequence, $Q_{\tilde{b}}^{\bar{w}}$ has no zeros at strictly negative $%
\tilde{r}$. A priori $Q_{\tilde{b}}^{\bar{w}}$ could have 4 zeros at $\tilde{%
r}>0$. This case is already excluded, however, since we saw that the
function has at most one maximum and one minimum at strictly positive $%
\tilde{r}$. Therefore, as far as zeros at $\tilde{r}>0$ are concerned, only
the following 3 cases can occur: (a) 2 simple zeros, (b) 1 double zero, (c)
no zero at all. 

In Fig. 3 we show an example of each case.
In this figure and all similar diagrams the notation $\tilde{r}_{\tilde{b}_{\pm} }^{\text{I}}$ stands for either $\tilde{r}_{H_{\pm} }^{\text{I}}\equiv \tilde{r}%
_{\pm }^{\text{I}}$ or $\tilde{r}_{S_{\pm} }^{\text{I}}$, depending on the
interpretation of $\tilde{b}$. The superscript ``I'' indicates that the respective
radii refer to the improved metric.
\newpage
\centerline{}
\centerline{}
\centerline{}
\begin{minipage}[t]{.55\linewidth}
\begin{center}
\begin{pspicture}(2.3,-4)(12.7,2)
%\begin{pspicture}(0.3,-4)(10.7,2)
\includegraphics[width=\linewidth]{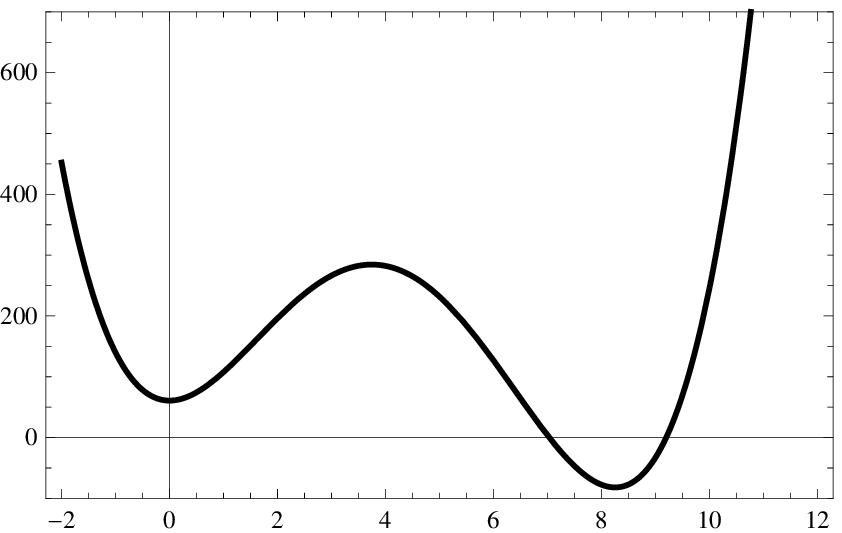}
%\rput[l]{90}(-8.57,0.5){$\tilde{Q}(\tilde{r})$}
\rput[l]{90}(-9.1,2.5){$\tilde{Q}(\tilde{r})$}
\rput[l](-4.25,-0.1){$\tilde{r}$ }
\rput[l](-2.55,0.72){$\tilde{r}_2$ }
%\rput[l](-4.7,2.35){$\tilde{r}_1$}
\rput[l](-5,3.25){$\tilde{r}_1$}
\rput[l](-3,4){$\tilde{b}=7.8$}
%\rput[l](-5.5,0){$\bar{w}=1$}
\rput[l](-3,3.5){$\tilde{m}=8$}
\rput[l](-3,1.35){$\tilde{r}^{\text{I}}_{\tilde{b}-}$}
\rput[l](-1.35,1.35){$\tilde{r}^{\text{I}}_{\tilde{b}+}$}
\rput[l](-7,-0.3){a)}
\end{pspicture}
\end{center}
\end{minipage}\hfill
\begin{minipage}[t]{.55\linewidth}
\begin{center}
\begin{pspicture}(1.5,0.07)(12,2.07)
%\begin{pspicture}(1.5,2.07)(12,4.07)
%\begin{pspicture}(0.5,1.75)(11,3.75)
%\begin{pspicture}(-0.5,4.75)(10,6.75)
\includegraphics[width=\linewidth]{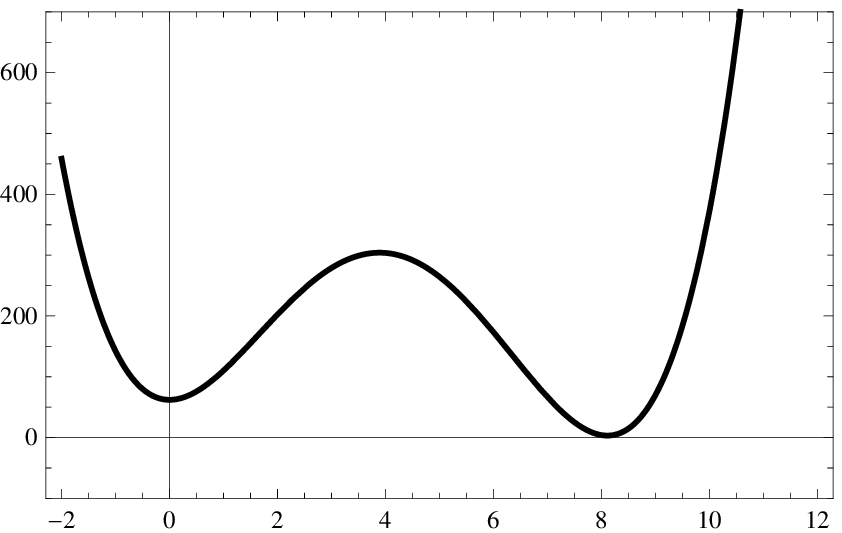}
%\rput[l]{90}(-8.57,2.5){$\tilde{Q}(\tilde{r})$}
\rput[l]{90}(-9.1,2.5){$\tilde{Q}(\tilde{r})$}
%\rput[l](-3.3,-0.3){$\tilde{r}$}
\rput[l](-4.25,-0.1){$\tilde{r}$}
\rput[l](-3,4){$\tilde{b}=7.88$}
%\rput[l](-2.5,3.5){$\bar{w}=1$}
\rput[l](-3,3.5){$\tilde{m}=8$}
\rput[l](-1.95,1.4){$=\tilde{r}^{\text{I}}_{\tilde{b}-}$}
\rput[l](-3.2,1.4){$=\tilde{r}^{\text{I}}_{\tilde{b}+}$}
\rput[l](-3.5,1.4){$\tilde{r}_2$}
%\rput[l](-5.45,2){$\tilde{r}_1$}
\rput[l](-5,3.25){$\tilde{r}_1$}
\rput[l](-7,-0.3){b)}
%\rput[l](-4,-0.7){b)}
\end{pspicture}
\end{center}
\end{minipage}
%\centerline{}
%\centerline{}
%\centerline{}
%\centerline{}
%\centerline{}
%\centerline{}
%\centerline{}
%\centerline{}
%\centerline{}
\begin{minipage}[t]{.55\linewidth}
\begin{center}
%\begin{pspicture}(-3.4,7.6)(2.4,2.6)
%\begin{pspicture}(-3.4,6.6)(2.4,1.6)
\begin{pspicture}(-3.4,2.6)(2.4,-2.4)
\includegraphics[width=\linewidth]{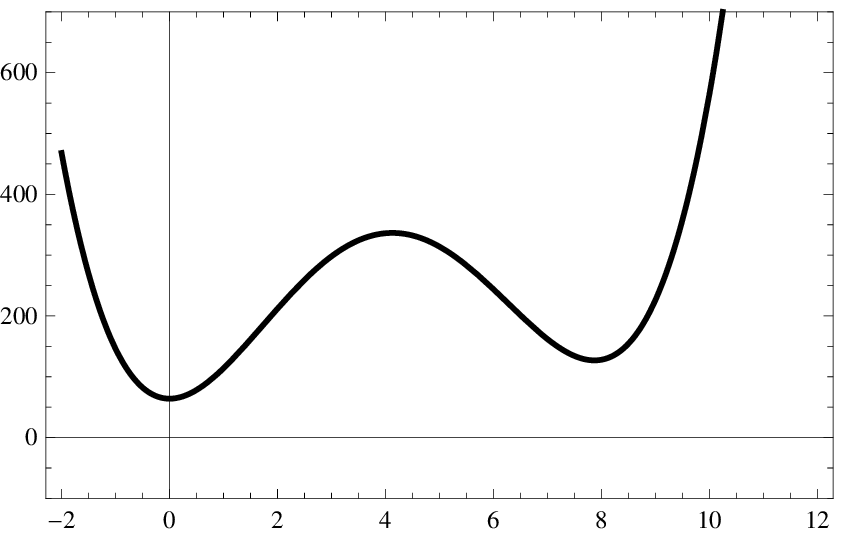}
\rput[l]{90}(-9.1,2.5){$\tilde{Q}(\tilde{r})$}
\rput[l](-3,4){$\tilde{b}=8$}
\rput[l](-3,3.5){$\tilde{m}=8$}
\rput[l](-7,-0.3){c)}
%\rput[l](-4.5,-0.7){c)}
%\rput[l](-3.3,-0.2){$\tilde{r}$}
\rput[l](-4.25,-0.1){$\tilde{r}$ }
%\rput[l](-4.7,-1.2){Fig. 3.}
\rput[l](-11,-1.7){Fig. 3:}
\rput[l](-9.3,-1.7){The figures show examples of the 3 possible configurations}
\rput[l](-9.3,-2.2){the function $Q^{\bar{w}}_{\tilde{b}}$ of eq. (4.6) can assume, with two, one, and}
\rput[l](-9.3,-2.7){no zero on the positive real axis.}
%\rput[l](-5,-1.2){Fig. 3.}
%\rput[l](-9.8,-1.7){The figures show examples of the 3 possible configurations}
%\rput[l](-9.8,-2.2){the function $Q^{\bar{w}}_{\tilde{b}}$ of eq. (4.6) can assume, with two, one, and}
%\rput[l](-9.8,-2.7){no zero on the positive real axis.}

%%\rput[l](-9.8,-1.2){For $\tilde{a} \to 0$, $\tilde{m}$ assumes its minimum value at $\sqrt{\bar{w}}$, and for $\tilde{a} \to \infty$ $\tilde{m}$ a-}
%%\rput[l](-9.8,-1.7){pproaches the classical behavior, namely $\tilde{m}\to \tilde{a}$.}
\end{pspicture}
\end{center}
\end{minipage}
\centerline{}
\centerline{}
\centerline{}
%\centerline{}

From the definition of $Q_{\tilde{b}}^{\bar{w}}$, eq. (\ref{4.6}), it is
obvious that the occurrence of zeros is the more likely the larger is $%
\tilde{m}$ and the smaller are $\tilde{b}$ and $\bar{w}$. The reason is that
for $\tilde{m}$ large and $\tilde{b}$, $\bar{w}$ small the second term on
the RHS of (\ref{4.6}) $-2\tilde{m}\tilde{r}^{3}<0$, becomes very negative
and the positive terms $\left( \tilde{b}^{2}+\bar{w}\right) \tilde{r}^{2}>0$
and $\bar{w}\tilde{b}^{2}>0$ are small which favors zeros. Therefore we
expect that, for $\tilde{a}$ (and $\bar{w}$) fixed, there are two zeros for
large $\tilde{m}$ ( case (a) ) and no zero for small $\tilde{m}$ ( case (c) ). In between there is a
critical mass at which the extremal situation of a simple double zero is
realized ( case (b) ). 

In Fig. 4 we show that this is indeed the case. Here the radii of both
horizons and critical limit surfaces are displayed; this amounts to $\tilde{b%
}=\tilde{a}$ and $\tilde{b}=\tilde{a}\cos \theta $ in the formulas above. In
all diagrams we fixed $\tilde{a}=5$ (and $\bar{w}=1$), and plotted the
classical and improved radii as a function of $\tilde{m}$. The 4 diagrams
correspond to different values of $\theta $. Generically ( cases (b) and (c) )
we find 4 different improved radii $\tilde{r}_{\pm }^{\text{I}}$, $\tilde{r}%
_{S_{\pm} }^{\text{I}}\left( \theta \right) $ when $\tilde{m}$ is very large.
When we lower $\tilde{m}$ we reach a point at which the two horizons
coalesce, $\tilde{r}_{+}^{\text{I}}=\tilde{r}_{-}^{\text{I}}$, and below
which there is no horizon any longer, but there still exist two critical
limit surfaces. Lowering $\tilde{m}$ even further the two static limit
surfaces coalesce at a certain critical mass, $\tilde{r}_{S_{+}}^{\text{I}%
}\left( \theta \right) =\tilde{r}_{S_{-}}^{\text{I}}\left( \theta \right) $,
and for even smaller $\tilde{m}$ there exist neither a horizon nor a static
limit surface.

Fig. 4a) applies to the poles \ $\left( \theta =0,\pi \right) $ where the
event horizons and static limit surfaces touch, $\tilde{r}_{+}^{\text{I}}=%
\tilde{r}_{S_{+}}^{\text{I}}\left( \theta \right) $, $\tilde{r}_{-}^{\text{I}}=%
\tilde{r}_{S_{-}}^{\text{I}}\left( \theta \right) $. Fig. 4d) refers to the
equatorial plane ($\theta =\pi /2$) in which, classically, $r_{S_{-}}=0$, $%
r_{S_{+}}=2m$.

In the 2-dimensional diagrams of Fig. 5 we display the $\theta $-dependence
of the various radii. Here we picked the parameter values $\tilde{m}=6$, $%
\tilde{a}=5$ for which there exist two horizons $H_{\pm }$ and two static
limits $S_{\pm }$. (For the constant $\bar{w}$ we chose $\bar{w}=4$.)

Both Fig. 4 and 5 show that the quantum effects are the larger the smaller is $%
\tilde{m}$. For $\tilde{m}\equiv M/m_{\rm Pl}\gg 1$ the critical surfaces of the
improved black hole coincide essentially with those of the classical one.
Lowering $M$ we find that the radius of the outer horizon $H_{+}$ is always
smaller than in the classical case, while the radius of $H_{-}$ is always
larger than classically. Similarly we see that $\tilde{r}_{S_{+}}^{\text{I}%
}\left( \theta \right) <\tilde{r}_{S_{+}}\left( \theta \right) $ whereas $\tilde{r}_{S_{-}}^{\text{I}}\left( \theta \right)
>\tilde{r}_{S_{-}}\left( \theta \right) $. Both for horizons and static limits the extremal points where the
upper and the lower branch of the curves meet are shifted towards larger
masses by the quantum corrections.

\newpage
\begin{minipage}[t]{.55\linewidth}
\begin{center}
\begin{pspicture}(2.5,0.5)(5.5,6)
%\begin{pspicture}(1,0.5)(4,6)
\includegraphics[width=\linewidth]{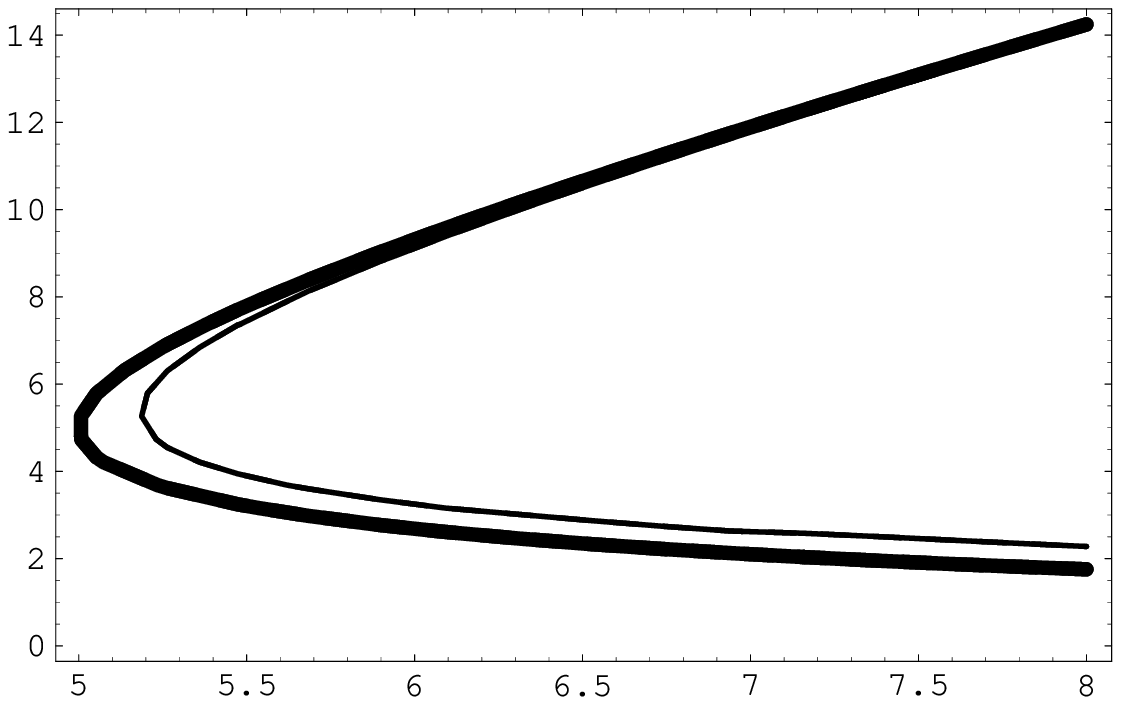}
\rput[l]{90}(-8.9,4){$r_{b\pm}$}
%\rput[l](-7,6.3){$a=5\ ,\ \bar{w}=1\ ,\ \theta=0,\pi$}
\rput[l](-7,6.3){$\theta=0,\pi$}
\rput[l](-0.5,1.35){$m$}
\rput[l](-7.5,1.35){a)}
\end{pspicture}
\end{center}
\end{minipage}\hfill
\begin{minipage}[t]{.55\linewidth}
\begin{center}
\begin{pspicture}(5.3,0.5)(1.3,6)
%\begin{pspicture}(4,0.5)(0,6)
\includegraphics[width=\linewidth]{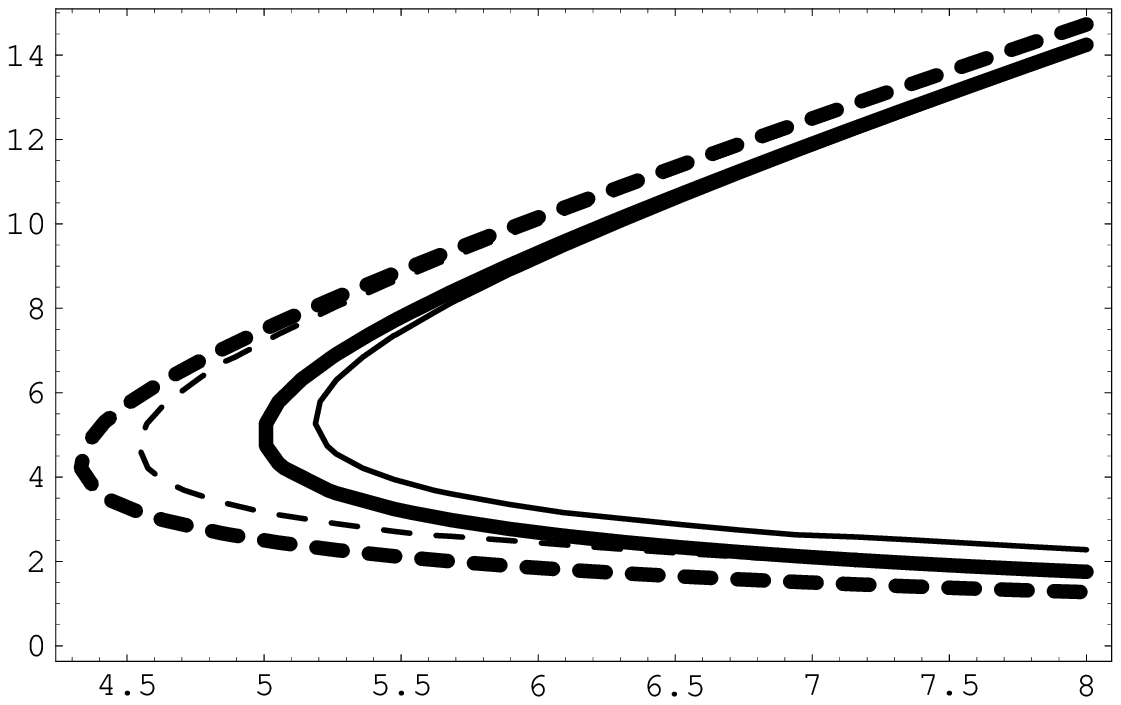}
\rput[l]{90}(-8.9,4){$r_{b\pm}$}
\rput[l](-7,6.3){$\theta=\frac{\pi}{6},\frac{5\pi}{6}$}
\rput[l](-0.5,1.35){$m$}
\rput[l](-7.5,1.35){b)}
\end{pspicture}
\end{center}
\end{minipage}\hfill
\begin{minipage}[t]{.55\linewidth}
\begin{center}
%\begin{pspicture}(1.5,1)(4.7,6.5)
%\begin{pspicture}(0.5,1)(3.7,6.5)
%\includegraphics[width=\linewidth]{Graf_d_Eq_r_3.tex_gr1.eps}
%\includegraphics[width=\linewidth]{Graf_d_Eq_r_3.tex_gr1_II_09.eps}
%\begin{pspicture}(1.5,0)(4.7,5.5)
%\begin{pspicture}(1.5,-0.5)(4.7,5)
\begin{pspicture}(1.5,0.4)(4.7,5.9)
\includegraphics[width=\linewidth]{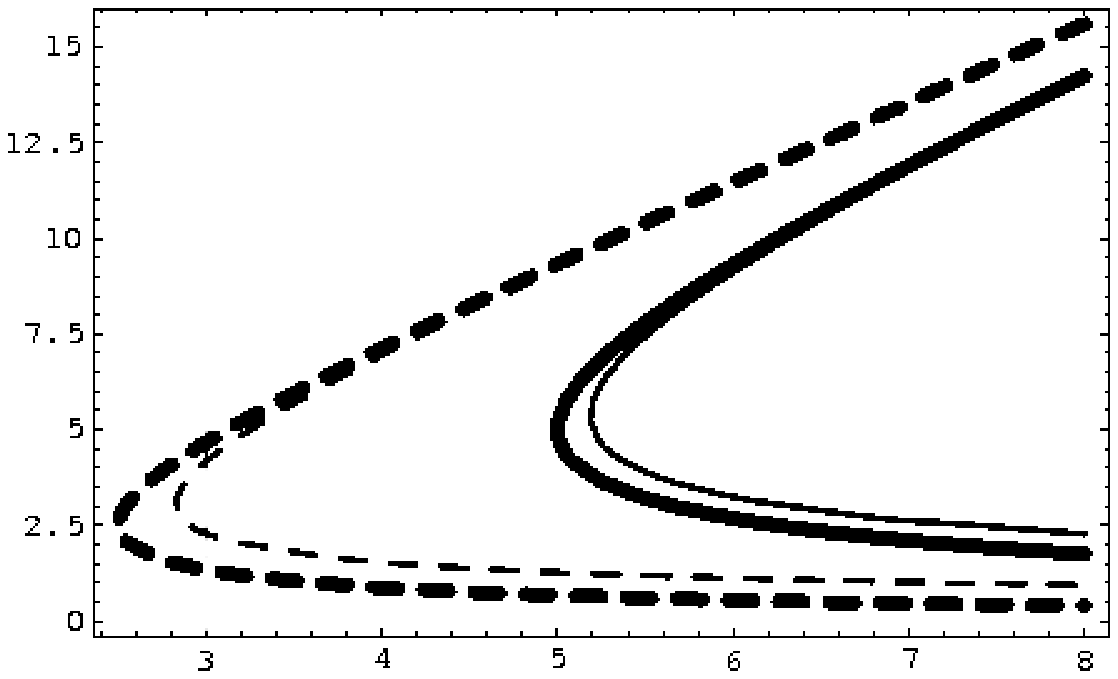}
\rput[l]{90}(-8.9,3){$r_{b\pm}$}
%\rput[l](-7,6){$a=5\ ,\ \bar{w}=1\ ,\ \theta=\frac{2\pi}{6},\frac{4\pi}{6}$}
%\rput[l](-7,6){$\theta=\frac{2\pi}{6},\frac{4\pi}{6}$}
\rput[l](-7,5){$\theta=\frac{2\pi}{6},\frac{4\pi}{6}$}
\rput[l](-0.5,0.1){$m$}
\rput[l](-7.5,0.1){c)}
\end{pspicture}
\end{center}
\end{minipage}\hfill
\begin{minipage}[t]{.55\linewidth}
\begin{center}
\begin{pspicture}(0.5,1.5)(3.5,7)
%\begin{pspicture}(-0.5,1)(2.5,6.5)
\includegraphics[width=\linewidth]{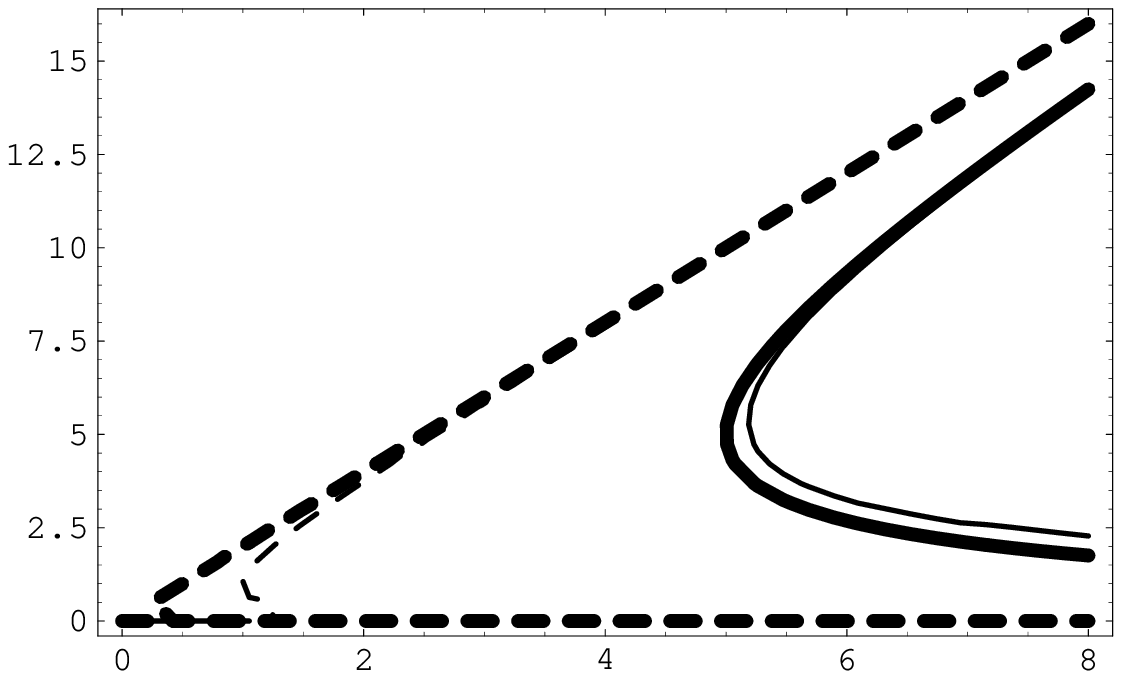}
\rput[l]{90}(-8.9,4){$r_{b\pm}$}
%\rput[l](-6.5,6){$a=5\ ,\ \bar{w}=1\ ,\ \theta=\frac{\pi}{2}$}
\rput[l](-6.5,6){$\theta=\frac{\pi}{2}$}
\rput[l](-0.5,1.2){$m$}
\rput[l](-7.5,1.2){d)}
\end{pspicture}
\end{center}
\end{minipage}
\centerline{}
\centerline{}
\centerline{}
\centerline{}
\centerline{}
%\centerline{}
%\centerline{}
\begin{center}
%\rput[l](-0.3,3){Fig. 4.}
\rput[l](-9.4,2.5){Fig. 4:}
\rput[l](-7.7,2.5){The figures show the $m$-dependence of the radii $r_{b\pm}$ (thick lines) and $r^{\text{I}}_{b\pm}$ (thin}
\rput[l](-7.7,2){lines) for $a=5$ and several values of $\theta$. The continuous lines represent $r_{\pm}$ and $r^{\text{I}}_{\pm}$.}
\rput[l](-7.7,1.5){The dashed lines represent $r_{S\pm}$ and $r^{\text{I}}_{S\pm}$.}
%\rput[l](-7.7,1.5){provement with $d\left(r\right)=r$ shifts $r_{b\pm}$ smoothly to the radii $r^{\text{I}}_{b\pm}$ for
%$m_{\text{pl}} \ll m$. The extre-}
%\rput[l](-7.7,1){mal points are also moved to higher values of $m$.}
%\rput[l](-7.7,0){Fig. 5.12: For $\theta= 0,\ \pi$, event horizons and static limits coincide.}
%\rput[l](-7.7,-0.5){Figs. 5.13 and 5.14: For arbitrary values of $\theta \neq \frac{\pi}{2}$, there are
%eight different radii,}
%\rput[l](-5.92,-1){the four classical $r_{b \pm}$, and the four quantum corrected $r^{\text{I}}_{b \pm}$.}
%\rput[l](-7.7,-1.5){Fig. 5.15: At the equatorial plane, $r_{S-}=0$ and $r_{S+}=2m$.}
\end{center}

\newpage
\centerline{}
\centerline{}
\centerline{}
\centerline{}
\begin{minipage}[t]{.55\linewidth}
\begin{center}
%\begin{pspicture}(1.8,-1.8)(12.2,3.8)
%\begin{pspicture}(1,-1.8)(11.4,3.8)
\begin{pspicture}(2.3,-1.8)(12.7,3.8)
\includegraphics[width=\linewidth]{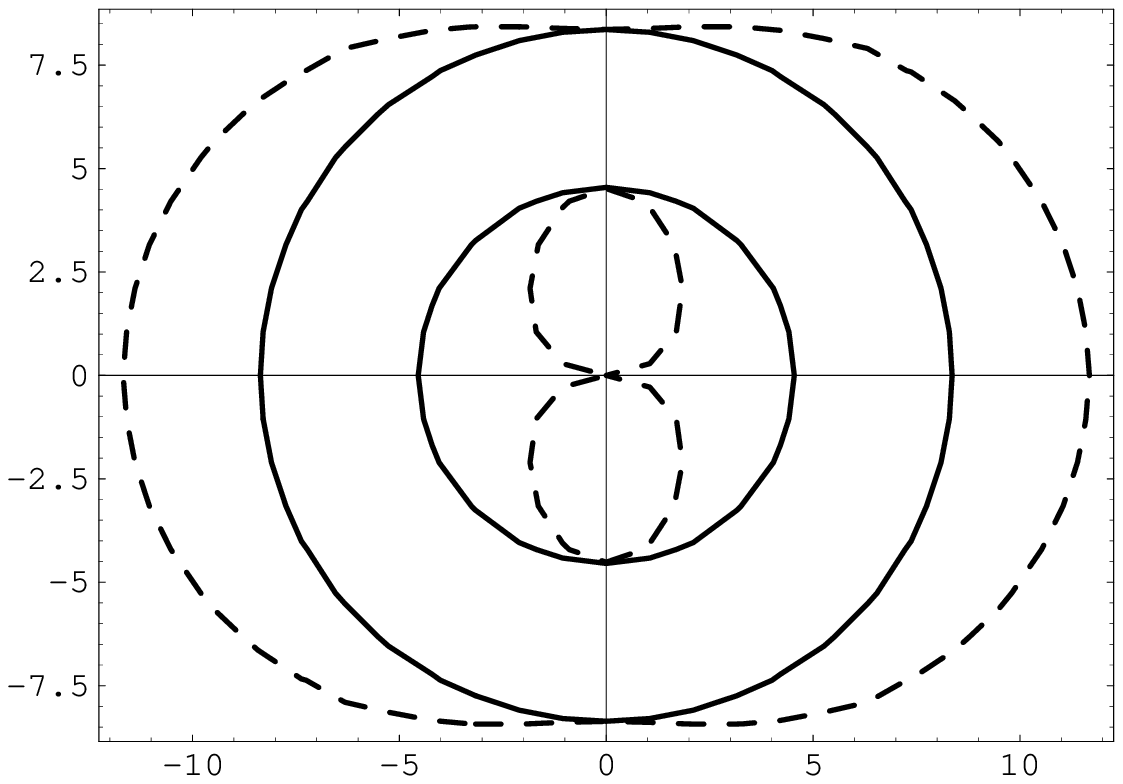}
%\rput[l](-8,4){$y$}
%\rput[l](-0.5,1.7){$x$}
\rput[l](-8,0.5){a)}
%\rput[l](0,-0.5){Fig. 5.}
%\rput[l](-8,-0.5){Fig. 5:}
\rput[l](-8,-0.5){Fig. 5: The figures show a cross section through the event horizons (continuous lines) and}
\rput[l](-8,-1.0){static limit surfaces (dashed lines) in the $xz$-plane 
for a quantum black hole with $\tilde{m}=6$}
\rput[l](-8,-1.5){and $\tilde{a}=5$. To facilitate the comparison with the classical case, in Fig. 5b) the corresponding}
\rput[l](-8,-2.0){classical surfaces are superimposed (thick lines).}
\end{pspicture}
\end{center}
\end{minipage}\hfill
\begin{minipage}[t]{.55\linewidth}
\begin{center}
%%\begin{pspicture}(0.5,1.75)(11,3.75)
%\begin{pspicture}(-.3,1.75)(10.2,3.75)
\begin{pspicture}(2,1.75)(12.5,3.75)
\includegraphics[width=\linewidth]{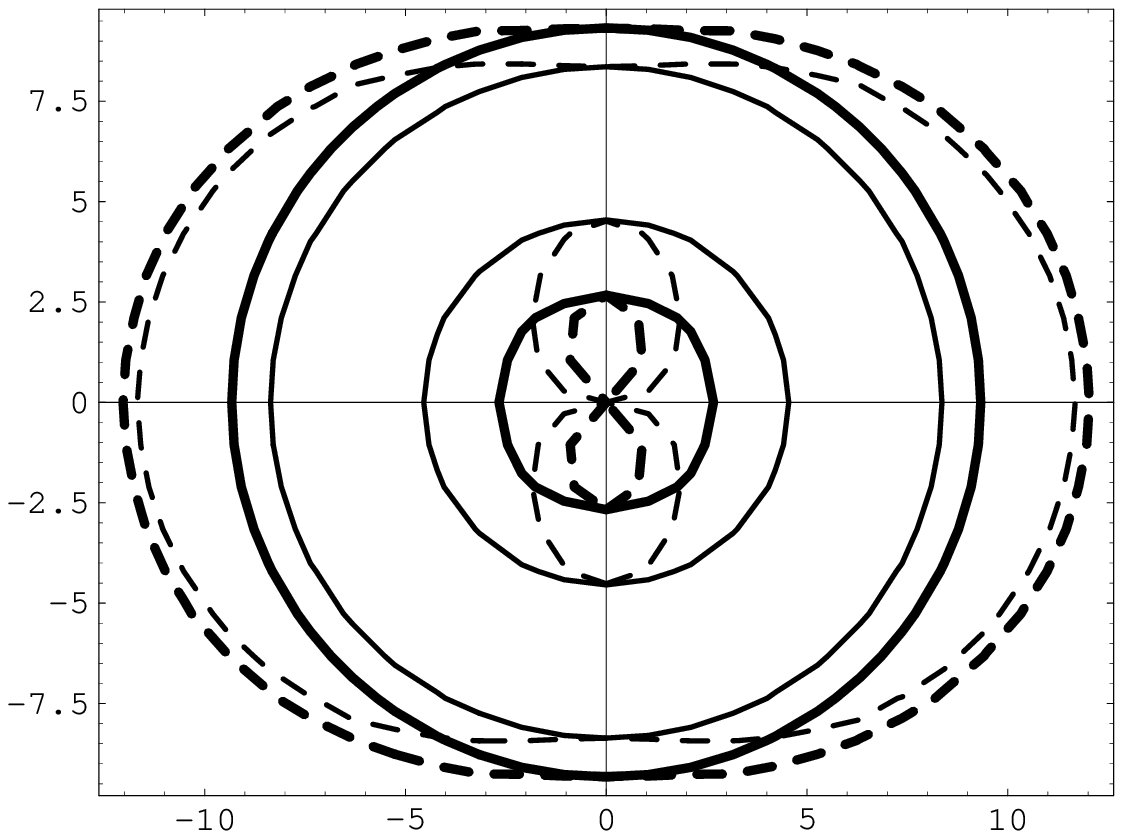}
%\rput[l](-8,4){$y$}
%\rput[l](-0.4,1.8){$x$}
\rput[l](-8,0.5){b)}
%\rput[l](-7,0){Improved critical surfaces for $m=6$,}
%\rput[l](-7,-0.5){$a=5$, $\bar{w}=4$.}
\end{pspicture}
\end{center}
\end{minipage}
%\begin{minipage}[t]{.55\linewidth}
%\begin{center}
%%\begin{pspicture}(-3.4,6.6)(2.4,1.6)
%\begin{pspicture}(-4.2,6.6)(1.6,1.6)
%\includegraphics[width=\linewidth]{2DPlotBH_Compo.tex_gr1.eps}
%\rput[l]{90}(-8.3,3.8){$y$}
%\rput[l](-0.4,1.8){$x$}
%\rput[l](-4.5,0.8){Fig. 5.26.}
%\rput[l](-7,0){Classical plus improved critical surfaces}
%\rput[l](-7,-.5){for $m=6$, $a=5$, $\bar{w}=4$.}
%\end{pspicture}
%\end{center}
%\end{minipage}
%\newpage
\centerline{}
\centerline{}
\subsection{The quantum extremality condition}

Let us determine the condition on $\tilde{m}$ and $\tilde{b}$ which implies
a double zero of $Q_{\tilde{b}}^{\bar{w}}\left( \tilde{r}\right) $. If $b=a$
this is the condition for the two horizons $H_{+}$ and $H_{-}$ to coincide,
i.e. for the quantum black hole to be extremal. When $Q_{\tilde{b}}^{\bar{w}%
}\left( \tilde{r}\right) $ has a double zero at some value of $\tilde{r}$,
the function must have a (local) minimum there. Since $\tilde{r}=\tilde{r}%
_{2}$ of (\ref{4.17}) is the only minimum it has for $\tilde{r}>0$, it
follows that the extremal case is realized precisely if $Q_{\tilde{b}}^{\bar{%
w}}$ vanishes at $\tilde{r}_{2}$: $\left. Q_{\tilde{b}}^{\bar{w}}\left(
\tilde{r}_{2}\right) \right| _{\text{extremal}}=0$. Inserting (\ref{4.17})
into (\ref{4.6}) we obtain
\begin{equation}
Q_{\tilde{b}}^{\bar{w}}\left( \tilde{r}_{2}\right) =-\frac{27\tilde{m}^{4}}{%
32}\left[ \left( 1-\frac{8}{9}\frac{\tilde{b}^{2}+\bar{w}}{\tilde{m}^{2}}%
\right) ^{\frac{3}{2}}+\frac{8}{27}\frac{\left( \tilde{b}^{2}-\bar{w}\right)
^{2}}{\tilde{m}^{4}}-\frac{4}{3}\frac{\tilde{b}^{2}+\bar{w}}{\tilde{m}^{2}}+1%
\right]  \label{4.19}
\end{equation}
As a result, setting $b=a$, the condition for $H_{+}=H_{-}$ reads
\begin{equation}
\left( 1-\frac{8}{9}\frac{\tilde{a}^{2}+\bar{w}}{\tilde{m}^{2}}\right) ^{%
\frac{3}{2}}+\frac{8}{27}\frac{\left( \tilde{a}^{2}-\bar{w}\right) ^{2}}{%
\tilde{m}^{4}}-\frac{4}{3}\frac{\tilde{a}^{2}+\bar{w}}{\tilde{m}^{2}}+1=0
\label{4.20}
\end{equation}
We shall refer to (\ref{4.20}) as the ``quantum extremality condition''. If $%
\bar{w}=0$ it reduces to $\tilde{m}=\tilde{a}$ for the classical Kerr
metric, and if $\tilde{a}=0$ to $\tilde{m}=\sqrt{\bar{w}}$ which is the
correct result for the extremal version of the improved Schwarzschild black
hole, see ref. \cite{bh2}. In the general case $\bar{w}%
\neq 0$, $\tilde{a}\neq 0$ the condition (\ref{4.20}) can be solved for $%
\tilde{m}=\tilde{m}\left( \tilde{a}\right) $ only numerically. The result is
shown in Fig. 6. We observe that $\tilde{m}\left( \tilde{a}\right) $
approaches the classical $\tilde{m}=\tilde{a}$ for large $a$, but deviates
significantly for $\tilde{a}\rightarrow 0$.

\begin{minipage}[t]{.55\linewidth}
\begin{center}
\begin{pspicture}(-2.4,7.6)(3.4,2.6)
%\begin{pspicture}(-3.4,7.6)(2.4,2.6)
\includegraphics[width=\linewidth]{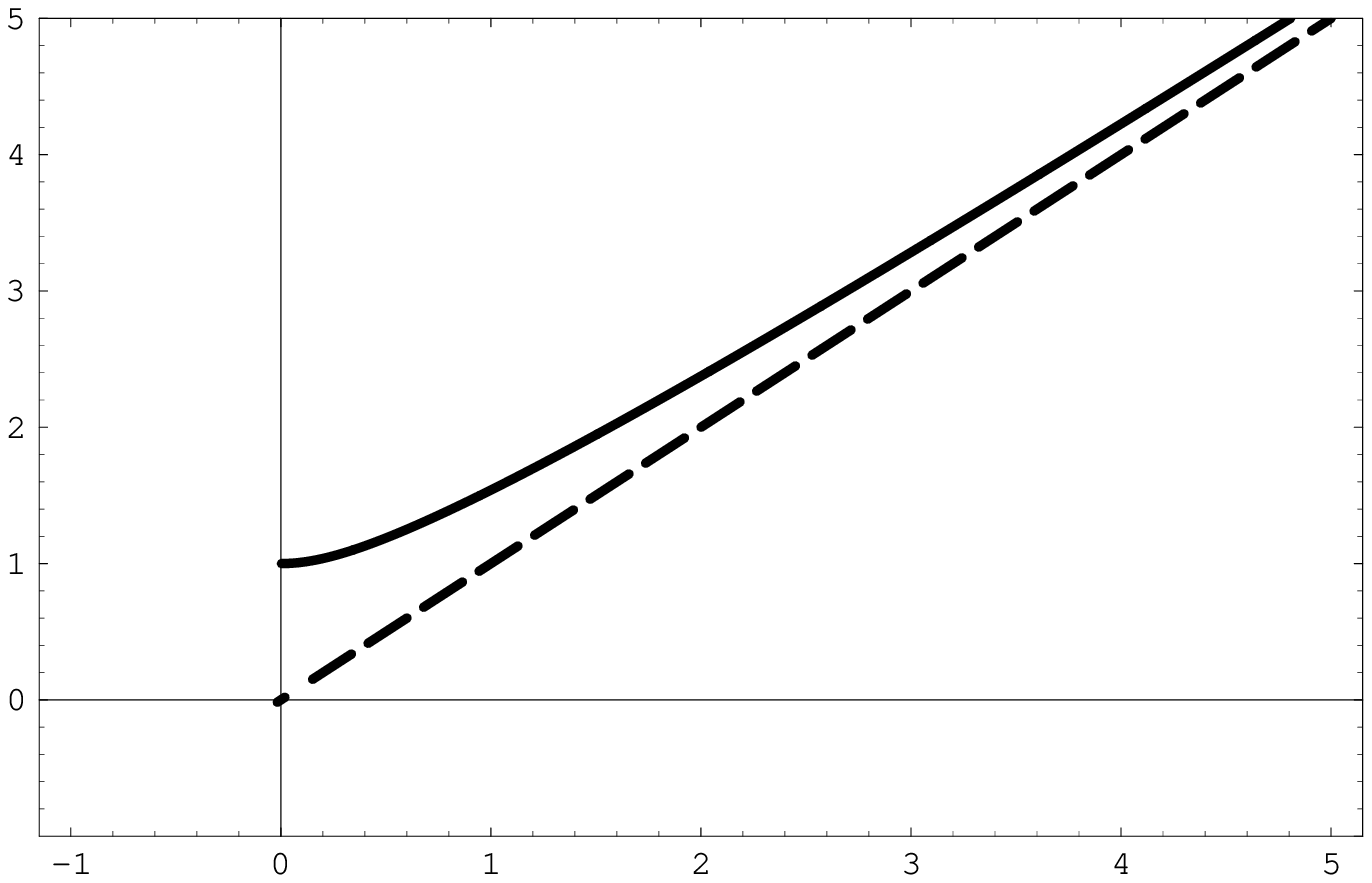}
\rput[l]{90}(-9.2,4){$\tilde{m}(\tilde{a})$}
\rput[l](-1,2.4){$\tilde{a}$ }
\rput[l](-8,3.7){$\sqrt{\bar{w}}$}
%\psline[linestyle=dashed,linewidth=0.5pt](-13.35,5.7)(-10.7,5.7)
%\rput[l](-5,0.8){Fig. 6.}
\rput[l](-11.5,0.3){Fig. 6:}
\rput[l](-9.8,0.3){The solution $\tilde{m}\left(\tilde{a}\right)$ of the ``quantum extremality condition''}
\rput[l](-9.8,-0.2){for the improved Kerr black hole (with $d(r)=r$ and $\bar{w}=1$).}
\rput[l](-9.8,-0.7){The dashed line represents the $\tilde{m}\left(\tilde{a}\right)=\tilde{a}$
dependence of the}
\rput[l](-9.8,-1.2){classical Kerr spacetime. 
For $\tilde{a} \to 0$, $\tilde{m}$ assumes its minimum}
\rput[l](-9.8,-1.7){value at $\sqrt{\bar{w}}$, while it approaches the classical behavior for}
\rput[l](-9.8,-2.2){$\tilde{a} \to \infty$.}
\end{pspicture}
\end{center}
\end{minipage}

\centerline{}
\centerline{}
\centerline{}
\centerline{}
\centerline{}
\centerline{}
\centerline{}
\subsection{Exact distance function}

Up to now we employed the simplified distance function $d\left( r\right) =r$
which has the virtue that all calculations can be performed analytically.
Using numerical techniques we have repeated the above analysis for the
``exact'' distance function (\ref{2.5}), (\ref{2.6}). It turns out that,
qualitatively, the results found with the ``exact'' $d\left( r\right) $ are
exactly the same as those from the $d\left( r\right) =r$ approximation. This
concerns in particular the number of horizons and critical surfaces, the
systematics of their mass and angular momentum dependence, and their
dissappearing at extremal configurations. (See Fig. 7 for an example.)

Thus one of the main results is that the classical and the improved Kerr
metric, sufficiently far away from extremality, have \textit{the same} number of horizons and static limit surfaces. This was
different for the Schwarzschild metric: the classical spacetime has 1
horizon, but the improved spacetime has 2. So, a priori one might have
expected a similar doubling in the case of the Kerr metric. Actually this is
not what happens: The quantum corrections do not generate new critical
surfaces but rather smoothly deform the classical ones.

In the language of ``catastrophe theory'' \cite{Poston,Milnor} this can be
understood from the structural stability properties of the zeros and
critical points of $Q_{\tilde{b}}^{\bar{w}}$. The corresponding function for
the classical Schwarzschild metric has a ``structurally unstable'' triple
zero at $\tilde{r}=0$; giving a nonzero value to $\bar{w}$ it dissolves
into a double zero at $\tilde{r}=0$ and a simple one at $\tilde{r}>0$. The
very same transition from a triple to a double plus a simple zero happens to
the Kerr metric already classically by a nonzero value of $\tilde{a}$. If,
in addition, the quantum parameter $\bar{w}\propto O\left( \hbar \right) $
is given a nonzero value, no further zero is generated. It is easy to
formally prove the structural stability of the classical Kerr zeros \cite{Tesis}.
\newpage
\begin{minipage}[t]{.55\linewidth}
\begin{center}
\begin{pspicture}(2,0.5)(5,6)
%\begin{pspicture}(1,0.5)(4,6)
\includegraphics[width=\linewidth]{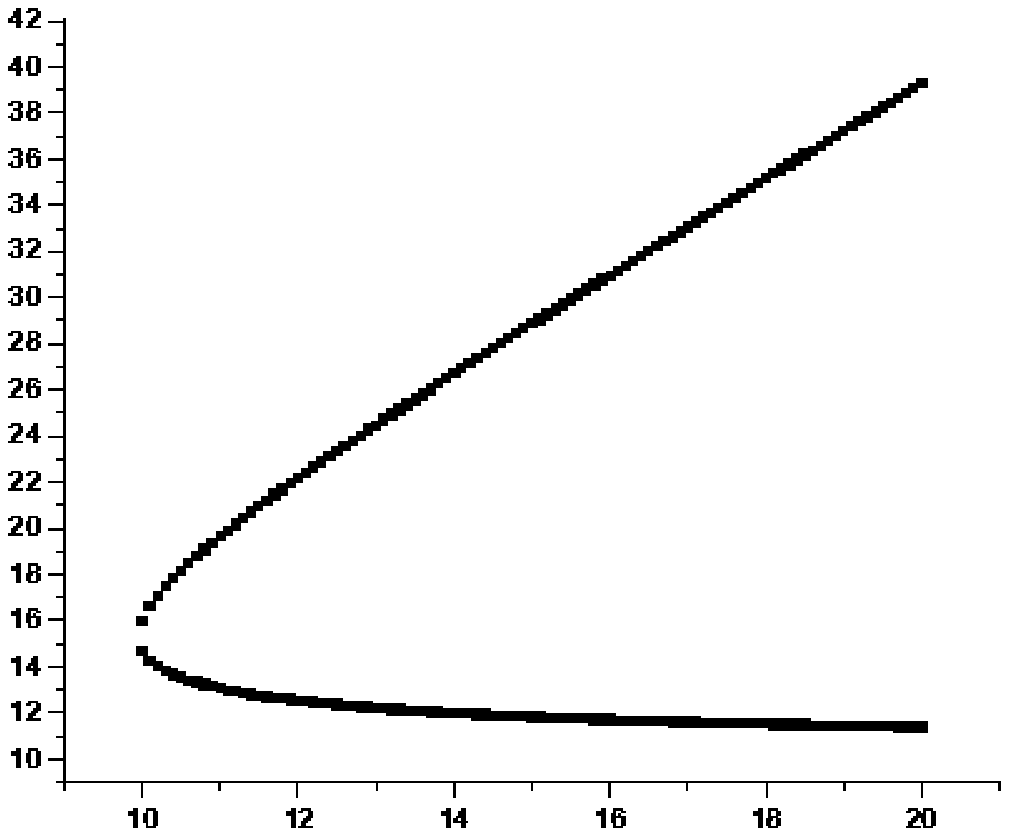}
%\fbox{\includegraphics[width=\linewidth]{/home/etuiran/Promo/Progs/Data/set5/q0s5.eps}
%%\includegraphics[width=\linewidth]{/home/etuiran/Promo/Progs/Data/set6/a0CQs6.eps}
%}

\rput[l](-1,0.2){$m$}
\rput[l]{90}(-8,3.2){$r\left(m\right)$}
%\rput[l](-6.5,5.5){$a=0\ ,\ \bar{w}=5\ ,\ \theta=\frac{\pi}{2}$}
%\rput[l]{90}(-8.9,4){$r\left(m\right)$}
%\rput[l](-0.5,1.35){$m$}
%\rput[l](-7.5,1.35){a)}
\rput[l](-6.5,5.5){$a=0$}
\rput[l](-7.5,0.2){a)}

\end{pspicture}
\end{center}
\end{minipage}\hfill
\begin{minipage}[t]{.55\linewidth}
\begin{center}
%\begin{pspicture}(4,0.5)(0,6)

\begin{pspicture}(6.1,0.5)(2.1,6)

\includegraphics[width=\linewidth]{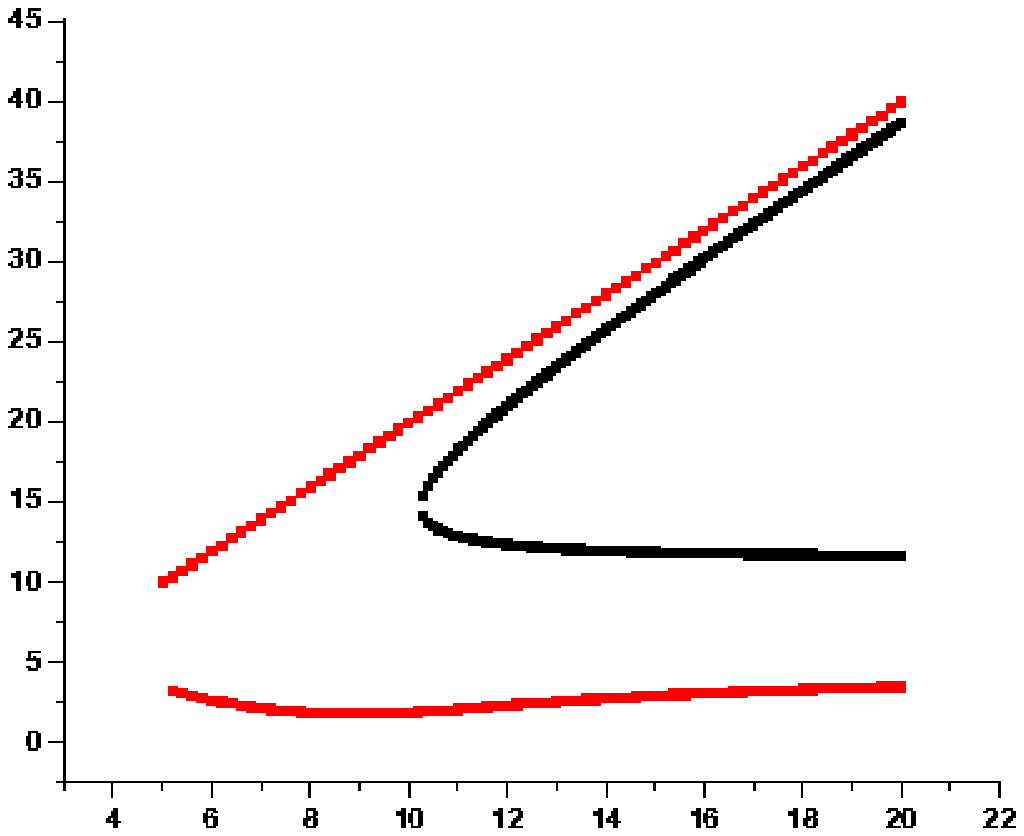}
%\fbox{
%\includegraphics[width=\linewidth]{/home/etuiran/Promo/Progs/Data/set5/q5s5.eps}
%%\includegraphics[width=\linewidth]{/home/etuiran/Promo/Progs/Data/set6/a5CQs6.eps}
%}
%\rput[l](-6.5,5.5){$a=5\ ,\ \bar{w}=5\ ,\ \theta=\frac{\pi}{2}$}

%\rput[l](-6.5,5.5){$a=5$}
%\rput[l](-4.5,-0.5){b)}

%\rput[l](-1,0.2){$m$}
%\rput[l]{90}(-7.3,2.7){$r\left(m\right)$}

\rput[l](-1,0.2){$m$}
\rput[l]{90}(-8,3.2){$r\left(m\right)$}
\rput[l](-6.5,5.5){$a=5$}
\rput[l](-7.5,0.2){b)}

\end{pspicture}
\end{center}
\end{minipage}\hfill
\begin{minipage}[t]{.55\linewidth}
\begin{center}
%\begin{pspicture}(1.2,2.5)(4.4,8)
%\begin{pspicture}(1,2.5)(4.2,8)

\begin{pspicture}(1,2.5)(4.5,8.3)

\includegraphics[width=\linewidth]{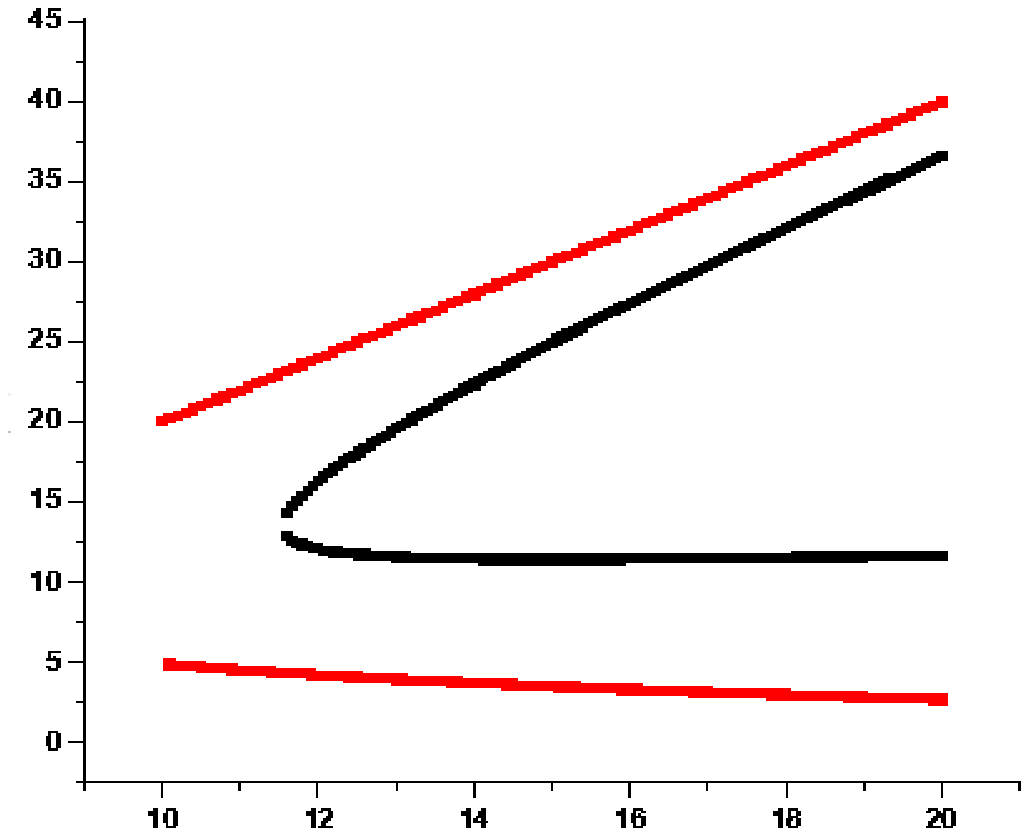}
%\includegraphics[width=\linewidth]{/home/etuiran/Promo/Progs/Data/set5/q10s5_II_09.eps}
%\fbox{
%\includegraphics[width=\linewidth]{/home/etuiran/Promo/Progs/Data/set5/q10s5.eps}
%%\includegraphics[width=\linewidth]{/home/etuiran/Promo/Progs/Data/set6/a10CQs6.eps}
%}
%\rput[l](-6.5,5.5){$a=10\ ,\ \bar{w}=5\ ,\ \theta=\frac{\pi}{2}$}

%\rput[l](-6.5,5.5){$a=10$}
%\rput[l](-4.5,-0.5){c)}

%\rput[l](-1,0.2){$m$}
%\rput[l]{90}(-7.3,2.7){$r\left(m\right)$}

\rput[l]{90}(-8,3.2){$r\left(m\right)$}
\rput[l](-6.5,5.5){$a=10$}

\rput[l](-7.5,0.2){c)}
\rput[l](-1,0.2){$m$}

\end{pspicture}
\end{center}
\end{minipage}\hfill
\begin{minipage}[t]{.55\linewidth}
\begin{center}
%\begin{pspicture}(-.7,2.5)(2.3,8)
\begin{pspicture}(1.1,2.8)(3.6,8.3)

\includegraphics[width=\linewidth]{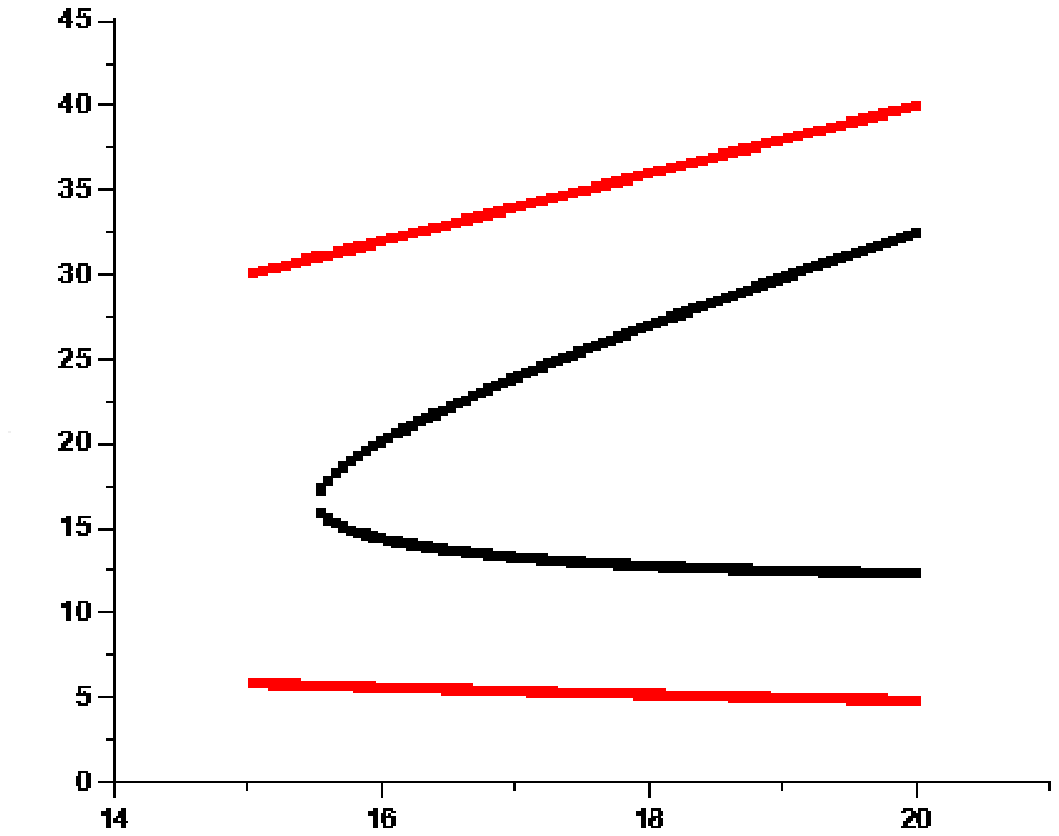}

%\includegraphics[width=\linewidth]{/home/etuiran/Promo/Progs/Data/set5/q15s5_II_09.eps}
%\fbox{
%\includegraphics[width=\linewidth]{/home/etuiran/Promo/Progs/Data/set5/q15s5.eps}
%%\includegraphics[width=\linewidth]{/home/etuiran/Promo/Progs/Data/set6/a15CQs6.eps}
%}
%\rput[l](-1,0.2){$m$}
%\rput[l]{90}(-7.3,2.7){$r\left(m\right)$}
%\rput[l](-6.5,5.5){$a=15\ ,\ \bar{w}=5\ ,\ \theta=\frac{\pi}{2}$}

%\rput[l](-6.5,5.5){$a=15$}
%\rput[l](-4.5,-0.5){d)}

\rput[l](-1,0.2){$m$}
\rput[l]{90}(-8,3.2){$r\left(m\right)$}
\rput[l](-6.5,5.5){$a=15$}
\rput[l](-7.5,0.2){d)}

\end{pspicture}
\end{center}
\end{minipage}
%\centerline{}
%\centerline{}
%\centerline{}
%\centerline{}
%\centerline{}
\centerline{}
\centerline{}
\centerline{}
\centerline{}
\centerline{}
\centerline{}
\centerline{}
\centerline{}
\centerline{}
\begin{center}
\rput[l](-9.2,3){Fig. 7:}
\rput[l](-7.5,3){The figures show the $m$-dependence of the improved  radii $r^{\text{I}}_{b\pm}$ obtained from the}
\rput[l](-7.5,2.5){exact $d\left(r\right)$ given in (\ref{2.5}), (\ref{2.6})
at $\theta=\frac{\pi}{2}$, for $\bar{w}=5$ and several values of $a$. The}
\rput[l](-7.5,2){outer curves are the improved static limits $r^{\text{I}}_{S\pm}$, the inner ones the improved event}
\rput[l](-7.5,1.5){horizons $r^{\text{I}}_{\pm}$. The structure of the curves is essentially the same as in the $d\left(r\right)=r$}
\rput[l](-7.5,1){approximation.}
%\rput[l](-7.7,0){An extremal configuration is visible in the improved Schwarzschild spacetime.}
%\rput[l](-7.7,-0.5){Figs. 5.21 to 5.23: The range from $m=a$ to $m=20$ is shown for the improved $H$'s and $S$'s.}
%\rput[l](-7.7,-1){The pattern of curves is similar as in the $d\left(r\right)=r$ approximation.}
\end{center}
\newpage

\section{Penrose process}

One of the most remarkable features of rotating black holes is the
possibility of extracting energy from them, by means of the Penrose process
for instance \cite{Taylor}. This is possible since under certain kinematical conditions
test particles in the Kerr metric can be in a state of negative energy. In
fact, let us consider a composite system \rm{A}, consisting of two particles \rm{B}
and C, which crosses the static limit. It disintegrates into B and C near
the event horizon whereby particle B is in a state of negative energy.
Subsequently B falls through the horizon, thus making a negative
contribution to the black hole\'{}s internal energy. The other particle, C, leaves the ergosphere and reaches
its final state at infinity. The conservation of the total energy for the
black hole and the test particles implies an increased energy for the test
particle C. The energy it gains equals minus the change in the internal
energy of the black hole.

As the possibility of energy extraction is intimately limited to the
existence of negative energy states we shall now analyze this issue for the
improved Kerr metric in order get a first impression of the impact the
quantum gravity corrections have on the region of the test particle phase
space with $E<0$.

The conserved energy of a point particle is given by eq. (\ref{3.6}). If we
use BL coordinates and parametrize its trajectory by the proper time $\tau $
we have explicitly, with the angular velocity $\Omega \equiv d\varphi /dt$,
\begin{eqnarray}
E =-mt^{\mu }g_{\mu \nu }\frac{dx^{\nu }}{d\tau }  \label{5.1}
=-m\left[ g_{tt}\frac{dt}{d\tau }+g_{\varphi t}\frac{d\varphi }{d\tau }
\right]=-m\left[ g_{tt}+g_{\varphi t}\Omega \right] \frac{dt}{d\tau }
\end{eqnarray}
Using the explicit form of the improved Kerr metric the negative energy
constraint $E\leq 0$ boils down to
\begin{equation}
\Omega \leq \Omega _{0}\equiv -\frac{g_{tt}}{g_{\varphi t}}=\frac{2MG\left(
r\right) r-\rho ^{2}}{2MG\left( r\right) ra\sin ^{2}\theta }  \label{5.2}
\end{equation}
Following \cite{Taylor} it is convenient to reexpress the inequality (\ref{5.2})
in terms of the tangential ``bookkeeper velocity''
\begin{equation}
v_{\text{tan}} \equiv R\left( r,\theta \right) \frac{d\varphi }{dt}=R\left(
r,\theta \right) \Omega  \label{5.3}
\end{equation}
with the reduced circumference
\begin{equation}
R\left( r,\theta \right) \equiv \sqrt{g_{\varphi \varphi }}=\sqrt{\frac{%
\Sigma _{I}\sin ^{2}\theta }{\rho ^{2}}}  \label{5.4}
\end{equation}
(The reduced circumference is defined such that $ds^{2}=R^{2}d\varphi ^{2}$
if $dt=dr=d\theta =0$.) In terms of $v_{\text{tan}}$ the negative energy condition $\Omega
\leq \Omega _{0}$ becomes $v_{\text{tan}}\leq R\Omega _{0}$, or
\begin{equation}
v_{\text{tan}}\left( r\right) \leq v_{0}\equiv R\left( r,\theta \right)
\left( \frac{2MG\left( r\right) r-\rho ^{2}}{2MG\left( r\right) ra\sin
^{2}\theta }\right)  \label{5.5}
\end{equation}

In the following we shall restrict our analysis to the equatorial plane, $%
\theta =\pi /2$. In this case the condition (\ref{5.5}) assumes the form
\begin{equation}
v_{\text{tan}}\left( r\right) \leq \frac{1}{a}\sqrt{r^{2}+a^{2}+\frac{%
2Ma^{2}G\left( r\right) }{r}}\left( 1-\frac{r}{2MG\left( r\right) }\right)
=v_{0}^{\rm{eq}}\left( r\right)  \label{5.6}
\end{equation}
Here $v_{0}^{\rm{eq}}$ denotes the bookkeeper tangential
velocity, i.e. the velocity refering to the \textit{coordinate} time $t$ of a
particle which moves in the equatorial plane and has vanishing energy, $E=0$.

The phase space for the rotational motion of a massive test particle is
bounded by the $v\left( r\right) $-curves for co- and counter-rotating light
rays:
\begin{equation}
v_{-}^{\rm{light}}\left( r\right) <v_{\text{tan}}\left(
r\right) <v_{+}^{\rm{light}}\left( r\right)  \label{5.7}
\end{equation}
The bookkeeper tangential velocities for light follow from (\ref{3.12}):
\begin{equation}
v_{\pm }^{\rm{light}}\left( r\right) =R\left( r,\theta
\right) \Omega _{\pm }=R\left( r,\theta \right) \left( \omega \pm \sqrt{%
\omega ^{2}-\frac{g_{tt}}{g_{\varphi \varphi }}}\right)  \label{5.8}
\end{equation}
In the $\left( r,v_{\text{tan}}\right) $-plane, the part of the test
particle phase space corresponding to $E<0$ is obtained by intersecting the
regions defined by the inequalities (\ref{5.6}) and (\ref{5.7}),
respectively.

The situation is sketched qualitatively in Fig. 8. Besides $v_{\pm }^{\rm{light}}$ and $v_{0}^{\rm{eq}}$ 
the figure shows also the $r$-dependence of the dragging velocity $v_{\text{dragging}}=R\left(
r,\theta \right) \omega $. It is not difficult to prove that for any 
function $G\left( r\right)$, the $v_{0}^{\rm{eq}}$- and $v_{-}^{\rm{light}}$-curves intersect at the static limit $\left( r=r_{S_{+}}\right) $, and that $v_{0}^{\rm{eq}}=v_{+}^{\rm{light}}=v_{-}^{\rm{light}}=v_{\text{dragging}}$ at the horizon $\left( r=r_{+}\right) $.

In Figs. 9 and 10 we show the corresponding realistic plots which were
obtained numerically. Fig. 9 corresponds to the classical, and Fig. 10 to the
improved case. All plots refer to the equator, $\theta =\pi /2$, and in the
improved case the function $G\left( r\right) $ was taken as in eq. (\ref{4.3}%
) with $d\left( r\right) =r$. Next to each $\left( r,v\right)$-plot we
display the $M$-dependence of the radii $r_{\pm }$, $r_{S\pm }$ and indicate
by a dashed vertical bar the $M$-value used in the corresponding plot on the
LHS. This presentation makes it obvious if, and how many critical surfaces
exist for the corresponding $M$-value.

When varying the mass $m=MG$ in the Figs. 9 and 10 we keep the ratio $a/m$
fixed. The reason is that, classically, $r_{\pm }$ and $r_{S\pm }$ are
\textit{linear} functions of $m$ if we readjust $a$ such that
$a/m=const$; see eqs. (\ref{1.8}), (\ref{1.9}). As a consequence, the
negative energy region for the classical metric changes its size with $m$, but not its shape. This
can be seen in Fig. 9. Hence changes of the shape are entirely due to the
quantum effects.

Fig. 10a) shows the region of negative energy for $M=5m_{\text{pl}}$, $a=4.5$%
. Since we are still sufficiently away from the Planck region the shape of
the improved negative energy region is not too different from the
classical one. In Fig. 10b) we have changed $M$ from 5 to 4 Planck masses
for which the shape of the negative energy region is almost
unchanged. Besides the $E<0$ region discussed above figures 10a) and 10b) show an internal negative
energy region bounded by $r_{S_-}^{\text{I}}$ and $r_{-}^{\text{I}}$. Since
the possibility of extraction of energy relies on the existence of
stationary states with negative energy outside $r_{+}^{\text{I}}$, the
internal region cannot be considered physically relevant.

Figures 10c) to 10f) were obtained for the regime $M\approx m_{\text{pl}}$.
Drastic changes in the shape of the negative energy regions are visible. Since
the reliability of our method is questionable in this regime, conclusions about this region have to be considered 
with some care. We analyse
these cases nevertheless since they hint at the possibility of interesting
new features.

In Fig. 10c) the quantum extremal black hole with $M=M_{cr}$ and $r_{-}^{%
\text{I}}=r_{+}^{\text{I}}=r_{\text{extr}}^{\text{I}}$ has been reached. The
internal and external negative energy regions touch at $r_{\text{extr}}^{%
\text{I}}$.

Fig. 10d) shows a hypothetical configuration for $M<M_{\text{cr}}$ with two
static limits $S_{\pm }^{\text{I}}$ and no event horizon. The internal and
external negative energy regions merged into just one. This region is
bounded by the static limit surfaces at $r_{S_-}^{\text{I}}$ and $r_{S_+}^{\text{I}}$.
In this case there exists an ergosphere from where energy can be
extracted, but no horizons.

Figures 10e) and f) show configurations in which no extraction of energy is
possible. At the extremal static limit configuration shown in figure 10e)
the negative energy region is reduced to zero size.

This analysis suggests that, while it is possible to extract energy from
classical black holes with arbitrarily small masses and angular momenta,
there exists a lowest mass for the Penrose mechanism in the improved Kerr
spacetime. It is close to the Planck mass and defined by the extremal static
limit. However, since the reliability of our method is questionable in the
regime $M\approx m_{\rm{Pl}}$, it would be desirable to investigate this
possibility by independent methods.

%\centerline{}
%\centerline{}
%\centerline{}
%\centerline{}
%\centerline{}
%\centerline{}
%\centerline{}
%\begin{minipage}[t]{.65\linewidth}
%\begin{center}
%\begin{pspicture}(-1.5,1.5)(1.5,8)
%\includegraphics[width=\linewidth]{Penrose_Scheme_II_09_Ver_2.eps}
%\rput[l](0,4.9){\mbox{$r$}}
%\rput[l](-10.8,8){\mbox{$v(r)$}}
%\rput[l](-2.1,6.8){\mbox{$v_+^{\rm light}$}}
%\rput[l](-2.1,3.87){\mbox{$v_-^{\rm light}$}}
%\rput[l](-2.1,5.7){\mbox{$v_{\rm dragging}$}}
%\rput[l](-4.5,3.2){\mbox{$v_0^{\rm eq}$}}
%\rput[l](-7,4.9){\mbox{$r_{+}$}}
%\rput[l](-5.8,4.9){\mbox{$r_{S+}$}}
%\end{pspicture}
%\end{center}
%\end{minipage}

\centerline{}

\begin{minipage}[t]{.65\linewidth}
\begin{center}
%\begin{pspicture}(-1.5,1.5)(1.5,8)
\begin{pspicture}(-1.5,0.1)(1.5,6.6)
\includegraphics[width=\linewidth]{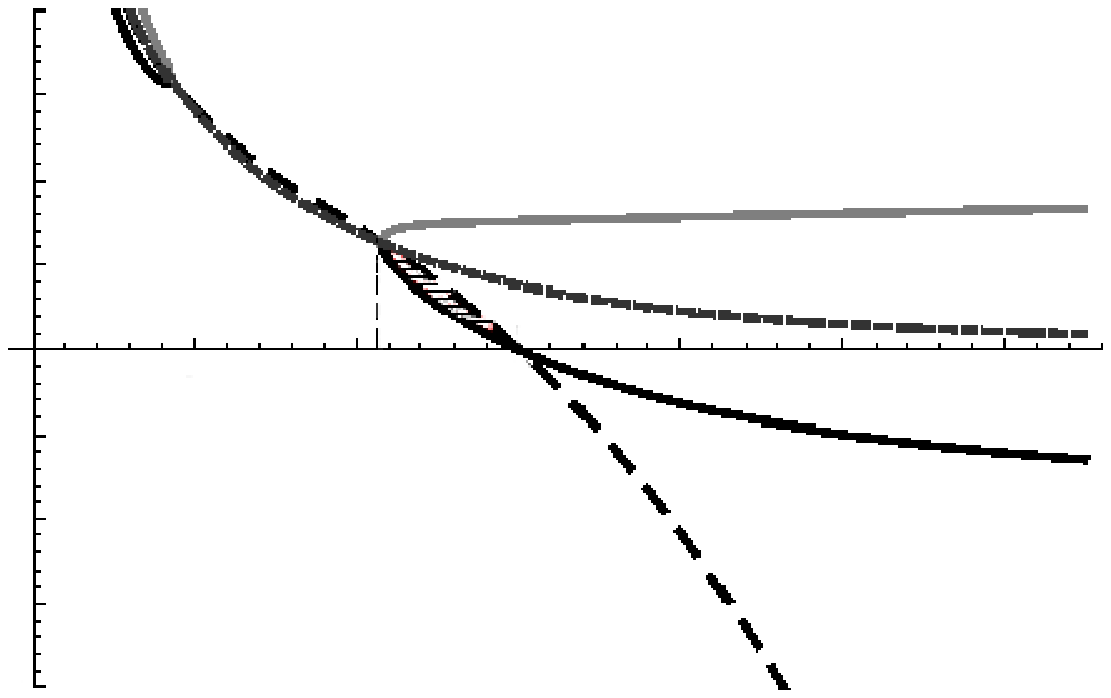}
\rput[l](0,3.3){\mbox{$r$}}
\rput[l](-10.8,6.8){\mbox{$v(r)$}}
\rput[l](-2.1,5.2){\mbox{$v_+^{\rm light}$}}
\rput[l](-2.1,2.27){\mbox{$v_-^{\rm light}$}}
\rput[l](-2.1,4.1){\mbox{$v_{\rm dragging}$}}
\rput[l](-4.5,1.68){\mbox{$v_0^{\rm eq}$}}
\rput[l](-7,3.3){\mbox{$r_{+}$}}
\rput[l](-5.8,3.3){\mbox{$r_{S+}$}}
\end{pspicture}
\end{center}
\end{minipage}
%\centerline{}
\centerline{}
\centerline{}
\centerline{}
\centerline{}
\centerline{}
\centerline{}
\begin{center}
\rput[l](-9.2,3){Fig. 8:}
\rput[l](-7.5,3){The figure shows schematically the $r$-dependence of $v_{\pm}^{\rm light}$, $v_0^{\rm eq}$ and $v_{\rm dragging}$}
\rput[l](-7.5,2.5){at the equatorial plane. The hatched region corresponds to pairs $\left(r,v\right)$ for which}
\rput[l](-7.5,2){the test particle has negative energy.}
%\rput[l](-7.5,1.5){horizons $r^{\text{I}}_{\pm}$. The structure of the curves is essentially the same as in the $d\left(r\right)=r$}
%\rput[l](-7.5,1){approximation.}
\end{center}

\newpage

\begin{minipage}[t]{.55\linewidth}
\begin{center}
%\begin{pspicture}(2,0.5)(5,6)
%\begin{pspicture}(1,0.5)(4,6)
%\begin{pspicture}(4,0.5)(6,6)
\begin{pspicture}(2.5,0.5)(5.5,6)
\includegraphics[width=\linewidth]{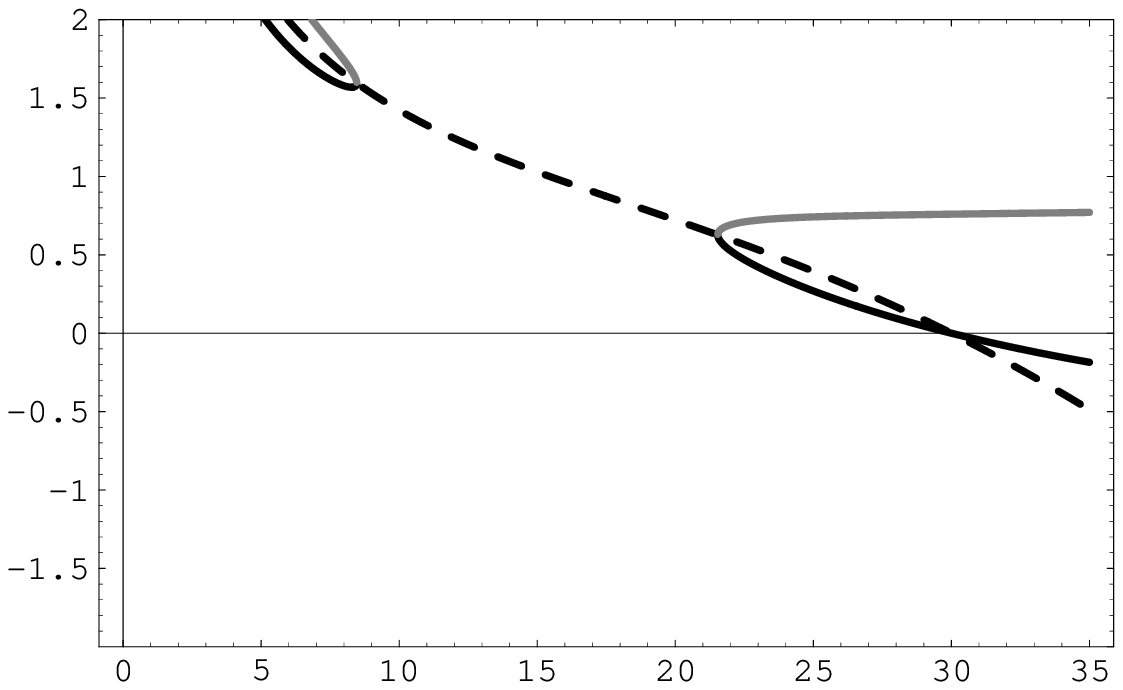}
%\rput[l](-6,7){M=15\ ,\ a=13.5\ ,\ $\bar{w}$=0}
\rput[l](-6,7.3){M=15\ ,\ a=13.5\ ,\ $\bar{w}$=0}
\rput[l](-.8,1.35){\mbox{$r$}}
%\rput[l](-9,4.2){\mbox{$v(r)$}}

\rput[l](-9,4.5){\mbox{$v(r)$}}

%\rput[l](-6.2,4.2){\mbox{$r_{-}$}}
%\rput[l](-3.1,4.2){\mbox{$r_{+}$}}
%\rput[l](-1.5,4.2){\mbox{$r_{S+}$}}

\rput[l](-6.2,4.2){\mbox{$r_{-}$}}
\rput[l](-3.3,4.2){\mbox{$r_{+}$}}
\rput[l](-1.5,4.2){\mbox{$r_{S+}$}}

%\psline[linestyle=dashed,linewidth=0.5pt](-4.9,3.08)(-7.05,3.08)    %h-
%\psline[linestyle=dashed,linewidth=0.5pt](-4.9,6.13)(-5.7,6.13)  %h+
\rput[l](-3.1,6.3){\tiny{Allowed}}
\rput[l](-3.1,6){\tiny{Negative Energy}}
\rput[l](-3.1,5.7){\tiny{Region}}
%\psline{->}(-2,5.8)(-2.3,4.8)

\psline{->}(-2,5.8)(-2.3,5)

\rput[l](-8,1.35){a)}
\end{pspicture}
\end{center}
\end{minipage}\hfill
\begin{minipage}[t]{.55\linewidth}
\begin{center}
%\begin{pspicture}(5,0.5)(1,6)
%\begin{pspicture}(4,0.5)(-0,6)
\begin{pspicture}(5.3,0.5)(1.3,6)
\includegraphics[width=\linewidth]{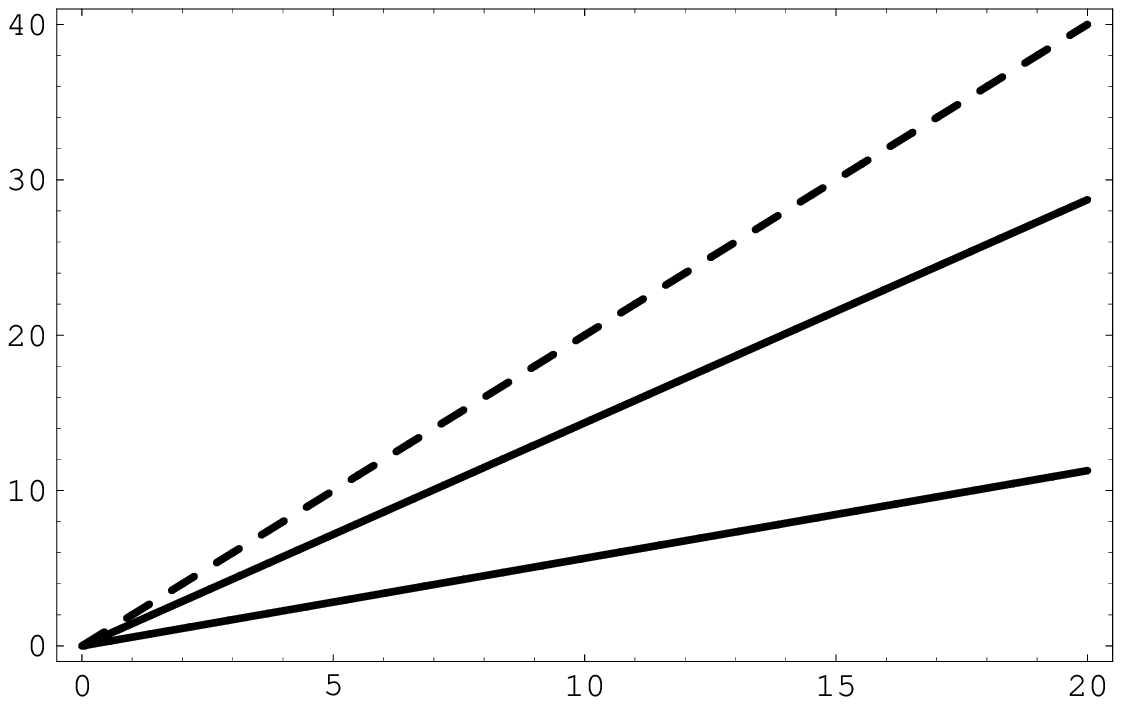}
\rput[l](-.8,1.35){\mbox{$M$}}
\rput[l](-9.5,4.1){\mbox{$r(M)$}}

\rput[l](-2.77,5.95){\mbox{$r_{S+}$}}
\rput[l](-2.63,5){\mbox{$r_+$}}
\rput[l](-2.63,3.3){\mbox{$r_-$}}

%\rput[l](-2.67,2.2){\mbox{$r_{S-}$}}
%\psline[linestyle=dashed,linewidth=0.5pt](-2.2,1.98)(-2.18,5.9)
\psline[linestyle=dashed,linewidth=0.5pt](-2.2,1.98)(-2.24,5.83)
%\rput[l](-8,1.35){b)}
\end{pspicture}
\end{center}
\end{minipage}\hfill
\begin{minipage}[t]{.55\linewidth}
\begin{center}
%\begin{pspicture}(0.5,1)(3.7,6.5)
%\begin{pspicture}(0.5,2)(3.7,7.5)
%\begin{pspicture}(1.5,0.4)(4.7,5.9)
\begin{pspicture}(1.5,1.4)(4.7,6.9)
\includegraphics[width=\linewidth]{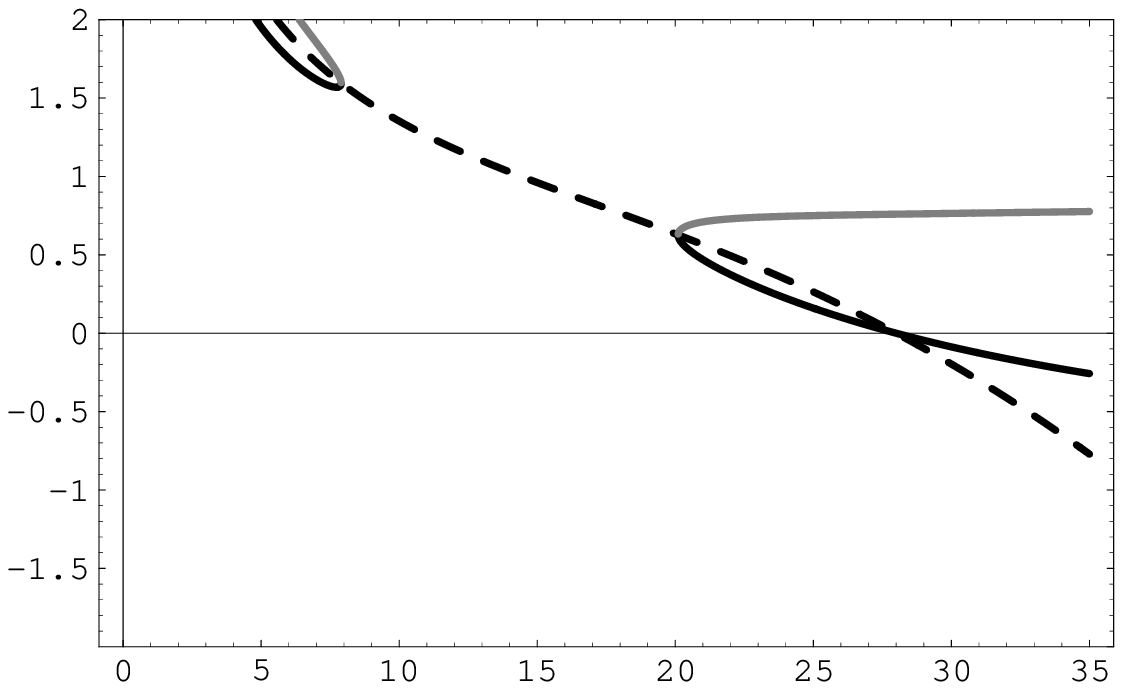}
\rput[l](-6,7.3){M=14\ ,\ a=12.6\ ,\ $\bar{w}$=0}
\rput[l](-.8,1.35){\mbox{$r$}}
\rput[l](-9,4.5){\mbox{$v(r)$}}

\rput[l](-3.1,6.3){\tiny{Allowed}}
\rput[l](-3.1,6){\tiny{Negative Energy}}
\rput[l](-3.1,5.7){\tiny{Region}}
\psline{->}(-2,5.8)(-2.4,4.9)

%\rput[l](-7.03,7.3){\tiny{Allowed}}
%\rput[l](-6.83,7.3){\tiny{Negative Energy}}
%\rput[l](-6.63,7.3){\tiny{Region}}
%\psline{->}(-6.5,7.7)(-5.40008,7.317079)

%\rput[l](-6.2,4.2){\mbox{$r_{-}$}}
%\rput[l](-3.3,4.2){\mbox{$r_{+}$}}
%\rput[l](-1.5,4.2){\mbox{$r_{S+}$}}

\rput[l](-6.2,4.2){\mbox{$r_{-}$}}
\rput[l](-3.6,4.2){\mbox{$r_{+}$}}
\rput[l](-1.92,4.2){\mbox{$r_{S+}$}}

%\psline[linestyle=dashed,linewidth=0.5pt](-4.9,2.95)(-7,2.95)    %h-
%\psline[linestyle=dashed,linewidth=0.5pt](-4.9,5.81)(-5.7,5.81)  %h+
\rput[l](-8,1.35){b)}
\end{pspicture}
\end{center}
\end{minipage}\hfill
\begin{minipage}[t]{.55\linewidth}
\begin{center}
%\begin{pspicture}(-.5,1)(2.5,6.5)
%\begin{pspicture}(-.5,2)(2.5,7.5)
\begin{pspicture}(0.5,1.5)(3.5,7)
\includegraphics[width=\linewidth]{Penrose_Poster8_1.tex_gr1.eps}
%\rput[l](-3.04,5.7){\mbox{$r_{S+}$}}
%\rput[l](-2.95,4.77){\mbox{$r_+$}}
%\rput[l](-2.95,3.22){\mbox{$r_-$}}
\rput[l](-3.20,5.77){\mbox{$r_{S+}$}}
\rput[l](-3.08,4.77){\mbox{$r_+$}}
\rput[l](-3.08,3.22){\mbox{$r_-$}}
%\psline[linestyle=dashed,linewidth=0.5pt](-2.64,1.98)(-2.64,5.25)
\psline[linestyle=dashed,linewidth=0.5pt](-2.62,1.98)(-2.62,5.6)
%\rput[l](-3.7,6.25){\mbox{$h_-$}}
%\rput[l](-2.3,6.25){\mbox{$s_-$}}
%\rput[l](-5.2,6.25){\mbox{$h_+$}}
%\rput[l](-6.2,6.25){\mbox{$s_+$}}
%\psline[linestyle=dashed,linewidth=0.5pt](-2.25,6.76)(-6.1,6.76)
\rput[l](-0.8,1){\mbox{$M$}}
\rput[l](-9.5,4.1){\mbox{$r(M)$}}
%\rput[l](-8,1.35){d)}
\end{pspicture}
\end{center}
\end{minipage}
\centerline{}
\centerline{}
\centerline{}
\centerline{}
\centerline{}
\centerline{}
\centerline{}
\begin{center}
\rput[l](-9.2,3){Fig. 9:}
\rput[l](-7.5,3){The plots on the LHS of this figure display the $r$-dependence of $v_{\pm}^{\rm light}$ and $v_0^{\rm eq}$ 
and}
\rput[l](-7.5,2.5){are analogous to Fig. 8. They refer to classical Kerr black holes with $M=15m_{\rm pl}$,}
\rput[l](-7.5,2){$a=13.5m_{\rm pl}$ and $M=15m_{\rm pl}$, $a=12.6m_{\rm pl}$, respectively. 
They have the same ratio}
\rput[l](-7.5,1.5){$a/m=0.9$. On the RHS the radius of the critical surfaces is displayed for all ma-}
\rput[l](-7.5,1){sses up to $20m_{\rm pl}$, for the constant ratio $a/m=0.9$ as in the corresponding plots on}
\rput[l](-7.5,0.5){the LHS. The dashed vertical line in the plots on the RHS symbolizes the mass va-}
\rput[l](-7.5,0){lues used on the LHS.}
%\rput[l](-7.5,1.5){horizons $r^{\text{I}}_{\pm}$. The structure of the curves is essentially the same as in the $d\left(r\right)=r$}
%\rput[l](-7.5,1){approximation.}
\end{center}

\newpage

\begin{minipage}[t]{.55\linewidth}
\begin{center}
%\begin{pspicture}(1,0.5)(4,6)
\begin{pspicture}(2.5,0.5)(5.5,6)
\includegraphics[width=\linewidth]{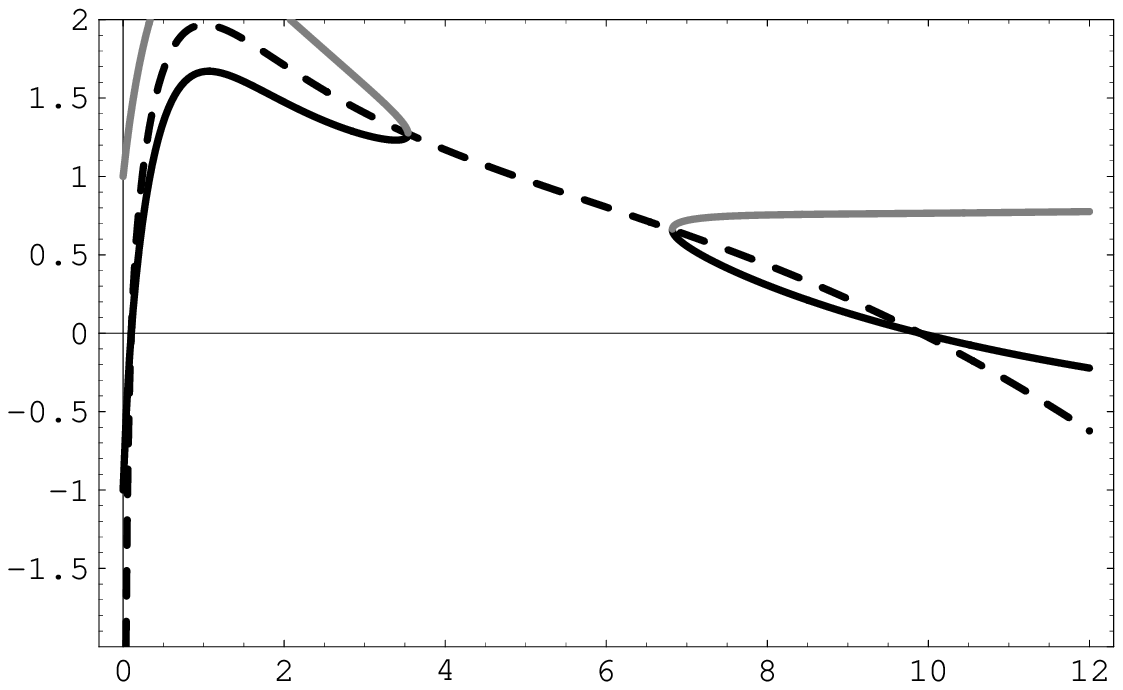}
\rput[l](-6,7.3){M=5\ ,\ a=4.5\ ,\ $\bar{w}$=1}

%\rput[l](-5.67,4.15){\mbox{$r^{\text{I}}_{-}$}}
%\rput[l](-7.74,4.15){\mbox{$r^{\text{I}}_{S-}$}}
%\rput[l](-3.52,4.15){\mbox{$r^{\text{I}}_{+}$}}
%\rput[l](-1.75,4.15){\mbox{$r^{\text{I}}_{S+}$}}

\rput[l](-5.7,4.15){\mbox{$r^{\text{I}}_{-}$}}
\rput[l](-7.74,4.15){\mbox{$r^{\text{I}}_{S-}$}}
\rput[l](-3.57,4.15){\mbox{$r^{\text{I}}_{+}$}}
\rput[l](-1.75,4.15){\mbox{$r^{\text{I}}_{S+}$}}

%\psline[linestyle=dashed,linewidth=0.5pt](-4.93,3.50)(-6.6,3.50) %h-
%\psline[linestyle=dashed,linewidth=0.5pt](-4.93,5.75)(-5.8,5.75) %h+
\rput[l](-.8,1.35){\mbox{$r$}}
\rput[l](-9,4.5){\mbox{$v(r)$}}
\rput[l](-3.1,6.3){\tiny{Allowed}}
\rput[l](-3.1,6){\tiny{Negative Energy}}
\rput[l](-3.1,5.7){\tiny{Region}}
\psline{->}(-2,5.8)(-2.5,5)
%\rput[l](-7.03,7.3){\tiny{Allowed}}
%\rput[l](-6.83,7.3){\tiny{Negative Energy}}
%\rput[l](-6.63,7.3){\tiny{Region}}
%\psline{->}(-6.5,7.7)(-5.60008,6.817079)
\rput[l](-8,1.35){a)}
\end{pspicture}
\end{center}
\end{minipage}\hfill
\begin{minipage}[t]{.55\linewidth}
\begin{center}
%\begin{pspicture}(4,0.5)(-0,6)
\begin{pspicture}(5.3,0.5)(1.3,6)
\includegraphics[width=\linewidth]{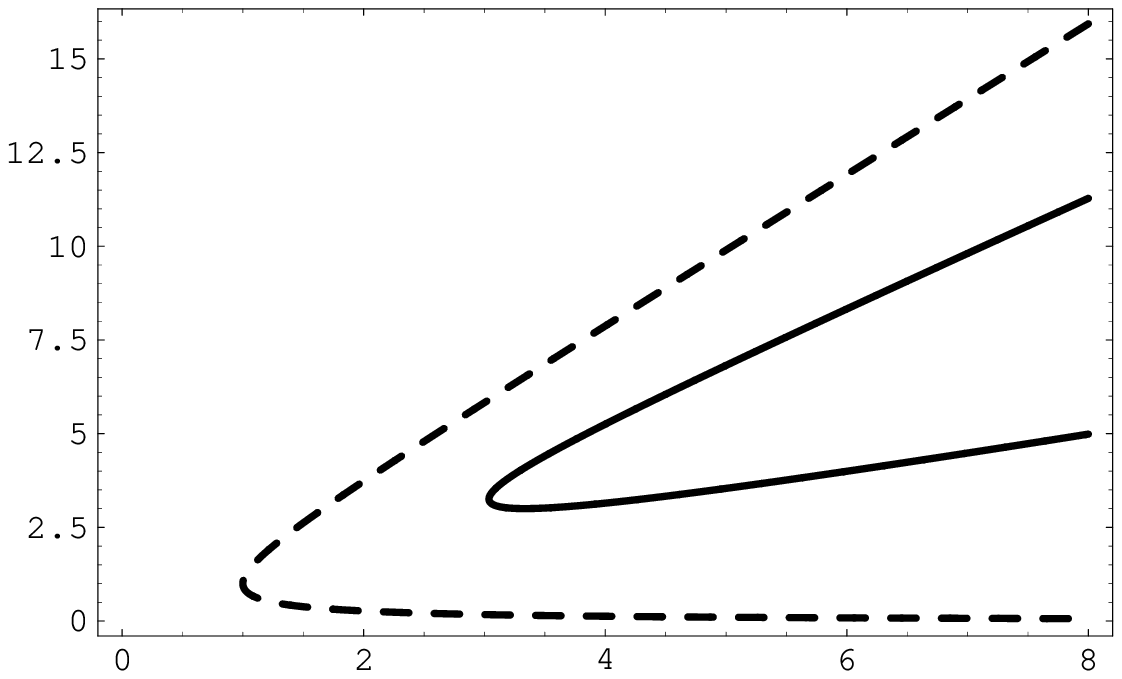}

%\rput[l](-.8,1.35){\mbox{$M$}}
%\rput[l](-9.5,4.1){\mbox{$r(M)$}}
%\rput[l](-3.44,5.1){\mbox{$r^{\text{I}}_{S+}$}}
%\rput[l](-3.35,4.25){\mbox{$r^{\text{I}}_+$}}
%\rput[l](-3.35,3.35){\mbox{$r^{\text{I}}_-$}}
%\rput[l](-3.54,2.4){\mbox{$r^{\text{I}}_{S-}$}}

\rput[l](-.8,1.35){\mbox{$M$}}
\rput[l](-9.5,4.1){\mbox{$r(M)$}}
\rput[l](-3.64,5.37){\mbox{$r^{\text{I}}_{S+}$}}
\rput[l](-3.52,4.44){\mbox{$r^{\text{I}}_+$}}
\rput[l](-3.48,3.45){\mbox{$r^{\text{I}}_-$}}
\rput[l](-3.64,2.49){\mbox{$r^{\text{I}}_{S-}$}}

\psline[linestyle=dashed,linewidth=0.5pt](-3.08,2.1)(-3.08,5.15)
\end{pspicture}
\end{center}
\end{minipage}\hfill
\begin{minipage}[t]{.55\linewidth}
\begin{center}
%\begin{pspicture}(0.5,2)(3.7,7.5)
%\begin{pspicture}(1.5,0.4)(4.7,5.9)
\begin{pspicture}(1.5,1.4)(4.7,6.9)
\includegraphics[width=\linewidth]{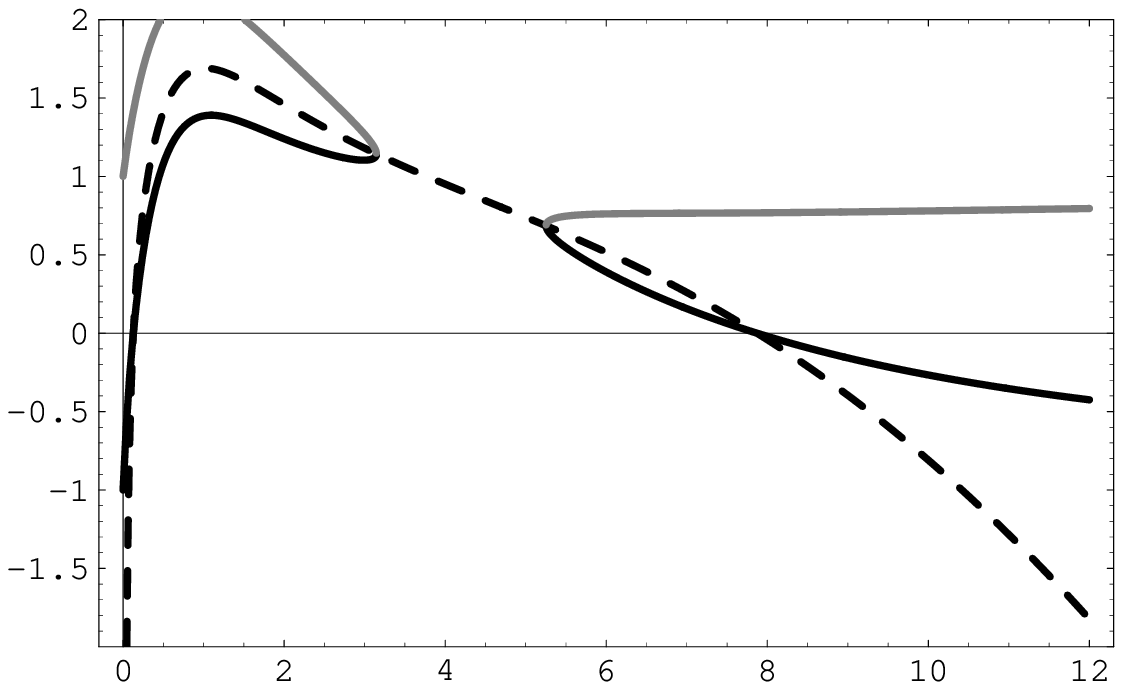}
\rput[l](-6,7.3){M=4\ ,\ a=3.6\ ,\ $\bar{w}$=1}
\rput[l](-.8,1.35){\mbox{$r$}}
\rput[l](-9,4.5){\mbox{$v(r)$}}

%\rput[l](-5.75,4.2){\mbox{$r^{\text{I}}_{-}$}}
%\rput[l](-7.74,4.2){\mbox{$r^{\text{I}}_{S-}$}}
%\rput[l](-4.4,4.2){\mbox{$r^{\text{I}}_{+}$}}
%\rput[l](-2.9,4.2){\mbox{$r^{\text{I}}_{S+}$}}

\rput[l](-6,4.2){\mbox{$r^{\text{I}}_{-}$}}
\rput[l](-7.74,4.2){\mbox{$r^{\text{I}}_{S-}$}}
\rput[l](-4.6,4.2){\mbox{$r^{\text{I}}_{+}$}}
\rput[l](-3.2,4.2){\mbox{$r^{\text{I}}_{S+}$}}

%\rput[l](-5.7,4.15){\mbox{$r^{\text{I}}_{-}$}}
%\rput[l](-7.74,4.15){\mbox{$r^{\text{I}}_{S-}$}}
%\rput[l](-3.57,4.15){\mbox{$r^{\text{I}}_{+}$}}
%\rput[l](-1.75,4.15){\mbox{$r^{\text{I}}_{S+}$}}

\rput[l](-4.1,6.3){\tiny{Allowed}}
\rput[l](-4.1,6){\tiny{Negative Energy}}
\rput[l](-4.1,5.7){\tiny{Region}}
\psline{->}(-3,5.8)(-3.5,5)
%\psline[linestyle=dashed,linewidth=0.5pt](-4.9,3.25)(-6.42,3.25) %h-
%\psline[linestyle=dashed,linewidth=0.5pt](-4.9,4.65)(-5.85,4.65)   %h+
%\rput[l](-7.03,7.3){\tiny{Allowed}}
%\rput[l](-6.83,7.3){\tiny{Negative Energy}}
%\rput[l](-6.63,7.3){\tiny{Region}}
%\psline{->}(-6.5,7.7)(-5.60008,6.817079)
\rput[l](-8,1.35){b)}
\end{pspicture}
\end{center}
\end{minipage}\hfill
\begin{minipage}[t]{.55\linewidth}
\begin{center}
%\begin{pspicture}(-.5,2)(2.5,7.5)
\begin{pspicture}(0.5,1.5)(3.5,7)
\includegraphics[width=\linewidth]{Penrose_Poster7_1.tex_gr1.eps}
\rput[l](-.8,1.35){\mbox{$M$}}
\rput[l](-9.5,4.1){\mbox{$r(M)$}}
%\rput[l](-4.34,4.54){\mbox{$r^{\text{I}}_{S+}$}}
%\rput[l](-4.25,3.74){\mbox{$r^{\text{I}}_+$}}
%\rput[l](-4.18,3.15){\mbox{$r^{\text{I}}_-$}}
%\rput[l](-4.38,2.38){\mbox{$r^{\text{I}}_{S-}$}}

\rput[l](-4.58,4.8){\mbox{$r^{\text{I}}_{S+}$}}
\rput[l](-4.5,3.97){\mbox{$r^{\text{I}}_+$}}
\rput[l](-4.48,3.33){\mbox{$r^{\text{I}}_-$}}
\rput[l](-4.58,2.5){\mbox{$r^{\text{I}}_{S-}$}}

%\psline[linestyle=dashed,linewidth=0.5pt](-3.78,1.98)(-3.78,4.25)
\psline[linestyle=dashed,linewidth=0.5pt](-4.03,2.1)(-4.03,4.58)
\end{pspicture}
\end{center}
\end{minipage}

\newpage

\begin{minipage}[t]{.55\linewidth}
\begin{center}
%\begin{pspicture}(1,0.5)(4,6)
\begin{pspicture}(2.5,0.5)(5.5,6)
\includegraphics[width=\linewidth]{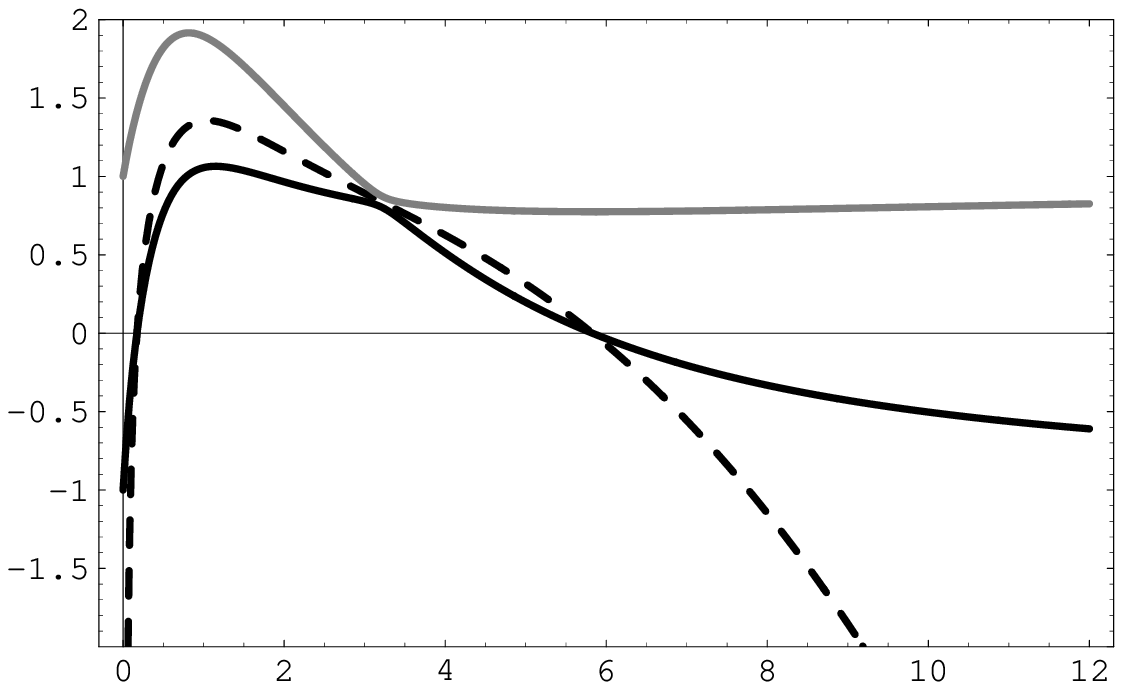}
\rput[l](-6,7.3){M=3\ ,\ a=2.7\ ,\ $\bar{w}$=1}

\rput[l](-5.9,4.15){\mbox{$r^{\text{I}}_{-}$}}
\rput[l](-5.9,3.55){\mbox{$r^{\text{I}}_{+}$}}
\rput[l](-7.6,4.15){\mbox{$r^{\text{I}}_{S-}$}}
\rput[l](-4.35,4.15){\mbox{$r^{\text{I}}_{S+}$}}

\rput[l](-5.1,6.3){\tiny{Allowed}}
\rput[l](-5.1,6){\tiny{Negative Energy}}
\rput[l](-5.1,5.7){\tiny{Region}}
\psline{->}(-4,5.8)(-4.55,4.9)

%\psline[linestyle=dashed,linewidth=0.5pt](-4.91,3.3)(-6,3.3)
\rput[l](-.8,1.35){\mbox{$r$}}
\rput[l](-9,4.5){\mbox{$v(r)$}}
%\rput[l](-7.03,3.3){\tiny{Allowed}}
%\rput[l](-6.83,3.3){\tiny{Negative Energy}}
%\rput[l](-6.63,3.3){\tiny{Region}}
%\psline{->}(-6.5,3.7)(-5.60008,4.1)
\rput[l](-8,1.35){c)}
\end{pspicture}
\end{center}
\end{minipage}\hfill
\begin{minipage}[t]{.55\linewidth}
\begin{center}
\begin{pspicture}(5.3,0.5)(1.3,6)
\includegraphics[width=\linewidth]{Penrose_Poster7_1.tex_gr1.eps}
\rput[l](-.8,1.35){\mbox{$M$}}
\rput[l](-9.5,4.1){\mbox{$r(M)$}}

%\rput[l](-5.2,4){\mbox{$r^{\text{I}}_{S+}$}}
%\rput[l](-5.4,3.1){\mbox{$r^{\text{I}}_{\text{extr}}$}}
%\rput[l](-5.33,2.45){\mbox{$r^{\text{I}}_{S-}$}}

\rput[l](-5.53,4.18){\mbox{$r^{\text{I}}_{S+}$}}
\rput[l](-5.73,3.3){\mbox{$r^{\text{I}}_{\text{extr}}$}}
\rput[l](-5.53,2.49){\mbox{$r^{\text{I}}_{S-}$}}

%\psline[linestyle=dashed,linewidth=0.5pt](-4.68,1.98)(-4.68,3.71)
\psline[linestyle=dashed,linewidth=0.5pt](-5,2.1)(-5,3.81)

\end{pspicture}
\end{center}
\end{minipage}\hfill
\begin{minipage}[t]{.55\linewidth}
\begin{center}
\begin{pspicture}(1.5,1.4)(4.7,6.9)
\includegraphics[width=\linewidth]{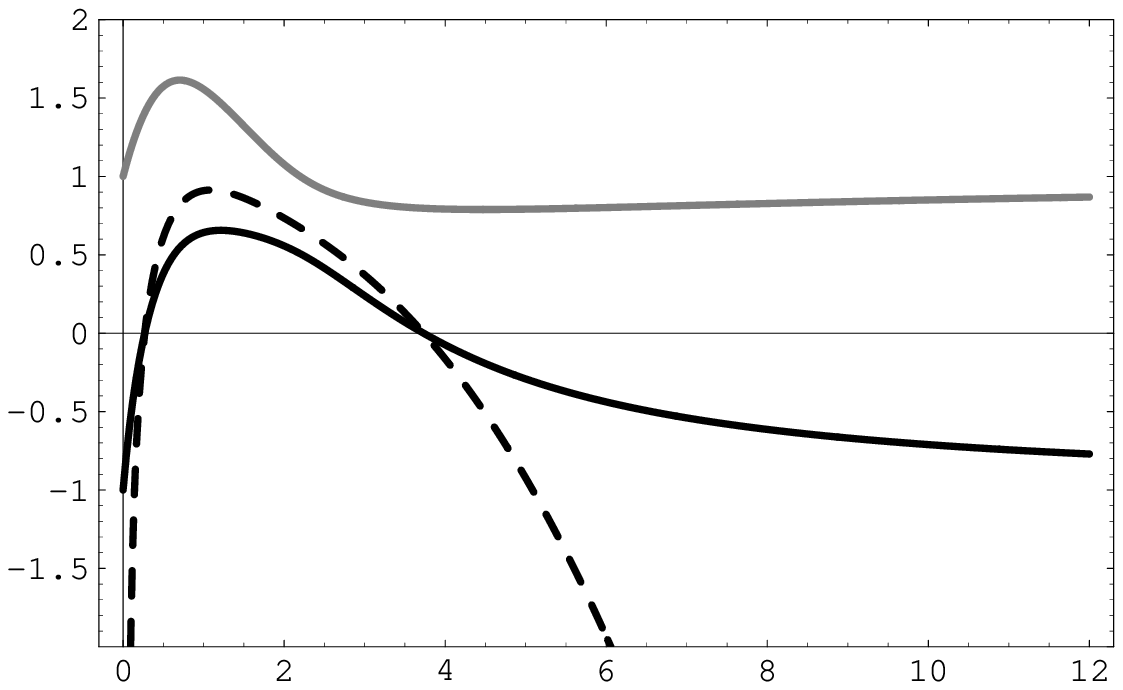}
\rput[l](-6,7.3){M=2\ ,\ a=1.8\ ,\ $\bar{w}$=1}

\rput[l](-7.62,4.15){\mbox{$r^{\text{I}}_{S-}$}}
\rput[l](-5.77,4.15){\mbox{$r^{\text{I}}_{S+}$}}
\rput[l](-.8,1.35){\mbox{$r$}}
\rput[l](-9,4.5){\mbox{$v(r)$}}

\rput[l](-6.1,6.3){\tiny{Allowed}}
\rput[l](-6.1,6){\tiny{Negative Energy}}
\rput[l](-6.1,5.7){\tiny{Region}}
\psline{->}(-5,5.8)(-6,5.15)

%\rput[l](-7.03,3.3){\tiny{Allowed}}
%\rput[l](-6.83,3.3){\tiny{Negative Energy}}
%\rput[l](-6.63,3.3){\tiny{Region}}
%\psline{->}(-6.5,3.7)(-5.60008,3.1)
\rput[l](-8,1.35){d)}
\end{pspicture}
\end{center}
\end{minipage}\hfill
\begin{minipage}[t]{.55\linewidth}
\begin{center}

\begin{pspicture}(0.5,1.5)(3.5,7)
\includegraphics[width=\linewidth]{Penrose_Poster7_1.tex_gr1.eps}
\rput[l](-6.5,3.61){\mbox{$r^{\text{I}}_{S+}$}}
\rput[l](-6.56,2.6){\mbox{$r^{\text{I}}_{S-}$}}
\psline[linestyle=dashed,linewidth=0.5pt](-5.95,2.1)(-5.95,3.19)
\rput[l](-.8,1.35){\mbox{$M$}}
\rput[l](-9.5,4.1){\mbox{$r(M)$}}
\end{pspicture}
\end{center}
\end{minipage}
\newpage

\begin{minipage}[t]{.55\linewidth}
\begin{center}
%\begin{pspicture}(1,0.5)(4,6)
\begin{pspicture}(2.5,0.5)(5.5,6)
\includegraphics[width=\linewidth]{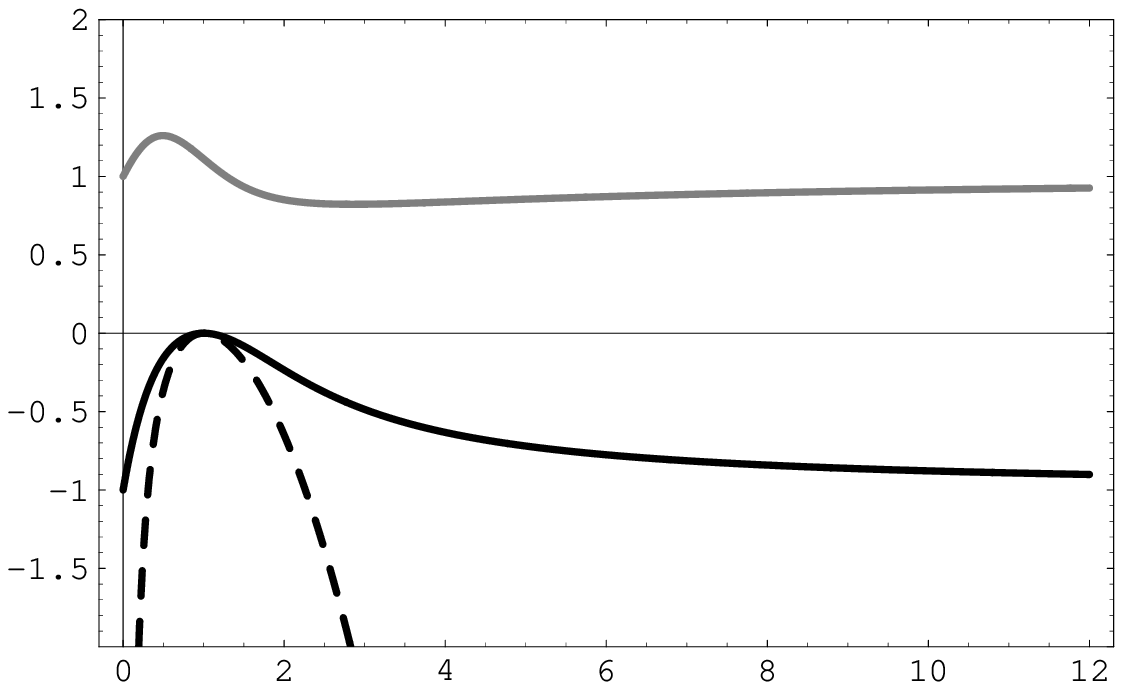}
\rput[l](-6,7.3){M=1\ ,\ a=0.9\ ,\ $\bar{w}$=1}
\rput[l](-7.37,4.05){\mbox{$r^{\text{I}}_{S-}$}}
\rput[l](-7.37,3.5){\mbox{$r^{\text{I}}_{S+}$}}
\rput[l](-.8,1.35){\mbox{$r$}}
\rput[l](-9,4.5){\mbox{$v(r)$}}

\rput[l](-5.1,6.3){\tiny{No Allowed}}
\rput[l](-5.1,6){\tiny{Negative Energy}}
\rput[l](-5.1,5.7){\tiny{Region}}

%\rput[l](-7.03,3.3){\tiny{No Allowed}}
%\rput[l](-6.83,3.3){\tiny{Negative Energy}}
%\rput[l](-6.63,3.3){\tiny{Region}}
\rput[l](-8,1.35){e)}
\end{pspicture}
\end{center}
\end{minipage}\hfill
\begin{minipage}[t]{.55\linewidth}
\begin{center}
%\begin{pspicture}(4,0.5)(-0,6)
\begin{pspicture}(5.3,0.5)(1.3,6)
\includegraphics[width=\linewidth]{Penrose_Poster7_1.tex_gr1.eps}
\rput[l](-.8,1.35){\mbox{$M$}}
\rput[l](-9.5,4.1){\mbox{$r(M)$}}

%\rput[l](-6.8,3.43){\mbox{$r^{\text{I}}_{S+}$}}
\rput[l](-7.85,2.65){\mbox{$r^{\text{I}}_{S\text{extr}}$}}
\psline[linestyle=dashed,linewidth=0.5pt](-6.9,2.1)(-6.9,2.48)
%\rput[l](-4.5,1.2){Fig. 6.17.}
%\psline[linestyle=dashed,linewidth=0.2pt](-2.3,2.08)(-2.7,2.08)
%\rput[l](-2.5,3.8){\mbox{$s_-$}}
%\rput[l](-2.88,3.8){\mbox{$s_+$}}
\end{pspicture}
\end{center}
\end{minipage}\hfill
\begin{minipage}[t]{.55\linewidth}
\begin{center}
%\begin{pspicture}(0.5,2)(3.7,7.5)
%\begin{pspicture}(1.5,0.4)(4.7,5.9)
\begin{pspicture}(1.5,1.4)(4.7,6.9)
\includegraphics[width=\linewidth]{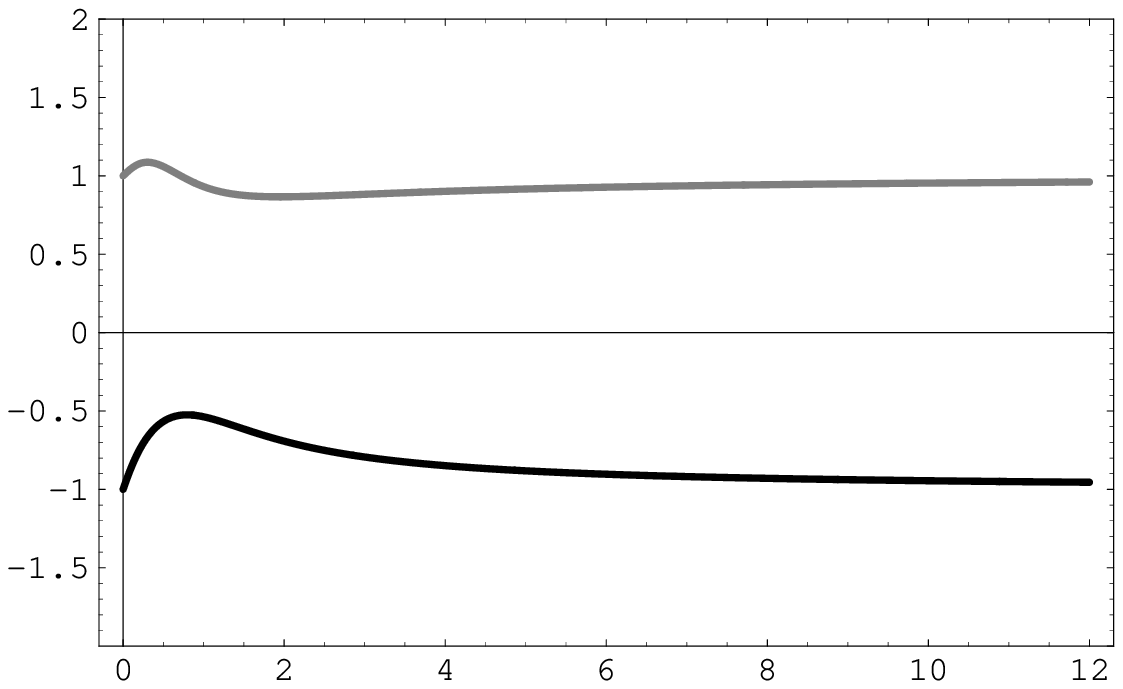}
\rput[l](-6,7.3){M=0.5\ ,\ a=0.45\ ,\ $\bar{w}$=1}
\rput[l](-.8,1.35){\mbox{$r$}}
\rput[l](-9,4.5){\mbox{$v(r)$}}
\rput[l](-8,1.2){f)}

\rput[l](-5.1,6.6){\tiny{No Allowed}}
\rput[l](-5.1,6.3){\tiny{Negative Energy}}
\rput[l](-5.1,6){\tiny{Region}}

%\rput[l](-7.03,3.3){\tiny{No Allowed}}
%\rput[l](-6.83,3.3){\tiny{Negative Energy}}
%\rput[l](-6.63,3.3){\tiny{Region}}
\end{pspicture}
\end{center}
\end{minipage}\hfill
\begin{minipage}[t]{.55\linewidth}
\begin{center}
%\begin{pspicture}(-.5,2)(2.5,7.5)
\begin{pspicture}(0.5,1.5)(3.5,7)
\includegraphics[width=\linewidth]{Penrose_Poster7_1.tex_gr1.eps}
%\psline[linestyle=dashed,linewidth=0.5pt](-2.28,1.57)(-7.68,1.57)
\psline[linestyle=dashed,linewidth=0.5pt](-7.37,2.1)(-7.37,7.03)
\rput[l](-.8,1){\mbox{$M$}}
\rput[l](-9.5,4.1){\mbox{$r(M)$}}
%\rput[l](-4.5,1.2){Fig. 6.19.}
\end{pspicture}
\end{center}
\end{minipage}
\begin{center}
\centerline{}
\centerline{}
\centerline{}
\centerline{}
\centerline{}
\centerline{}
\centerline{}
\rput[l](-9.2,3){Fig. 10:}
\rput[l](-7.5,3){The same type of plots as in Fig. 9, but now for the improved Kerr black hole, with masses}
\rput[l](-7.5,2.5){ranging from $M=5m_{\rm pl}$ down to $M=0.5m_{\rm pl}$. All examples considered have an identical}
\rput[l](-7.5,2){ratio $a/m=0.9$.}
%\rput[l](-7.5,1.5){horizons $r^{\text{I}}_{\pm}$. The structure of the curves is essentially the same as in the $d\left(r\right)=r$}
%\rput[l](-7.5,1){approximation.}
\end{center}

\newpage

\section{Vacuum energy-momentum tensor and \\ energy conditions} \label{Sec Vac}

We may reinterpret the RG improved vacuum Kerr metric $g_{\mu \nu }^{\text{%
imp}}$ as a \textit{classical} spacetime in presence of
matter. Knowing $g_{\mu \nu }^{\text{imp}}$ explicitly, we can compute its
Einstein tensor and insist on the validity of the classical field equation
\begin{equation}
G_{\mu \nu }\left( g^{\text{imp}}\right) =8\pi G_{0}\;T_{\mu \nu }^{\text{Q}}  \label{6.1}
\end{equation}
This equation then defines a vacuum energy momentum tensor which describes the
energy and momentum of a fictitious ``pseudo matter'' which reproduces the
quantum corrections found by the RG improvement by means of the conventional
Einstein equation. The explicit calculation yields, after a fair amount of algebra,
\begin{equation}
T_{\mu \nu }^{\text{Q}}\left( r,\theta \right) =\frac{M}{32\pi G_{0}\rho
^{6}\Delta }\left[
\begin{array}{cccc}
q_{1} & 0 & 0 & v \\
0 & q_{2} & 0 & 0 \\
0 & 0 & q_{3} & 0 \\
v & 0 & 0 & q_{4}
\end{array}
\right]  \label{6.2}
\end{equation}
with the entries ($n=$1,2,3,4)
\begin{eqnarray}
q_{n}\left( r,\theta \right) &\equiv &\alpha _{n}\left( r,\theta \right)
G^{\prime }\left( r\right) +\beta _{n}\left( r,\theta \right) G^{\prime
\prime }\left( r\right) \;  \label{6.3} \\
v\left( r,\theta \right) &\equiv &\alpha _{\nu }\left( r,\theta \right)
G^{\prime }\left( r\right) +\beta _{\nu }\left( r,\theta \right) G^{\prime
\prime }\left( r\right)  \notag
\end{eqnarray}
Here the coefficient functions are given by
\begin{eqnarray}
\alpha_{1}\left( r,\theta \right) &\equiv &-\left( a^{2}+r^{2}\right) \left[
8r^{2}\left( a^{2}+r^{2}\right) -a^{4}\left( \sin 2\theta \right) ^{2}\right]
\nonumber \\&&-16ra^{2}MG\sin ^{2}\theta \cos ^{2}\theta  \label{6.4} \\
\alpha _{2} &\equiv &8r^{2}\Delta ^{2}\;,\;\alpha _{3}\equiv 8\Delta
a^{2}\cos ^{2}\theta \\
\alpha _{4} &\equiv &\csc ^{2}\theta \alpha _{3}-8a^{2}r^{2}\;,\;\alpha
_{\nu }\equiv 8ar^{2}\left( r^{2}+a^{2}\right) -a\alpha _{3} \\
\beta _{1}\left( r,\theta \right) &\equiv &4\Delta r\rho ^{2}a^{2}\sin
^{2}\theta \;,\;\beta _{2}\left( r,\theta \right) \equiv 0
\end{eqnarray}
\begin{eqnarray}
\beta _{3}\left( r,\theta \right) &\equiv &4\Delta r\rho ^{2}\;,\;\beta
_{4}\left( r,\theta \right) \equiv 4\Delta r\rho ^{2}\csc ^{2}\theta \\
\beta _{\nu }\left( r,\theta \right) &\equiv &-4a\Delta r\rho ^{2}
\end{eqnarray}
The rows and columns of the $T_{\mu \nu }^{\text{Q}}$ matrix above are
ordered in the sequence $t-r-\theta -\varphi $. The matrix is diagonal
except for the $t\varphi $ entry. A nonzero value of $T_{t \varphi }^{\text{Q}}$ was to be expected, of course, since this corresponds precisely
to matter rotating about the $z$-axis.

It is not difficult to diagonalize $T_{\mu \nu }^{\text{Q}}.$ In its
eigenbasis it reads
\begin{equation}
T_{\mu \nu }^{\text{Q}}\left( r,\theta \right) =\frac{M}{32\pi G_{0}\rho
^{6}\Delta }\;\ \rm{diag}\left[
\begin{array}{cccc}
l_{1}, & l_{2}, & l_{3}, & l_{4} \end{array} \right]  \label{6.5}
\end{equation}
with the diagonal matrix elements
\begin{eqnarray}
l_{1} &\equiv &\frac{1}{2}\left[ q_{1}+q_{4}+\sqrt{q_{1}^{2}-2q_{1}q_{4}+q_{4}^{2}+4v^{2}}\right]  \label{6.6} \\
l_{2} &\equiv &q_{2}\;,\;l_{3}\equiv q_{3}  \notag \\
l_{4} &\equiv &\frac{1}{2}\left[q_{1}+q_{4}-\sqrt{q_{1}^{2}-2q_{1}q_{4}+q_{4}^{2}+4v^{2} }\right]  \notag
\end{eqnarray}

Despite the formal analogy it would be premature to conclude that the
vacuum quantum effects can be mimicked by the presence of matter. The reason
is that $T_{\mu \nu }^{\text{Q}}$ turns out to violate all the
positivity conditions which are usually assumed to be satisfied by
physically realizable matter \cite{Hawking-Ellis}. For a diagonalized energy momentum tensor 
$T_{\mu }^{\;\nu }$ $=\rm{diag}\left[
\begin{array}{cccc}
-\rho, & p_{1}, & p_{2}, & p_{3} \end{array}
\right] $ one distiguishes the following ``energy conditions'' \cite{Poisson,Hawking-Ellis}:
\begin{eqnarray}
\text{weak energy condition} &:&\text{ }\rho \geqslant 0\;,\;\rho +p_{i}>0
\label{6.7} \\
\text{null energy condition} &:&\;\rho +p_{i}\geqslant 0  \notag \\
\text{dominant energy condition} &:&\text{ }\rho \geqslant 0\;,\;\rho
\geqslant \left| p_{i}\right|  \notag \\
\text{strong energy condition} &:&\;\rho +p_{i}\geqslant 0\;,\;\rho
+\sum_{i}p_{i}\geqslant 0\;  \notag
\end{eqnarray}
From (\ref{6.5}) with (\ref{6.6}) we can read off the energy density $\rho $
and the pressures $p_{i},\;i=1,2,3,$ corresponding the energy momentum
tensor $T_{\mu \nu }^{\text{Q}}$. It is then straightforward to check
numerically whether or not the energy conditions (\ref{6.7}) are satisfied.
The result is that \textit{all four energy conditions are violated, at least in a part of the improved Kerr spacetime.} 

This result does not come completely unexpected; also the vacuum expectation value of
energy momentum operators (as in the case of the Casimir effect, for
instance) typically violates the energy conditions. As a consequence, the quantum gravity effects are qualitatively different
from those due to ordinary matter. From the practical point of view this
means that the analysis of the improved black hole does not reduce to
applying the many known results and theorems which are available for
classical black holes with matter. The reason is that in most cases their
derivation assumes the validity of one or the other of the conditions (\ref
{6.7}). For instance, for deriving the focusing theorem for timelike
geodesic congruences from Raychaudhuri's equation one needs the strong
energy condition \cite{Poisson}. Furthermore, the thermodynamics of the improved black
holes is \textit{not} a special case of the familiar (semi-) classical black
hole thermodynamics with matter.

\section{Dressing of mass and angular momentum}

The improved Kerr metric describes an isolated object in an asymptotically
flat spacetime. As this spacetime posesses the two Killing vectors $\boldsymbol{t}$ and $\boldsymbol{\varphi}$ we can
ascribe a mass and an angular momentum to this object by means of the Komar integrals \cite{Komar,Poisson}:

\begin{equation}
M_{\text{Komar}}=-\frac{1}{8\pi G_{0}}\int_{S}\nabla ^{\alpha }t^{\beta
}dS_{\alpha \beta }  \label{7.1}
\end{equation}
\begin{equation}
J_{\text{Komar}}=\frac{1}{16\pi G_{0}}\int_{S}\nabla ^{\alpha }\varphi
^{\beta }dS_{\alpha \beta }  \label{7.2_rep}
\end{equation}
Here $S$ is a two-sphere at spatial infinity. Its surface element $%
dS_{\alpha \beta }$ is given by $dS_{\alpha \beta }=-2n_{\left[ \alpha
\right. }r_{\left. \beta \right] }\sqrt{\sigma }d^{2}\theta $ where $%
n_{\alpha }$ and $r_{\alpha }$ are the timelike and spacelike normals to $S$%
. Here $\sigma $ is the determinant of $\sigma _{ab}$, the metric induced
from $g_{\alpha \beta }$ in the 2-d surface $S$, and $d^{2}\theta
\equiv d\theta ^{1}d\theta ^{2}$ with $\theta ^{a}$ angular coordinates on $%
\mathrm{S}$. The integrals for $M_{\mathrm{Komar}}$ and $J_{\mathrm{Komar}}$
probe the metric only at spatial infinity. Since the improved Kerr metric
equals the classical one far away from the black hole, the values of $M_{%
\mathrm{Komar}}$ and $J_{\mathrm{Komar}}$ are not changed by the RG
improvement. It is well known \cite{Poisson} that for the classical Kerr
metric they coincide with the mass and angular momentum parameters which it
contains:
\begin{equation}
M_{\mathrm{Komar}}=M\ ,\ J_{\mathrm{Komar}}=J  \label{7.3}
\end{equation}
Thus, for $S$ a surface at spatial infinity, (\ref{7.3}) holds true also
in the improved case.

The mass and angular momentum of the object as measured at infinity receives
a contribution from the pseudo-matter mimicking the quantum effects. To
identify it we break up $M_{\mathrm{Komar}}$ and $J_{\mathrm{Komar}}$ into
two pieces, one which contains only the effect of the pseudo-matter within
the outer horizon $\mathrm{H}\equiv H_+$, and one which is due to the matter
distribution outside $\mathrm{H}$. The first contribution yields quantities $%
M_{H}$ and $J_{H}$ which we refer to as the mass and angular momentum of the
black hole, meaning here only the portion of space bounded by $\mathrm{H}$. 
The second contribution describes the ``dressing'' of this intrinsic mass and angular momentum 
by matter external to the black hole.

The relation between the parameters $M$ and $J$ calculated at the spatial
infinity and the quantities $M_{\text{H}}$ and $J_{\text{H}}$ calculated at
the event horizon can be derived if we consider a 3-d spacelike hypersurface $\Sigma $ extending from the event horizon to spatial infinity. Its inner
boundary is $\text{H}$, a two dimensional cross section of the event
horizon, and its outer boundary is $S$. Using Gauss' theorem and the field
equation (\ref{6.1}) we find that $M$ and $J$ can be decomposed as:
\begin{eqnarray}
M &=&M_{\text{H}}+2\int_{\Sigma }\left( T^{\text{Q}}_{\alpha \beta }-\frac{1}{2}%
T^{\text{Q}}g_{\alpha \beta }\right) n^{\alpha }t^{\beta }\sqrt{h}d^{3}y  \label{7.4} \\
J &=&J_{\text{H}}-\int_{\Sigma }\left( T^{\text{Q}}_{\alpha \beta }-\frac{1}{2}%
T^{\text{Q}}g_{\alpha \beta }\right) n^{\alpha }\varphi ^{\beta }\sqrt{h}d^{3}y
\label{7.5}
\end{eqnarray}
Here $h_{ab}$ is the metric induced in $\Sigma $ and $y^{a}\ \left(
a=1,2,3\right) $ are coordinates intrisic to this hypersurface. $M_{\text{H}%
} $ and $J_{\text{H}}$ are the ``genuine'' black-hole mass and angular
momentum, respectively. They are given by surface integrals over $\text{H}$:
\begin{equation}
M_{\mathrm{H}}=-\frac{1}{8\pi G_{0}}\int_{\mathrm{H}}\nabla ^{\alpha
}t^{\beta }ds_{\alpha \beta }  \label{7.6}
\end{equation}
\begin{equation}
J_{\mathrm{H}}=\frac{1}{16\pi G_{0}}\int_{\mathrm{H}}\nabla ^{\alpha
}\varphi ^{\beta }ds_{\alpha \beta }  \label{7.7}
\end{equation}
The surface element $ds_{\alpha \beta }=2\xi _{\left[ \alpha \right.
}N_{\left. \beta \right] }\sqrt{\sigma }d^{2}\theta =\left( \xi _{\alpha
}N_{\beta }-\xi _{\beta }N_{\alpha }\right) \sqrt{\sigma }d^{2}\theta $
involves an auxiliary null vector $N_{\alpha }$ which satisfies $N_{\alpha
}\xi ^{\alpha }=-1$ and $N_{\alpha }N^{\alpha }=0$ \cite{Poisson}.

The relations (\ref{7.4}) and (\ref{7.5}) can be interpreted as follows: The
total mass $M$ (angular momentum $J$) is given by a contribution $M_{H}$ ($%
J_{H}$) from the black hole, plus a contribution from the matter
distribution outside. If the black hole is in vacuum, then $M=M_{H}$ and $%
J=J_{H}$. According to the discussion of section \ref{Sec Vac} we expect that $M_{H}\neq
M$ and $J_{H}\neq J$ when the contributions of the ``quantum fluid'' are
taken into account, i.e. that the mass and the angular momentum of the black
hole get ``renormalized'' or ``dressed'' by the matter surrounding it.
This interpretation is confirmed by an explicit evaluation of the integrals (%
\ref{7.6}) and (\ref{7.7}). The calculation is somewhat lengthy but similar
to the classical one. The final answer reads \cite{Tesis}
\begin{equation}
M_{H}=M\frac{G\left( r_{+}\right) }{G_{0}}\left\{ 1-\left[ \frac{\left(
r_{+}^{2}+a^{2}\right) G^{\prime }\left( r_{+}\right) }{aG\left(
r_{+}\right) }\right] \arctan \left( \frac{a}{r_{+}}\right) \right\}
\label{7.8}
\end{equation}
\begin{equation}
J_{H}=\left\{ J+\left[ 1-\frac{2MG\left( r_{+}\right) }{a}\arctan \left(
\frac{a}{r_{+}}\right) \right] \left[ \frac{M^{2}G^{\prime }\left(
r_{+}\right) r_{+}^{2}}{a}\right] \right\} \frac{G\left( r_{+}\right) }{G_{0}%
}  \label{7.9}
\end{equation}

These results have a number of remarkable properties: \\
\textbf{(A) }One can verify that for any pair of black hole parameters, $%
\left( M,J\right) $, the ratio $M_{H}/M$ is always \textit{smaller }than
unity; it approaches unity only asymptotically, for $M\rightarrow \infty $,
when the quantum effects become insignificant. The interpretation is that
the black hole posseses a ``genuine'' (positive) mass $M_{H}$ to which the
quantum matter adds another \textit{positive} contribution to make up the mass
measured at infinity, $M$. Given the fact that the pseudo matter satisfies no
standard positivity condition it is by no means trivial that $M$ is larger
than $M_{H}$. However this is exactly what one would expect if quantum
gravity is \textit{antiscreening:} the metric fluctuations dress any test
mass (here the black hole interior) in such a way that the mass increases
with the distance \cite{mr}. The same is found to hold true for the angular
momentum: $J_{H}/J$ is always smaller than unity, i.e. the pseudo matter
increases the spin of the test mass. \\
\textbf{(B) }Despite their somewhat complicated structure, the results (\ref{7.8}) and (\ref{7.9}) satisfy the same Smarr formula which is valid for
classical black holes \cite{Tesis,Larry}:
\begin{equation}
M_{\text{H}}=2\Omega _{\text{H}}J_{\text{H}}+\frac{\kappa \mathcal{A}}{4\pi
G_{0}}  \label{7.10}
\end{equation}
Here $\Omega _{\text{H}}$ and $\kappa $ are given by the ``improved''
equations (\ref{3.18}) and (\ref{3.28}), respectively, and $\mathcal{A}$
denotes the surface of the outer horizon H. Both in the classical and the
improved case it can be written as
\begin{equation}
\mathcal{A}=4\pi \left( r_{+}^{2}+a^{2}\right)  \label{7.11}
\end{equation}
but for improved black holes the dependence of $r_{+}$ on $M$ and $J$ (or $a$%
) is much more complicated. \\
\textbf{(C)} The results (\ref{7.8}), (\ref{7.9}) are strikingly similar to
the corresponding formulas for the \textit{classical} Kerr-Newman spacetime \cite{Poisson} which, besides mass and angular momentum, is characterized by an electric charge $Q$. The expressions coincide \textit{exactly} if we identify
\begin{equation}
Q^{2}\;\hat{=}\;2Mr_{+}^{2}G^{\prime }\left( r_{+}\right) /G_{0}  \label{7.12}
\end{equation}
This coincidence does not come completely unexpected. In \cite{bh1} where
the $a=0-$case had been analysed it turned out that the improved
Schwarzschild metric has many features in common with the classical
Reissner-Nordstr\"{o}m metric (a minimum of the lapse function $f\left(
r\right) $, causal structure, etc.). For $a\neq 0$ there is still a
corresponding similarity between the improved Kerr metric and the classical
Kerr-Newman spacetime. The exact coincidence of the Komar integrals is
somewhat surprising though. It is intriguing to speculate that it might
have a deeper meaning.

\section{A modified first law of black hole thermodynamics} \label{Seccion 8}

The first law of \textit{classical} black hole thermodynamics states that
the one-form $2\pi \left( \delta M-\Omega _{\text{H}}\delta J\right) /\kappa$ is exact, i.e. that it can be written as the differential of a state function $%
S=S\left( M,J\right) .$ Hence
\begin{equation}
\delta M-\Omega _{\text{H}}\delta J=T\delta S,  \label{8.1}
\end{equation}
where one interpretes
\begin{equation}
T\left( M,J\right) =\frac{\kappa \left( M,J\right) }{2\pi }  \label{8.2}
\end{equation}
and $S$ as the black hole temperature and entropy, respectively \cite{Bek1,Hawking-Bardeen-C}. In terms of
its surface area $\mathcal{A}$ the latter is given by $S=\mathcal{A}/4G_{0}$ \cite{Hawking Area}. For these results to hold the functions (zero forms) $\kappa $ and $\Omega_{\text{H}}$ must have a very special $M-$ and $J-$ dependence. In section 3
we found the corresponding relations for the improved case, namely
\begin{eqnarray}
\kappa \left( M,J\right) &=&\frac{r_{+}^{\text{I}}-M\left[ r_{+}^{\text{I}%
}G^{\prime }\left( r_{+}^{\text{I}}\right) +G\left( r_{+}^{\text{I}}\right) %
\right] }{\left( r_{+}^{\text{I}}\right) ^{2}+\left( J/M\right) ^{2}}
\label{8.3} \\
\Omega _{\text{H}}\left( M,J\right) &=&\frac{\left( J/M\right) }{\left(
r_{+}^{\text{I}}\right) ^{2}+\left( J/M\right) ^{2}}
\label{8.4}
\end{eqnarray}
Here $r_{+}^{\text{I}}\equiv r_{+}^{\text{I}}\left( M,J\right) $, but this
relationship cannot be written down in closed form.

In this section we analyze whether the RG improved black holes satisfy a
quantum corrected version of the first law (\ref{8.1}), and if so, how the
temperature and entropy get modified.

\subsection{Preliminaries}

The states an improved Kerr black hole can be in are labeled by the two
parameters $M$ and $J$. We visualize the corresponding state space as (part
of) the 2-dimensional euclidean plane with cartesian coordinates $x^{1}=M$, $%
x^{2}=J$. Using the convenient language of differential forms, state
functions are zero forms on this space, i.e. scalars $f=f\left( x\right)
\equiv f\left( M,J\right) $. Defining the exterior derivative as\footnote{%
To conform with the standard notation of thermodynamics we denote the
exterior derivative by $\delta $ rather than $d$.}
\begin{equation*}
\delta =\delta M\frac{\partial }{\partial M}+\delta J\frac{\partial }{%
\partial J}
\end{equation*}
a differential form $\boldsymbol{\alpha}$ is \textit{closed} if $\delta %
\boldsymbol{\alpha}=0$, and it is \textit{exact} if $\boldsymbol{\alpha}%
=\delta \boldsymbol{\beta}$ where $\boldsymbol{\beta}$ denotes a $\left(
p-1\right) $-form when $\boldsymbol{\alpha}$ is a $p$-form. The state space
being 2-dimensional, the only case of interest is $p=1$. A general 1-form
has the expansion $\boldsymbol{\alpha}=P\left( M,J\right) \delta M+N\left(
M,J\right) \delta J$. This 1-form is closed if
\begin{equation}
\frac{\partial P}{\partial J}=\frac{\partial N}{\partial M}  \label{8.5}
\end{equation}
and it is exact if there exists a zero-form $S\left( M,J\right) $ such that $%
\boldsymbol{\alpha}=\delta S$ or, in components, $P=\partial S/\partial
M\;,\;N=\partial S/\partial J$. We assume that the states $\left( M,J\right)
$ form a simply connected subset of the euclidean plane so that $\delta %
\boldsymbol{\alpha}=0$ is necessary and sufficient for the exactness of $%
\boldsymbol{\alpha}$.

If $\boldsymbol{\alpha}$ is not exact, one can try to find an integrating
factor $\mu \left( M,J\right) $ such that the product $\mu %
\boldsymbol{\alpha}$ is exact: $\mu \left( M,J\right) \boldsymbol{\alpha}%
=\delta S$. Hence $\delta \left( \mu \boldsymbol{\alpha}\right) =0$, or $%
\partial \left( \mu P\right) /\partial J=\partial \left( \mu N\right)
/\partial M$, which implies a quasi-linear partial differential equation for
the $0-$form $\mu \left( M,J\right)$ \cite{Simmons,Courant,Caratheodory}:
\begin{equation}
P\left( \frac{\partial \mu }{\partial J}\right) -N\left( \frac{\partial \mu
}{\partial M}\right) =\mu \left[ \left( \frac{\partial N}{\partial M}\right)
-\left( \frac{\partial P}{\partial J}\right) \right]  \label{8.6}
\end{equation}

\subsection{Does there exist an entropy-like state function?}

The $1-$form we are actually interested in is
\begin{equation}
\boldsymbol{\alpha}=\frac{2\pi }{\kappa \left( M,J\right) } \Bigl( \delta M-\Omega _{\text{H}}\left( M,J \right) \delta J\Bigr)  \label{8.7}
\end{equation}
with
\begin{equation}
P\equiv \frac{2\pi }{\kappa }\;,\;N\equiv -\frac{2\pi \Omega _{\text{H}}}{\kappa }  \label{8.8}
\end{equation}
involving the surface gravity and angular velocity of eqs. (\ref{8.3}) and (\ref{8.4}). The crucial question is whether $\boldsymbol{\alpha}$ is closed,
i.e. whether its components (\ref{8.8}) satisfy the integrability condition (%
\ref{8.5}). The explicit calculation reveals that for a generic $G\left(
r\right) $ this is actually \textit{not} the case: The $1-$form (\ref{8.7})
with the quantum corrected versions of $\kappa $ and $\Omega _{\text{H}}$ is
not closed and, as a consequence, not exact. (This calculation is
straightforward in principle, but rather tedious \cite{Tesis}. One has to be careful
about differentiating all the implicit $M-$ and $J-$ dependencies that enter
via $r_{+}^{\text{I}}\left( M,J\right) $. One does not need the explicit
form of this function; its partial derivatives can be expressed in terms of $%
r_{+}^{\text{I}}$ itself by differentiating the horizon condition $\Delta\left( r_{+}^{\text{I}}\right) =0$.)

As $\boldsymbol{\alpha}$ is not exact in the improved case we must conclude
that there does \textit{not} exist a differential relation of the type
\begin{equation}
\delta M-\Omega _{\text{H}}\delta J=\left( \frac{\kappa }{2\pi }\right)
\delta \left( \frac{\mathcal{A}}{4G_{0}}+\text{quantum corrections}\right)
\label{8.9}
\end{equation}
which could play the role of a modified first law for quantum black holes.
The interpretation of (\ref{8.9}) would have been clear: The
Bekenstein-Hawking temperature of the improved black holes is related to the
surface gravity by $T=\kappa /2\pi $, as in the classical case, and there
exists a state function $S\left( M,J\right) $ which equals the classical $%
\mathcal{A}/4G_{0}$ plus correction terms. Since $\boldsymbol{\alpha}$ is
actually not exact we must conclude that \textit{either} \textit{there
exists no entropy-like state function for the improved black holes or the
classical relation }$T=\kappa /2\pi $ \textit{does not hold true for them.}

We see that for quantum Kerr black holes even the very existence of an
entropy is a nontrivial issue. The situation was different for the improved
Schwarzschild black holes \cite{ bh2}. Since there the state space is
1-dimensional, $\boldsymbol{\alpha}\equiv \left( 2\pi /\kappa \right) \delta
M$ is trivially exact, $T=\kappa /2\pi $ continues to be valid and the
entropy one finds has indeed the structure $\mathcal{A}/4G_{0}+$quantum
corrections \cite{bh2,evap}.

Thus we are led to conclude that if there exists a modified, i.e. quantum
version of black hole thermodynamics which is accesible by RG improvement
then the temperature cannot be simply proportional to the surface gravity, $%
T\neq \kappa /2\pi $. While a priori it is perhaps not very surprising that
the semi-classical relation $T=\kappa /2\pi $ is subject to quantum gravity
correction this causes a difficulty of principle. Within the present
approach we were able to find the corrected $M-$ and $J-$ dependence of $%
\kappa $ and $\Omega _{\text{H}}$ and, as a result, we know the corrected $%
1- $form $\boldsymbol{\alpha}$. However without additional input, knowledge
of $\boldsymbol{\alpha}$ is not enough to deduce the two functions $T\left(
M,J\right) $ and $S\left( M,J\right) $. There exist infinitely many pairs $%
\left( T,S\right) $ such that $\boldsymbol{\alpha}=T\delta S$ for a
prescribed $\boldsymbol{\alpha}$. In a full-fledged quantum gravity version
of black hole thermodynamics it might be possible to find the ``correct'' one, presumably.

A general theory of this kind is beyond the scope of the present paper. Here
we only consider the possible structure of a modified first law. As we shall
see in the next subsection, progress can be made by restricting the
discussion to black holes of small angular momentum. To leading order in a $%
J^{2}$ expansion the corrections to the temperature and entropy are found to
be uniquely fixed.

\subsection{Temperature and entropy to order $J^{2}$}

By time reflection symmetry, the small$-J$ expansions of the temperature and
entropy read
\begin{equation}
T\left( M,J\right) =T_{0}\left( M\right) +T_{2}\left( M\right) J^{2}+O\left(
J^{4}\right)  \label{8.10}
\end{equation}

\begin{equation}
S\left( M,J\right) =S_{0}\left( M\right) +S_{2}\left( M\right) J^{2}+O\left(
J^{4}\right)  \label{8.11}
\end{equation}
The terms of lowest order, $T_{0}\left( M\right) $ and $S_{0}\left( M\right)
$, refer to the RG improved Schwarzschild spacetime \cite{bh2,evap}. They
satisfy $\delta M=T_{0}\delta S_{0}$ or $1/T_{0}\left( M\right)
=dS_{0}\left( M\right) /dM$. In \cite{bh2} this relation has been integrated
in order to find the entropy of the improved Schwarzschild black hole:
\begin{equation}
S_{0}=\int_{M_{\text{cr}}}^{M}\frac{dM^{\prime }}{T_{0}\left( M^{\prime
}\right) }  \label{8.12}
\end{equation}
In the approximation $d\left( r\right) =r$ the
temperature was found to be given by
\begin{eqnarray}
T_{0}\left( M\right) &=&\frac{1}{4\pi G_{0}M_{\text{cr}}}\frac{\sqrt{Y\left(
1-Y\right) }}{1+\sqrt{1-Y}}  \label{8.13} \\
&=&\frac{1}{8\pi G_{0}M}\left[ 1-\frac{1}{4}\left( \frac{M_{\text{cr}}}{M}%
\right) ^{2}-\frac{1}{8}\left( \frac{M_{\text{cr}}}{M}\right) ^{4}+O\left(
M^{-6}\right) \right] \nonumber
\end{eqnarray}
Here $Y\equiv M_{\text{cr}}^{2}/M^{2}$ and $M_{\text{cr}}\equiv \sqrt{\bar{w}}\;m_{\rm Pl}$. 
(The ``critical'' mass $M_{\text{cr}}$ is the smallest mass
for which the improved Schwarzschild spacetime has an event horizon \cite
{bh2}.) Using (\ref{8.13}) in (\ref{8.12}) yields
\begin{eqnarray}
S_{0}\left( M\right) &=&S_{0}\left( M_{\text{cr}}\right) +2\pi \bar{w}\left[
Y^{-1}\sqrt{1-Y}\left( 1+\sqrt{1-Y}\right) +\arctan \sqrt{1-Y}\right]
\notag \\
&=&S_{0}\left( M_{\text{cr}}\right) +\frac{\mathcal{A}_{\text{Class}}^{\text{%
Sch}}}{4G_{0}}+\notag \\ &&2\pi \bar{w}\left[ \ln \left( \frac{2M}{M_{\text{cr}}}\right)
-\frac{3}{2}-\frac{3}{8}\left( \frac{M_{\text{cr}}}{M}\right) ^{2}-\frac{5}{%
32}\left( \frac{M_{\text{cr}}}{M}\right) ^{4}+O\left( M^{-6}\right) \right] \label{8.14}
\end{eqnarray}
Here $\mathcal{A}_{\text{Class}}^{\text{Sch}}\equiv 4\pi \left(
2G_{0}M\right) ^{2}$ is the classical Schwarzschild surface area. The first
few terms of the large$-M$ expansions given in (\ref{8.13}) and (\ref{8.14})
are rather reliable predictions probably since for $M\gg m_{\rm Pl}$ the
classical spacetime is only weakly distorted by quantum effects.

Next we try to determine $T_{2}$ and $S_{2}$ such that $\delta M-\Omega _{%
\text{H}}\delta J=T\delta S$ is satisfied to order $J^{2}$. Inserting the
ans\"{a}tze for $T$ and $S$ we have
\begin{eqnarray}
\delta M-\Omega _{\text{H}}\delta J &=&\left[ T_{0}\left( M\right)
+T_{2}\left( M\right) J^{2}\right] \delta \left[ S_{0}\left( M\right)
+S_{2}\left( M\right) J^{2}\right]  \label{8.15} \\
&=&T_{0}\delta S_{0}+\delta S_{0}T_{2}J^{2}+\delta \left( S_{2}J^{2}\right)
T_{0}+O\left( J^{3}\right)  \notag
\end{eqnarray}
Exploiting that $T_{0}\delta S_{0}=\delta M$ we are left with
\begin{eqnarray}
-\Omega _{\text{H}}\delta J &=&\delta S_{0}T_{2}J^{2}+\delta \left(
S_{2}J^{2}\right) T_{0}+O\left( J^{3}\right)  \label{8.16} \\
&=&T_{2}J^{2}\left( \frac{dS_{0}}{dM}\right) \delta M+T_{0}\left[
J^{2}\left( \frac{dS_{2}}{dM}\right) \delta M+2JS_{2}\delta J\right]
+O\left( J^{3}\right)  \notag
\end{eqnarray}
Equating the coefficients of $\delta J$ and $\delta M$ we find the following
two coupled equations which determine $S_{2}$ and $T_{2}$:
\begin{equation}
T_{2}\left( \frac{dS_{0}}{dM}\right) +T_{0}\left( \frac{dS_{2}}{dM}\right)
=0+O\left( J^{4}\right)  \label{8.17}
\end{equation}
\begin{equation}
2JT_{0}S_{2}+\Omega _{\text{H}}=0+O\left( J^{3}\right)  \label{8.18}
\end{equation}
In eq. (\ref{8.18}) we need $\Omega _{\text{H}}$ to linear order in $J$
only. From (\ref{8.4}) we obtain
\begin{equation}
\Omega _{\text{H}}\left( M,J\right) =\frac{J}{Mr_{\text{Sch}+}^{\text{I}%
}\left( M\right) ^{2}}+O\left( J^{3}\right)  \label{8.19}
\end{equation}
where $r_{\text{Sch}+}^{\text{I}}\equiv r_{+}^{\text{I}}\left( J=0\right) $
refers to the improved Schwarzschild black hole. In the approximation $%
d\left( r\right) =r$ we are using here this radius is explicitly given by
\cite{bh2}
\begin{equation}
r_{\text{Sch}+}^{\text{I}}=G_{0}M\left[ 1+\sqrt{1-Y}\right]  \label{8.20}
\end{equation}
With (\ref{8.19}) in (\ref{8.18}) we can solve for the function $S_{2}$:
\begin{equation}
S_{2}\left( M\right) =-\Bigl[ 2M\;T_{0}\left( M\right) r_{\text{Sch}+}
^{\text{I}}\left( M\right) ^{2}\Bigr]^{-1} \label{8.21}
\end{equation}
Furthermore, taking advantage of $dS_{0}/dM=1/T_{0}$ again, we can solve (%
\ref{8.17}) for $T_{2}$ in terms of the, by now known, function $S_{2}$:
\begin{equation}
T_{2}\left( M\right) =-T_{0}\left( M\right) ^{2}\;\frac{dS_{2}\left( M\right)
}{dM}  \label{8.22}
\end{equation}
In deriving the relations (\ref{8.21}) and (\ref{8.22}) we were able to find
a well defined and \textit{unique} answer for the coefficients of the $%
J^{2}- $terms. Eq. (\ref{8.21}) for $S_{2}\left( M\right) $ involves only
the known Schwarzschild quantities $T_{0}$ and $r_{\text{Sch}+}^{\text{I}}$,
and once $S_{2}$ is known also $T_{2}$ is completely fixed by eq. (\ref{8.22}%
).

Using the results from the Schwarzschild case we obtain the following final
result for the temperature and entropy to order $J^{2}$:
\begin{eqnarray}
T\left( M,J\right) &=&\frac{1}{8\pi G_{0}M}\left[ 1-\frac{1}{4}\left( \frac{%
M_{\text{cr}}}{M}\right) ^{2}-\frac{1}{8}\left( \frac{M_{\text{cr}}}{M}%
\right) ^{4}+O\left( M^{-6}\right) \right]  \label{8.23} \\
&&-\frac{J^{2}}{32\pi M^{5}G_{0}^{3}}\left[ 1+\left( \frac{M_{\text{cr}}}{M}%
\right) ^{2}+\frac{15}{16}\left( \frac{M_{\text{cr}}}{M}\right) ^{4}+O\left(
M^{-6}\right) \right] +O\left( J^{4}\right) \notag
\end{eqnarray}
\begin{eqnarray}
S\left( M,J\right) &=&\frac{\mathcal{A}_{\text{class}}^{\text{Sch}}}{4G_{0}}%
+2\pi \bar{w}\left[ \ln \left( \frac{2M_{\text{cr}}}{M}\right) -\frac{3}{2}-%
\frac{3}{8}\left( \frac{M_{\text{cr}}}{M}\right) ^{2}-\frac{5}{32}\left(
\frac{M_{\text{cr}}}{M}\right) ^{4}+O\left( M^{-6}\right) \right] \notag \\
&&-\left( \frac{\pi J^{2}}{M^{2}G_{0}}\right) \left[ \allowbreak 1+\frac{3}{4%
}\left( \frac{M_{\text{cr}}}{M}\right) ^{2}+\frac{5}{8}\left( \frac{M_{\text{%
cr}}}{M}\right) ^{4}+O\left( M^{-6}\right) \right] +O\left( J^{4}\right) \label{8.24}
\end{eqnarray}
In writing down the result for the entropy we fixed the undetermined
constant of integration such that $S=0$ for $M=M_{\text{cr}}$ and $J=0$.

We observe that the angular momentum dependent terms in (\ref{8.23}) and (%
\ref{8.24}) \textit{decrease} both the black hole's temperature and entropy
as compared to the corresponding Schwarzschild quantities. We also see that
the size of the $J^{2}$-corrections increases with $M_{\text{cr}}/M$, i.e.
these corrections grow as the mass $M$ of the black hole becomes smaller
during the evaporation process.

In summarizing the most important aspects of the modified black hole
thermodynamics discussed in this section we recall that $2\pi T$ does not
agree with the surface gravity $\kappa $ here as it is the case in the familiar
(semi-) classical situation. We demonstrated that a modified first law can
exist only when we give up the relationship $T=\kappa /2\pi $. We also
showed that, to order $J^{2}$, there is a uniquely determined modification
of this relationship which allows for the existence of a state function $%
S\left( M,J\right) $ with the interpretation of an entropy.

\section{Summary and conclusion}

In this paper we tried to assess the impact of the leading quantum gravity
corrections on the properties of rotating black holes within the framework
of Quantum Einstein Gravity (QEG). Using the gravitational average action as
the basic tool we developed a scale-dependent picture of the spacetime
structure. We exploited that $\Gamma _{k}$ is a family of effective field
theories labeled by $k$. More precisely, to each point $\mathcal{P}$ we associated a
coarse-graining scale $k=k\left(\mathcal{P}\right) $ and then described a
neighborhood of $\mathcal{P}$ by the specific effective action $\Gamma _{k\left(\mathcal{P}\right) }$. In principle there could be several plausible choices of the
map $\mathcal{P}\rightarrow k\left( \mathcal{P}\right) $. They lead to different ``pictures'' of
the same physical system. Using the analogy of a microscope with a variable
resolving power \cite{avactrev} we are using a microscope with a position
dependent resolving power, and clearly the ``picture'' we see depends on how
we change the resolving power from point to point. For the black hole the
choice of $k\left( \mathcal{P}\right) $ is made less ambiguous than for a generic
spacetime since we would like the ``picture'', the improved metric, to have
the same symmetries as the classical metric. We chose $k\left( \mathcal{P}\right) $ to
be monotonically decreasing in the radial direction, giving the best
``resolution'' to points near the black hole and the worst to those
asymptotically far away. The experience with similar ``RG improvements''
indicates that in this way the improved metric encodes the leading quantum
corrections at least at a qualitative level.

The results we obtained can be summarized as follows. In general the quantum
corrections are small for heavy black holes ($M\gg m_{\rm Pl}$), but become
appreciable for light ones. Heavy quantum black holes have the same number
of critical surfaces as the classical ones, namely two static limit surfaces
and two horizons. (For $J=0$ the improvement had led to the formation of a
new horizon.) As one lowers $M$ towards the Planck mass, the two horizons
coalesce and then disappear. At an even smaller mass the static limit
surfaces coalesce and then disappear as well. 

Even though the reliability of the improvement method becomes questionable when the corrected metric is 
very different from the classical one we believe that the disappearence of
the horizons below a certain critical mass is a fairly reliable prediction.
In fact, this phenomenon has a very simple interpretation: The existence of
a horizon means that the gravitational field is so strong that it can trap
light; if, however, the strength of the gravitational interaction is reduced
at small distances by the RG running of $G$ then it is quite plausible that
very small objects with a low mass cannot prevent light from scaping. 

Whether or not these objects have a naked singularity remains an open question. The method used
here is likely to loose its validity close to the black hole's center.
Also on the basis of earlier investigations \cite{bh2}, it is
likely though that the quantum corrections soften the singularity (again
because $G$ is ``switched off'' at short distances); it is even conceivable
that it disappears altogether \cite{bh2}.

A particularly intriguing feature of the Kerr black hole is the possibility
of energy extraction. As the Penrose process is related to the existence of
negative energy states of test particles we analyzed the ``phase space'' of
such negative energy states in detail. In paticular we saw that, while it is
possible to extract energy from classical black holes of arbitrary small
mass and angular momentum, in the improved Kerr spacetime there exists a
minimum mass for energy extraction. It is defined by the extremal
configuration of the static limit surfaces.

We explained that even though the quantum black holes in the vacuum can be
reinterpreted as classical black holes in presence of a special kind of
matter mimicking the quantum fluctuations, many of their mechanical and in
particular thermodynamical properties are nevertheless nonstandard since
this ``pseudo matter'' does not satisfy any of the familiar energy
conditions.

As a first step towards an ``RG improved black hole thermodynamics'' we
analyzed the problem of identifying a state function which could possibly be
interpreted as an entropy. We saw that in the quantum case the 1-form $%
\left( \delta M-\Omega _{H}\delta J\right) /\kappa $ is no longer exact or,
stated differently, the surface gravity is not an integrating factor of $%
\delta M-\Omega _{H}\delta J$. We concluded that if an entropy is to exist
also for the improved black hole, their temperature cannot simply be
proportional to $\kappa $. We also saw that for small angular momentum, to
order $J^{2}$, there exist unambiguously defined modified relationships for
the $M$- and $J$-dependence of temperature and entropy. We hope to come back
to a more detailed discussion of these thermodynamical issues elsewhere.

\section{Acknowledgments}
E. T. Would like to thank the German Service of Academic Exchange (DAAD) and the Institute of Physics in Mainz for the financial support during the development of his \newline Ph. D. thesis.

%\pagebreak

\end{document}